\documentclass[12pt,onecolumn,notitlepage]{article}
\usepackage{amssymb}
\usepackage{amsfonts}
\usepackage{amsmath}
\usepackage{geometry}
\usepackage{setspace}
\usepackage{portland}
\usepackage{scalefnt}
\usepackage{lineno}
\usepackage{indentfirst}
\usepackage{apacite}
\usepackage{apalike}
\usepackage{fancyhdr}
\usepackage[font={small},labelfont={small}]{caption}
\usepackage{float}
\usepackage{color}
\usepackage[USenglish]{babel}
\usepackage{sectsty}

\allsectionsfont{\normalsize}
\input{tcilatex}
\begin{document}

\title{Fast Marginal Likelihood Estimation of the\\
Ridge Parameter(s) in Ridge Regression\\
and Generalized Ridge Regression for Big\ Data}
\author{George Karabatsos\thanks{%
1040 W. Harrison St. (MC\ 147), Chicago, Illinois, 60304, U.S.A. E-mail:\
gkarabatsos1@gmail.com. Phone:\ (312) 413-1816. FAX:\ (312) 996-5651.} \\
University of Illinois-Chicago}
\maketitle

\begin{abstract}
Unlike the ordinary least-squares (OLS) estimator for the linear model, a
ridge regression linear model provides coefficient estimates via shrinkage,
usually with improved mean-square and prediction error.\ This is true
especially when the observed design matrix is ill-conditioned or singular,
either as a result of highly-correlated covariates or the number of
covariates exceeding the sample size. This paper introduces novel and fast
marginal maximum likelihood (MML)\ algorithms for estimating the shrinkage
parameter(s) for the Bayesian ridge and power ridge regression models, and
an automatic plug-in MML\ estimator for the Bayesian generalized ridge
regression model. With the aid of the singular value decomposition of the
observed covariate design matrix, these MML estimation methods are quite
fast even for data sets where either the sample size ($n$) or the number of
covariates ($p$)\ is very large, and even when $p>n$. On several real data
sets varying widely in terms of $n$ and $p$, the computation times of the
MML\ estimation methods for the three ridge models, respectively, are
compared with the times of other methods for estimating the shrinkage
parameter in ridge, LASSO\ and Elastic Net (EN)\ models, with the other
methods based on minimizing prediction error according to cross-validation
or information criteria. Also, the ridge, LASSO, and EN\ models, and their
associated estimation methods, are compared in terms of prediction accuracy.
Furthermore, a simulation study compares the ridge models under MML
estimation, against the LASSO and EN\ models, in terms of their ability to
differentiate between truly-significant covariates (i.e., with non-zero
slope coefficients) and truly-insignificant covariates (with zero
coefficients).\newline
\end{abstract}

\newpage

\section{Introduction}

The linear regression model is ubiquitous in statistics. To begin to set the
notational scene for the rest of the paper, we first give a brief overview
of concepts related to this model. The linear model is defined by $\mathbf{y}%
=\mathbf{X}\boldsymbol{\beta }+\boldsymbol{\varepsilon }$, for observed
dependent variable observations $\mathbf{y}=(y_{1},\ldots ,y_{n})^{\intercal
}$, with $\mathbf{X}=(x_{ik})_{n\times p}$ the observed $n\times p$ design
matrix of $p$ covariates having sample covariance matrix $\frac{1}{n}\mathbf{%
X}^{\mathbf{\intercal }}\mathbf{X}$, with $\boldsymbol{\beta }=(\beta
_{1},\ldots ,\beta _{p})^{\intercal }\in 
\mathbb{R}
^{p}$ the $p$ slope coefficient parameters. Also, $\boldsymbol{\varepsilon }%
=(\varepsilon _{1},\ldots ,\varepsilon _{n})^{\intercal }$ is the vector of
regression errors with mean (expectation)\ \textrm{E}$(\varepsilon )=0$ and
variance \textrm{Var}$(\varepsilon )=\sigma ^{2}\in 
\mathbb{R}
_{+}$. The ordinary least-squares (OLS)\ estimator of the linear model is
given by $\widehat{\boldsymbol{\beta }}=(\mathbf{X}^{\intercal }\mathbf{X}%
)^{-1}\mathbf{X}^{\intercal }\mathbf{y}$ and $\widehat{\sigma }^{2}=\tfrac{1%
}{n-p}||\mathbf{y}-\mathbf{X}\widehat{\boldsymbol{\beta }}||^{2}$.
Throughout, with no loss of generality\footnote{%
The $p$ coefficient estimates $\widehat{\boldsymbol{\beta }}$ obtained from
the rescaled data $\mathcal{D}_{n}=(\mathbf{X},\mathbf{y})$ can be
transformed back to intercept and slope coefficient estimates $\widehat{%
\boldsymbol{\beta }}^{\ast }=(\widehat{\beta }_{0}^{\ast },\widehat{\beta }%
_{1}^{\ast },\ldots ,\widehat{\beta }_{p}^{\ast })^{\intercal }$ on the
scale of the original data set $\underline{\mathcal{D}}_{n}=(\underline{%
\mathbf{X}},\underline{\mathbf{y}})$, via $\widehat{\boldsymbol{\beta }}%
^{\ast }=\left( \frac{1}{n}\tsum\nolimits_{i=1}^{n}\overline{y}_{i}-\left( 
\widehat{\boldsymbol{\beta }}\circ \boldsymbol{\sigma }_{\underline{\mathbf{X%
}}}^{-1}\right) ^{\intercal },\left( \widehat{\boldsymbol{\beta }}\circ 
\boldsymbol{\sigma }_{\underline{\mathbf{X}}}^{-1}\right) ^{\intercal
}\right) ^{\intercal }$, where $(\boldsymbol{\mu }_{\underline{\mathbf{X}}},%
\boldsymbol{\sigma }_{\underline{\mathbf{X}}})$ give the $p$ column means
and standard deviations of the columns of $\underline{\mathbf{X}}=(%
\underline{x}_{ik})_{n\times p}$, respectively. The symbol $\circ $ refers
to the Hadamard product operator.}, it is assumed that all variables have
been rescaled to have zero-mean with $\tsum\nolimits_{i=1}^{n}y_{i}=0$, and
that $\mathbf{X}$ is in correlation form such that $\tsum%
\nolimits_{i=1}^{n}x_{ki}=0$ and $\tsum\nolimits_{i=1}^{n}x_{ki}^{2}=1$, for 
$k=1,\ldots ,p$. Then the estimate of the intercept parameter $\beta _{0}$
becomes zero and ignorable.

The linear model satisfies the equivalence relation $\mathbf{y}=\mathbf{X}%
\boldsymbol{\beta }+\boldsymbol{\varepsilon }=\mathbf{XW}\boldsymbol{\alpha }%
+\boldsymbol{\varepsilon }$, where the latter equation gives the canonical
linear model (Searle, 1982\nocite{Searle82}), defined on the orthogonalized
space, with coefficients $\boldsymbol{\alpha }=\left( \alpha _{1},\ldots
,\alpha _{q}\right) ^{\intercal }\in 
\mathbb{R}
^{q}$ and $q=\min (n,p)$. Specifically the orthogonalized space is based on
a singular value decomposition (s.v.d.)\ of $\mathbf{X}$, given by $\mathbf{X%
}=\mathbf{UDW}^{\intercal }$, where $\mathbf{U}$ and $\mathbf{W}$ are
orthogonal matrices of dimensions $n\times q$ and $p\times q$, respectively;
the columns of $\mathbf{XW}$ give the $q$ principal components of $\mathbf{X}
$; and $\mathbf{D}=\mathrm{diag}(d_{1},\ldots ,d_{q})$ is the diagonal
matrix of singular values $d_{1}>\cdots >d_{q}$, where $(d_{1}^{2},\ldots
,d_{q}^{2})$ provide at most the first $q\leq p$ non-zero eigenvalues $%
(d_{1}^{2},\ldots ,d_{p}^{2})$ of $\mathbf{X}^{\mathbf{\intercal }}\mathbf{X}
$. The OLS estimator of $\boldsymbol{\alpha }$ is given by $\widehat{%
\boldsymbol{\alpha }}=\mathbf{D}^{-1}\mathbf{U}^{\intercal }\mathbf{y}$,
giving the OLS estimator $\widehat{\boldsymbol{\beta }}=\mathbf{W}\widehat{%
\boldsymbol{\alpha }}$ for the original data space.

It is well-known that for the given population parameters $\left( 
\boldsymbol{\beta },\sigma ^{2}\right) $, the OLS estimator $\widehat{%
\boldsymbol{\beta }}$ is unbiased, and has mean-squared error (and hence
total variance)\ given by \textrm{MSE}$(\widehat{\boldsymbol{\beta }}\,|\,%
\boldsymbol{\beta })=\sigma ^{2}\tsum\nolimits_{k=1}^{p}d_{k}^{-2}$. Also, $%
\mathbf{X}\widehat{\boldsymbol{\beta }}$ has mean-squared error \textrm{MSE}$%
(\mathbf{X}\widehat{\boldsymbol{\beta }}\,|\,\boldsymbol{\beta })=\mathbb{E}%
\{(\mathbf{X}\boldsymbol{\beta }-\mathbf{X}\widehat{\boldsymbol{\beta }}%
)^{2}\}=(\widehat{\boldsymbol{\beta }}-\boldsymbol{\beta })^{\intercal }%
\mathbf{V}_{\mathbf{X}}(\widehat{\boldsymbol{\beta }}-\boldsymbol{\beta })$.
The prediction error is \textrm{PE}$(\mathbf{X}\widehat{\boldsymbol{\beta }}%
\,|\,\boldsymbol{\beta })=\mathrm{MSE}(\mathbf{X}\widehat{\boldsymbol{\beta }%
})+\sigma ^{2}$, with expected value ($\mathbb{E}$) taken over the joint
population distribution of $\left( \mathbf{X},\mathbf{y}\right) $ with $%
\widehat{\boldsymbol{\beta }}$ fixed, and $\mathbf{V}_{\mathbf{X}}$ is the
population covariance matrix of $\mathbf{X}$. The matrix $\mathbf{X}^{%
\mathbf{\intercal }}\mathbf{X}$ is positive-definite if and only all $p$ of
its eigenvalues are positive, in which case the OLS estimator $(\widehat{%
\boldsymbol{\beta }},\widehat{\sigma }^{2})$ clearly has finite \textrm{MSE}$%
(\widehat{\boldsymbol{\beta }})$. If also $\mathbf{X}^{\mathbf{\intercal }}%
\mathbf{X}=\mathbf{I}_{p}$ with $\mathbf{I}_{p}$ the $p$-dimensional
identity matrix, leading to determinant $|\mathbf{X}^{\mathbf{\intercal }}%
\mathbf{X}|$ of $1$, then $\mathbf{X}$ is called orthogonal (orthonormal).
The number $0\leq |\mathbf{X}^{\mathbf{\intercal }}\mathbf{X}|\,\leq 1$
measures the orthogonality of $\mathbf{X}$ (Cooley \& Lohnes, 1971\nocite%
{CooleyLohnes71}).

The OLS estimator $\widehat{\boldsymbol{\beta }}$ can be poorly determined
and exhibit high MSE when $\mathbf{X}$ exhibits multicollinearity due to the
presence of highly-correlated covariates. In this case, $\mathbf{X}^{\mathbf{%
\intercal }}\mathbf{X}$ is ill-conditioned, with $|\mathbf{X}^{\mathbf{%
\intercal }}\mathbf{X}|$ near $0$, and smallest eigenvalue $%
d_{p}^{2}\rightarrow 0$ leading to \textrm{MSE}$(\widehat{\boldsymbol{\beta }%
}\,|\,\boldsymbol{\beta })\rightarrow \infty $ and \textrm{PE}$(\mathbf{X}%
\widehat{\boldsymbol{\beta }}\,|\,\boldsymbol{\beta })\rightarrow \infty $.
The OLS estimator $\widehat{\boldsymbol{\beta }}$ does not even exist when
two or more covariates in $\mathbf{X}$ are perfectly correlated, or when the
number of covariates exceeds the number of observations (i.e., when $p>n$).
In this case, the matrix $\mathbf{X}^{\mathbf{\intercal }}\mathbf{X}$ is
singular, and is thus not of full rank (i.e., $\mathrm{rank}(\mathbf{X}^{%
\mathbf{\intercal }}\mathbf{X})<p$) with $|\mathbf{X}^{\mathbf{\intercal }}%
\mathbf{X}|\,=0$. However, these days, applied statistical research is often
concerned with the analysis of large data sets where either the number of
variables $p$ or observations $n$ is very large. Specifically, a data set
with a large number of covariates ($p$)\ can likely give rise to singular or
ill-conditioned $\mathbf{X}^{\mathbf{\intercal }}\mathbf{X}$, as a result of
the number of covariates exceeding the sample size (i.e., $p\geq n$), and/or
as a result of multicollinearity. For example, multicollinearity can likely
occur when the data consist of a large number of covariates ($p$), or can
occur when the data set is constructed by merging multiple data sets that
have one or more variables in common.

Ridge regression, which defines the alternative estimator $\overline{%
\boldsymbol{\beta }}_{\lambda }=(\mathbf{X}^{\intercal }\mathbf{X}+\lambda 
\mathbf{I}_{p})^{-1}\mathbf{X^{\intercal }y}$, provides a solution to the
inability of the OLS estimator $\widehat{\boldsymbol{\beta }}$ to handle
ill-conditioned or singular $\mathbf{X}^{\mathbf{\intercal }}\mathbf{X}$.
Here, $\lambda >0$ is the ridge parameter that shrinks the estimate of the
coefficients $\boldsymbol{\beta }$ towards zero, with the amount of
shrinking an increasing function of $\lambda $. Compared to the OLS
estimator $\widehat{\boldsymbol{\beta }}$, the ridge estimator $\overline{%
\boldsymbol{\beta }}_{\lambda }$ introduces some bias in exchange for lower
mean-squared \textrm{MSE}$(\widehat{\boldsymbol{\beta }}_{\lambda }\,|\,%
\boldsymbol{\beta })$ and prediction error \textrm{PE}$(\mathbf{X}\widehat{%
\boldsymbol{\beta }}_{\lambda }\,|\,\boldsymbol{\beta })$, especially when $%
\mathbf{X}^{\mathbf{\intercal }}\mathbf{X}$ is ill-conditioned or singular
(Hoerl \&\ Kennard, 1970\nocite{HoerlKennard70}). However, unlike the OLS
estimator, the ridge estimator $\overline{\boldsymbol{\beta }}_{\lambda }$
exists even when $\mathbf{X}^{\intercal }\mathbf{X}$ is singular, because $(%
\mathbf{X}^{\intercal }\mathbf{X}+\lambda \mathbf{I}_{p})$ is necessarily
non-singular when $\lambda >0$.

The ridge estimator $\overline{\boldsymbol{\beta }}_{\lambda }$ can be
characterized in at least four equivalent ways (e.g., Hastie, et al. 2009%
\nocite{HastieTibsFriedman09}). First, the ridge estimator is equivalent to
the penalized least-squares estimator, $\overline{\boldsymbol{\beta }}%
_{\lambda }=\arg \min_{\boldsymbol{\beta }}||\mathbf{y}-\mathbf{X}%
\boldsymbol{\beta }||^{2}+\lambda \tsum\nolimits_{k=1}^{p}\beta _{k}^{2}$.
Second, in terms of the orthogonalized space, the ridge estimator is given
by $\overline{\boldsymbol{\beta }}_{\lambda }=\mathbf{W}\widehat{\boldsymbol{%
\alpha }}_{\lambda }$, with $\widehat{\alpha }_{\lambda
,k}=\{d_{k}^{2}/(d_{k}^{2}+\lambda )\}\widehat{\alpha }_{k}$ for $k=1,\ldots
,q$. This shows that the ridge estimator $\overline{\boldsymbol{\beta }}%
_{\lambda }$ is obtained by shrinking the canonical OLS estimator $\widehat{%
\boldsymbol{\alpha }}=\left( \widehat{\alpha }_{1},\ldots ,\widehat{\alpha }%
_{q}\right) ^{\intercal }$ by the factors $\{d_{k}^{2}/(d_{k}^{2}+\lambda
)\} $ (resp.), and therefore applies a greater amount of shrinkage to OLS
estimates $\widehat{\alpha }_{k}$ having relatively small eigenvalues.
Third, from a Bayesian point of view, the ridge estimator $\overline{%
\boldsymbol{\beta }}_{\lambda }$ is the mean of the posterior distribution
of $\boldsymbol{\beta }$ under a $p$-variate normal prior distribution,
corresponding to normal probability density function (p.d.f.) \textrm{n}$%
_{p}(\boldsymbol{\beta }\,|\,\mathbf{0},\sigma ^{2}\lambda ^{-1}\mathbf{I}%
_{p})$ for the conditional random variable $\boldsymbol{\beta }\,|\,\sigma
^{2}$. Thus, from a Bayesian decision-theoretic perspective, $\overline{%
\boldsymbol{\beta }}_{\lambda }$ is the choice of point estimate of $%
\boldsymbol{\beta }$ that minimizes the posterior expected squared-error
loss. Fourth, when $\lambda =0$, the ridge estimator $\overline{\boldsymbol{%
\beta }}_{\lambda }$ becomes the OLS estimator $\widehat{\boldsymbol{\beta }}
$.

In ridge regression, the quality of coefficient estimates $\overline{%
\boldsymbol{\beta }}_{\lambda }$ and predictions hinge on the choice of the
ridge parameter, $\lambda $. As a result, several methods have been proposed
to estimate this parameter, based on either automatic plug-in estimation,
cross-validation, information criteria optimization, or Markov chain Monte
Carlo (MCMC), which we now briefly review. An enormous literature on these
methods has developed over the past 45 years, however, the vast majority of
them rely on the OLS estimates $(\widehat{\boldsymbol{\beta }},\widehat{%
\sigma }^{2})$ (Cule \&\ DeIorio, 2012\nocite{CuleDeIorio12}). In keeping
with the general spirit of this paper, we focus our review on ridge methods
that can handle data sets where $\mathbf{X}^{\mathbf{\intercal }}\mathbf{X}$
is either positive-definite, ill-conditioned, or singular (including when $%
p\geq n$), unless indicated otherwise.

The standard, Hoerl-Kennard-Baldwin (HKB)\ (1975\nocite%
{HoerlKennardBaldwin75}) plug-in estimator for the ridge regression model is
defined by $\widehat{\lambda }_{\text{HKB}}=p\widehat{\sigma }^{2}/\widehat{%
\boldsymbol{\beta }}^{\intercal }\widehat{\boldsymbol{\beta }}$. This
estimator is motivated by the fact that $\lambda _{\text{HKB}}=p\sigma ^{2}/%
\boldsymbol{\beta }^{\intercal }\boldsymbol{\beta }$ is the choice of $%
\lambda $ that minimizes the model's expected prediction error when $\mathbf{%
X}$ is orthonormal (Hoerl \&\ Kennard, 1970\nocite{HoerlKennard70}). The
HKB\ estimator relies on the OLS\ estimate $\widehat{\sigma }^{2}$, and
therefore does not exist when $\mathbf{X}^{\mathbf{\intercal }}\mathbf{X}$
is singular, including when $p\geq n$. However, this\ estimator was extended
to singular and $p\geq n$ settings (Cule \&\ DeIorio, 2013\nocite%
{CuleDeIorio13}), by replacing $\widehat{\sigma }^{2}$ with the error
variance estimate $\widehat{\sigma }_{r}^{2}=\tfrac{1}{n-\widehat{r}}||%
\mathbf{y}-\mathbf{X}\widehat{\boldsymbol{\alpha }}_{\widehat{r}}||^{2}$
obtained from a principal components regression having $\widehat{r}\leq
q=\min (n,p)$ components with coefficients $\boldsymbol{\alpha }=(\alpha
_{1},\ldots ,\alpha _{r})^{\intercal }$. Here, $\widehat{r}$ is the value of 
$r$ that minimizes $r-\tsum\nolimits_{k=1}^{q}d_{k}^{4}/(d_{k}^{2}+\lambda
_{r})^{2}$. Therefore $\widehat{r}$ best matches the degrees of freedom for
variance in the ridge regression model with ridge parameter $\lambda _{%
\widehat{r}}=\widehat{r}\widehat{\sigma }_{\widehat{r}}^{2}/\widehat{%
\boldsymbol{\alpha }}_{\widehat{r}}^{\intercal }\widehat{\boldsymbol{\alpha }%
}_{\widehat{r}}$. The HKB plug-in estimator is attractive because it quickly
obtains a ridge estimate $\widehat{\lambda }$ without iteration. However
this estimator may lead to an estimate of $\lambda $ that is not necessarily
optimal for the given data set at hand, especially when $\mathbf{X}$ is not
orthonormal.

The method of cross-validation, for ridge regression, involves first
specifying a grid of trial values of $\lambda $, and then selecting the
estimate $\widehat{\lambda }$ as the trial value that provides the smallest
prediction error for the ridge regression model in $K$-fold
cross-validation, usually $10$-fold or $n$-fold (e.g., Hastie et al. 2009%
\nocite{HastieTibsFriedman09}, Ch. 3, 7). The method takes $K$ (at least
nearly) equal-sized partitions of the $n$ observations at random, and then
for each trial value of $\lambda $, measures the model's overall
mean-squared predictive error over all $K$ partitions. Overall mean-squared
error is based on obtaining the estimate $\overline{\boldsymbol{\beta }}%
_{\lambda }^{(k)}$ for one partition and measuring the predictive error on
all the other $K-1$ partitions, for all $k=1,\ldots ,K$. However, $K$-fold
cross-validation is computationally expensive for data sets where either $n$
or $p$ is large, because this method requires $K$\ separate estimations of
the parameters of the ridge regression model for every trial value of $%
\lambda $. Typically in practice, at least 100 trial values of $\lambda $
are used.

Alternatively, the posterior distribution of the parameters $(\boldsymbol{%
\beta },\sigma ^{2},\lambda )$ of the ridge model, including their
(marginal)\ posterior mean and variances, may be estimated by MCMC sampling
of this distribution. This is a Bayesian ridge regression (RR)\ model, which
assumes normal likelihood density $\mathrm{n}_{n}(\mathbf{y}\,|\,\mathbf{X}%
\boldsymbol{\beta },\sigma ^{2}\mathbf{I}_{n})$ for the data. The
conditional random variables $\boldsymbol{\beta },\sigma ^{2}\,|\,\lambda $
may be assigned a conjugate multivariate normal (n)\ inverse-gamma (ig)
prior distribution with probability density function defined by:%
\begin{equation*}
\pi (\boldsymbol{\beta },\sigma ^{2}\,|\,\lambda )=\mathrm{n}(\boldsymbol{%
\beta }\,|\,\mathbf{0},\sigma ^{2}\lambda ^{-1}\mathbf{I}_{p})\mathrm{ig}%
(\sigma ^{2}\,|\,a,b),
\end{equation*}%
and $\lambda $ is assigned a gamma \textrm{ga}$(\lambda \,|\,a_{\lambda
},b_{\lambda })$ prior distribution (Denison et al. 2002\nocite%
{DenisonHolmesMallickSmith02}). Throughout, 
\begin{equation*}
\mathrm{n}(\boldsymbol{\beta }\,|\,\mathbf{\mu },\boldsymbol{\Sigma })=(2\pi
)^{-p/2}|\boldsymbol{\Sigma }|^{-1/2}\exp \left[ (-1/2)(\boldsymbol{\beta }%
\,-\mathbf{\mu })^{\intercal }\boldsymbol{\Sigma }^{-1}(\boldsymbol{\beta }%
\,-\mathbf{\mu })\right]
\end{equation*}%
denotes the ($p$) multivariate normal p.d.f., and the p.d.f.s of the inverse
gamma and gamma distributions are respectively given by: 
\begin{eqnarray*}
\mathrm{ig}(s\,|\,a,b) &=&(b^{a}/\Gamma (a))s^{-a-1}\exp (-b/s)\newline
\\
\mathrm{ga}(s\,|\,a,b) &=&(b^{a}/\Gamma (a))s^{a-1}\exp (-bs)
\end{eqnarray*}%
with each gamma distribution parameterized (resp.) by shape and rate
parameters $(a,b)$ (e.g., Johnson et al. 1994\nocite%
{JohnsonKotzBalkrishnan94}). Alternatively, the conditional random variable $%
\boldsymbol{\beta },\sigma ^{2}\,|\,\lambda $ may be assigned a
non-conjugate prior distribution (Griffin \&\ Brown, 2013\nocite%
{GriffinBrown13}; Tsonias \&\ Tassiopoulos, 2014\nocite%
{TsionasTassiopoulos14}). However, similar to the cross-validation method,
the MCMC estimation procedure is also computationally expensive when either $%
p$, $n$, or when the number of MCMC samples is large. This is true
especially when the procedure is used along with MCMC convergence
diagnostics such as trace plots and MCMC\ batch means analysis, as
recommended (e.g., Flegal \&\ Jones, 2011\nocite{FlegalJones11}). Besides,
it is impossible to fully confirm convergence of MCMC samples to the
posterior distribution (Cowles \&\ Carlin, 1996\nocite{CowlesCarlin96}).

Generalized cross-validation (GCV) (Golub, et al. 1979\nocite%
{GolubHeathWahba79}), an alternative method which approximates $n$-fold
cross-validation, can provide a computationally-faster method to estimate
the ridge parameter $\lambda $. The GCV criterion is defined by:%
\begin{equation}
\mathrm{GCV}(\lambda )=\frac{1}{n}\dsum\limits_{i=1}^{n}\left( \dfrac{y_{i}-%
\mathbf{x}_{i}^{\intercal }\overline{\boldsymbol{\beta }}_{\lambda }}{1-%
\mathrm{df}_{\lambda }/n}\right) ^{2},  \label{GCV}
\end{equation}%
where $\mathrm{df}_{\lambda }=\tsum\nolimits_{k=1}^{\min
(n,p)}d_{k}^{2}/(d_{k}^{2}+\lambda )$ is the effective degrees of freedom\
for a ridge model with parameter $\lambda $. Then an estimate $\widehat{%
\lambda }$ can be obtained as the value of $\lambda $ which minimizes $%
\mathrm{GCV}(\lambda )$. Usually in practice, the minimizing $\lambda $ is
found after evaluating $\mathrm{GCV}(\lambda )$ over a fine grid of
hypothesized trial values of $\lambda $. However, the $\mathrm{GCV}(\lambda
) $ criterion, like the Akaike Information Criterion defined by $\mathrm{AIC}%
(\lambda )=\frac{1}{n}\{||\mathbf{y}-\mathbf{X}\widehat{\boldsymbol{\beta }}%
_{\lambda }||^{2}+2\mathrm{df}_{\lambda }\}$ with penalty term $2\mathrm{df}%
_{\lambda }$ (AIC; Akaike, 1973\nocite{Akaike73}), is an inconsistent
selector of the true-value of $\lambda $ as $n\rightarrow \infty $ holding $%
p $ fixed (Shao, 1997\nocite{Shao93}); and tends to lead to a ridge estimate 
$\widehat{\lambda }$ that yields an over-fitted ridge regression model (see
Zhang, et al. 2009\nocite{ZhangLiTsai09}). A consistent estimator $\widehat{%
\lambda }$ of $\lambda $, if $p$ fixed as $n\rightarrow \infty $ (Zhang, et
al. 2009\nocite{ZhangLiTsai09}), can be defined as the minimizer of the
Bayesian information criterion (BIC; Schwarz, 1978\nocite{Schwarz78}) given
by%
\begin{equation}
\mathrm{BIC}(\lambda )=\frac{1}{n}\{||\mathbf{y}-\mathbf{X}\widehat{%
\boldsymbol{\beta }}_{\lambda }||^{2}+\log (n)\mathrm{df}_{\lambda }\},
\label{BIC}
\end{equation}%
which is based on a Laplace approximation of the marginal likelihood $\pi (%
\mathcal{D}_{n}\,|\smallskip \,\lambda )$ of the data $\mathcal{D}_{n}=(%
\mathbf{X},\mathbf{y})$ (Ripley, 1996, p.64\nocite{Ripley96}).

For the Bayesian RR\ model, a more principled approach to estimating the
ridge parameter $\lambda $ is by maximizing the marginal likelihood $\pi (%
\mathcal{D}_{n}\,|\smallskip \,\lambda )$ directly, instead of minimizing
some approximation of it. In particular the ratio $\pi (\mathcal{D}%
_{n}\,|\,\lambda )/\pi (\mathcal{D}_{n}\,|\,\lambda ^{\prime })$ gives a
Bayes factor describing how much the data has changed the odds for $\lambda $
versus $\lambda ^{\prime }$, with $\log \pi (\mathcal{D}_{n}\,|\,\lambda )$
the weight of evidence for $\lambda $ (Good, 1950\nocite{Good50}).

Under the theory of the Bayesian linear model (Denison et al., 2002\nocite%
{DenisonHolmesMallickSmith02}), the marginal density $\pi (\mathcal{D}%
_{n}\,|\,\lambda )$ of the Bayes RR\ model depends on the determinant $|%
\mathbf{X}^{\intercal }\mathbf{X}+\lambda \mathbf{I}_{p}|$, which is
computationally-demanding when $p$ is large. However, we will show in this
paper that after taking a s.v.d. of the design matrix $\mathbf{X}$, the
marginal density $\pi (\mathcal{D}_{n}\,|\lambda )$ for the Bayesian RR\
model can be re-expressed by a simpler equation that does not depend on any
computationally-expensive matrix operations, such as determinants or
inverses, whether or not $p>n$ (Karabatsos, 2014\nocite{Karabatsos14c}).
This means that the marginal density $\pi (\mathcal{D}_{n}\,|\smallskip
\,\lambda )$ for the model, which is a log-concave function of $\lambda $,
can be rapidly evaluated over many trial values of $\lambda $. Then a simple
optimization algorithm can be constructed, that quickly finds the estimate $%
\widehat{\lambda }$ that maximizes the marginal likelihood $\pi (\mathcal{D}%
_{n}\,|\smallskip \,\lambda )$, even for data sets where either $p$ or $n$
is very large. Then conditionally on $\widehat{\lambda }$, the ridge
estimate $\overline{\boldsymbol{\beta }}_{\widehat{\lambda }}$ can be
readily obtained to make inferences about $\boldsymbol{\beta }$, and to make
predictions with the Bayesian RR\ model. For $p>n$ settings, an alternative,
simplified equation of the marginal likelihood $\pi (\mathcal{D}%
_{n}\,|\lambda )$ was proposed for the same Bayesian RR model, where the
equation depends on the $n\times n$ orthogonal matrix $\mathbf{U}$ obtained
from the s.v.d. $\mathbf{X}=\mathbf{UDW}^{\intercal }$ of $\mathbf{X}$ (Neto
et al., 2014\nocite{NetoJangFriendMargolin14}). However, we will show later
with real data sets that this equation can be computationally-prohibitive
when $n$ is sufficiently large, especially when it is evaluated over
multiple trial values of $\lambda $.

The ordinary Bayesian (normal) RR model, despite or because of its
simplicity, often provides competitive predictive power, compared to linear
models that assign more complex prior distributions for $\boldsymbol{\beta }%
\,|\,\sigma ^{2}$ (e.g., Griffin \&\ Brown, 2013\nocite{GriffinBrown13}).
However the RR\ model is not a panacea because it occasionally exhibits
worse predictive performance, as a result of it over-shrinking the OLS\
coefficients $\widehat{\boldsymbol{\alpha }}=\left( \widehat{\alpha }%
_{1},\ldots ,\widehat{\alpha }_{q}\right) ^{\intercal }$ for components that
have relatively small eigenvalues (Polson \&\ Scott, 2012\nocite%
{PolsonScott12}; Griffin \&\ Brown, 2013\nocite{GriffinBrown13}). Therefore,
in addition to the\ Bayesian RR model, in this study we also consider two
more general Bayesian ridge regression models, each of which assigns a more
complex prior distribution to $\boldsymbol{\beta }\,|\,\sigma ^{2}$ that can
adaptively assign more weight to components that have small eigenvalues, as
needed.

The first is given by the Bayesian power ridge regression (PRR)\ model
(Frank \&\ Friedman, 1993\nocite{FrankFriedman93}), defined by a conjugate
multivariate normal inverse-gamma prior distribution, with probability
density function (p.d.f.):%
\begin{equation}
\pi (\boldsymbol{\beta },\sigma ^{2}\,|\,\lambda ,\delta )=\mathrm{n}(%
\boldsymbol{\beta \,}|\boldsymbol{\,}\mathbf{0},\sigma ^{2}\lambda ^{-1}(%
\mathbf{X}^{\intercal }\mathbf{X})^{\delta })\mathrm{ig}(\sigma ^{2}%
\boldsymbol{\,}|\boldsymbol{\,}a,b)  \label{PRR prior}
\end{equation}%
for $\boldsymbol{\beta },\,\sigma ^{2}\,|\,\lambda ,\delta $. Here, the
prior covariance matrix for the PRR\ model a function of two parameters, $%
(\lambda ,\delta )$, where $\lambda $ is a global shrinkage parameter and $%
\delta $ expresses eigenvalue preference. A value $\delta <0$ expresses
greater preference (less shrinkage) for principal components of $\mathbf{X}$
with smaller eigenvalues; the value $\delta =0$ corresponds to ordinary
ridge regression and no eigenvalue preference; and a value $\delta >0$
expresses greater preference (less shrinkage) for principal components with
higher eigenvalues, as in principle components regression and penalized
least-squares regression. When $\delta =-1$, the power ridge prior density $%
\mathrm{n}(\boldsymbol{\beta \,}|\boldsymbol{\,}\mathbf{0},\sigma
^{2}\lambda ^{-1}(\mathbf{X}^{\intercal }\mathbf{X})^{\delta })$ coincides
with the $g$-prior (Zellner, 1986\nocite{Zellner86}). The $g$-prior
replicates the covariance structure of the design matrix, provides an
automatic scaling based on the data, and is related to the Fisher
information matrix of the linear model up to the scalar $1/\lambda $ (Chen
\& Ibrahim, 2003\nocite{ChenIbrahim03}).

In this study, we show that for the Bayesian PRR model, after taking a
s.v.d.\ of $\mathbf{X}$, it is possible to re-express the marginal density $%
\pi (\mathcal{D}_{n}\,|\,\lambda ,\delta )$ of the model by a simpler
equation that require no computationally-expensive matrix operations, as in
the case of the Bayesian RR\ model that assumes conditional prior density 
\textrm{n}$_{p}(\boldsymbol{\beta }\,|\,\mathbf{0},\sigma ^{2}\lambda ^{-1}%
\mathbf{I}_{p})$, as mentioned earlier. Then for the Bayesian PRR\ model, it
is possible to develop a fast and simple optimization algorithm that can
quickly search over multiple trial values of the two dimensional space of $%
(\lambda ,\delta )$, to find the estimate $(\widehat{\lambda },\widehat{%
\delta })$ that maximizes the marginal likelihood $\pi (\mathcal{D}%
_{n}\,|\smallskip \,\lambda ,\delta )$. This is true even for very large
data sets where either $p$ or $n$ is very large.

We also consider the Bayesian generalized ridge regression (GRR)\ model,
which assigns a normal inverse-gamma prior distribution to $\boldsymbol{%
\beta },\sigma ^{2}\,|\,\boldsymbol{\lambda }$, with probability density
function%
\begin{equation}
\pi (\boldsymbol{\beta },\sigma ^{2}\,|\,\boldsymbol{\lambda })=\mathrm{n}(%
\boldsymbol{\beta \,}|\boldsymbol{\,}\mathbf{0},\mathbf{W}\mathrm{diag}%
(\lambda _{1},\ldots ,\lambda _{q})^{-1}\mathbf{W}^{\intercal })\mathrm{ig}%
(\sigma ^{2}\boldsymbol{\,}|\boldsymbol{\,}a,b),
\end{equation}%
based on multiple free shrinkage parameters $\boldsymbol{\lambda }=(\lambda
_{1},\ldots ,\lambda _{q})$ for the $q$ principal components of $\mathbf{X}$
obtained from the s.v.d. $\mathbf{X}=\mathbf{UDW}^{\intercal }$ of $\mathbf{X%
}$ (Hoerl \&\ Kennard, 1970\nocite{HoerlKennard70}). Special cases of the
GRR model includes the RR model which assumes a single shrinkage parameter
with $\lambda =\lambda _{1}=\cdots =\lambda _{q}$; and includes the power
ridge regression model which assumes that $\boldsymbol{\lambda }$ is a
function $\phi $ of two parameters and the $q$ eigenvalues of $\mathbf{X}%
^{\intercal }\mathbf{X}$, with $\boldsymbol{\lambda }=\phi (\lambda ,\delta
) $. In this study we also show that after taking a s.v.d.\ of $\mathbf{X}$,
the generalized ridge model also admits a marginal density $\pi (\mathcal{D}%
_{n}\,|\,\boldsymbol{\lambda })$ that can be computed by a simple equation
that requires no computationally-expensive matrix operations, including
matrix determinants and inverses.

Interestingly, we also show that for the Bayesian GRR model, the MML
estimate $\widehat{\boldsymbol{\lambda }}=(\widehat{\lambda }_{1},\ldots ,%
\widehat{\lambda }_{q})$ can be obtained through fast and direct calculation
of a set of simple closed-form equations, to provide an automatic plug-in
estimator of $\boldsymbol{\lambda }$ that does not require iteration. This
automatic plug-in estimator allows $\mathbf{X}^{\intercal }\mathbf{X}$ to be
singular and allows $\sigma ^{2}$ to be an unknown parameter with
inverse-gamma prior distribution. It is perhaps interesting that the fully
general ridge regression model admits faster computational speed in marginal
maximum likelihood estimation of $\boldsymbol{\lambda }$, compared to that
of the simpler, ordinary ridge regression and power ridge regression models
that have only 1 and 2 parameters for the prior distribution of $\boldsymbol{%
\beta }\,|\,\sigma ^{2}$, respectively. Previous research (Walker and Page,
2001\nocite{WalkerPage01}) proposed another set of closed-form equations to
provide a plug-in marginal maximum likelihood estimator $\widehat{%
\boldsymbol{\lambda }}$. However, these equations assumed that the error
variance $\sigma ^{2}$ is fixed and known, and that in applied settings, it
is provided by the OLS estimate $\widehat{\sigma }^{2}$ when $\sigma ^{2}$
is unknown. As mentioned, the OLS estimate $\widehat{\sigma }^{2}$ is
unavailable (unstable) when $\mathbf{X}^{\intercal }\mathbf{X}$ is singular
(ill-conditioned).

In general, the marginal maximum likelihood estimator $\widehat{\boldsymbol{%
\lambda }}$ has an obvious interpretation from a frequentist perspective,
for either the Bayesian RR\ model based on estimate $\widehat{\boldsymbol{%
\lambda }}=\widehat{\lambda }$, the Bayesian PRR\ model where $\widehat{%
\boldsymbol{\lambda }}$ is a function of the estimate $(\widehat{\lambda },%
\widehat{\delta })$, or the Bayesian GRR\ model based on the estimate $%
\widehat{\boldsymbol{\lambda }}=(\widehat{\lambda }_{1},\ldots ,\widehat{%
\lambda }_{q})$. Later we will show later that if the ridge parameter $%
\boldsymbol{\lambda }$ is assigned a uniform prior distribution, then the
estimator $\widehat{\boldsymbol{\lambda }}$ can be characterized as the
posterior mode of $(\boldsymbol{\beta },\sigma ^{2},\boldsymbol{\lambda })$
under the "Bayes empirical Bayes" framework, a fully Bayesian approach to
empirical Bayes (Deely \&\ Lindley, 1981\nocite{DeelyLindley81}), and is
preferred according to the Bayes factor (e.g., Kass \&\ Raftery, 1995\nocite%
{KassRaftery95}) over all pairwise comparisons of all possible ridge
parameter values $\{\boldsymbol{\lambda }\}$.

Conditionally on an obtained ridge marginal maximum likelihood estimate $%
\widehat{\boldsymbol{\lambda }}$, for either the Bayesian RR, PRR, or GRR
model, we will show that the marginal posterior covariance matrix of $%
\boldsymbol{\beta }$, posterior predictive quantities of $Y$\ given any
chosen $\mathbf{x}$, and various auxiliary statistics can be efficiently
calculated without needing to evaluate any matrix inverses and large-scale
matrix multiplications, which are known to be computationally-expensive when
either $p$ or $n$ is large. The subset of "significant"\ predictors, among
the $p$ given covariates, can be identified from Bayesian hypothesis tests
of the null hypothesis $H_{0k}:\beta _{k}=0$, against the general
alternative hypotheses $H_{1k}:\beta _{k}\neq 0$, for covariates $X_{k}$, $%
k=1,\ldots ,p$. Specifically, in standard practice, a null hypothesis $%
H_{0k}:\beta _{k}=0$ is rejected, thereby labelling the covariate $X_{k}$ as
a "significant predictor," when the 95\% posterior credible interval of $%
\beta _{k}$ excludes zero, with interval defined by the 2.5\% and 97.5\%\
percentiles of the marginal posterior distribution of $\beta _{k}$.
Alternatively, the hypothesis test may be based on whether or not zero is
included in the posterior interquartile range (50\% credible interval) of
the marginal posterior distribution of $\beta _{k}$ (Li\ \&\ Lin, 2010\nocite%
{LiLin10}). Yet another alternative statistic is provided by the scaled
neighborhood (SN) criterion. This criterion rejects the null hypothesis $%
H_{0k}:\beta _{k}=0$ when the posterior probability of $\beta _{k}$ being
within one marginal posterior standard deviation of zero is less than 1/2
(Berger, 1993\nocite{Berger93}; Li\ \&\ Lin, 2010\nocite{LiLin10}).

Conditioning posterior inferences of $\boldsymbol{\beta }$ on an estimate $%
\widehat{\boldsymbol{\lambda }}$ does not account for the extra variability
that is inherent in the estimation of the ridge parameter $\boldsymbol{%
\lambda }$. For practice involving real data, this extra variability is
negligible when $n$ is sufficiently large, and may not matter much when it
is simply of interest to identify significant predictors (covariates), as
either having slope coefficients with 95\% (or interquartile) posterior
credible intervals that exclude zero, or as having SN\ criteria below the
threshold of 1/2. Again, in this study we focus on estimation methods that
provide fast, deterministic, and approximate posterior estimation of model
parameters, which may be desired in practice for data sets where either $p$
or $n$ can be very large. For very large data sets and/or for data sets
having important policy implications, such a deterministic estimation method
may be preferred over slower and random-output MCMC posterior estimation
methods that aim to fully account for the uncertainty in all ridge model
parameters $(\boldsymbol{\beta },\sigma ^{2},\lambda )$. Moreover, for such
large data sets, Bayesian point estimation may be preferred over estimation
of the full posterior distribution. This is because it is difficult to fully
describe the full joint posterior distribution of more than two model
parameters, either graphically or analytically (Bernardo \&\ Ju\'{a}rez, 2003%
\nocite{BernardoJuarez03}), aside from the fact that MCMC\ may be too
computationally-inefficient for sufficiently large data sets. Indeed, many
applications of ridge regression models may involve hundreds or thousands of
slope coefficient parameters (i.e., covariates).

Ridge regression is unable shrink slope coefficient estimates to exactly
zero and hence is unable to eliminate (zero-out)\ "insignificant" predictors
(covariates) from the linear model. In contrast, popular "sparse" shrinkage
methods, such as the LASSO (Tibshirani, 1996\nocite{Tibshirani96}) and the
Elastic Net (EN; Zou \&\ Hastie, 2005\nocite{ZouHastie05}) can each shrink
coefficient estimates to exactly zero to eliminate (zero-out)\
"insignificant" predictors (covariates) in the linear model, according to
the penalized least squares estimator 
\begin{equation}
\overline{\boldsymbol{\beta }}_{\lambda ,\alpha }=\,\underset{\boldsymbol{%
\beta }}{\arg \min }||\mathbf{y}-\mathbf{X}\boldsymbol{\beta }||^{2}+\lambda
\tsum\nolimits_{k=1}^{p}\left( \tfrac{(1-\alpha )}{2}\beta _{k}^{2}+\alpha
|\beta _{k}|\right) .
\end{equation}%
Above, $0\leq \alpha \leq 1$, and the choice $\alpha =1$ defines the LASSO
estimator, and $\alpha =0$ defines a RR estimator, while the choice $\alpha
=1/2$ is viewed as a reasonable compromise (Zou \&\ Hastie, 2005\nocite%
{ZouHastie05}). For either the LASSO\ or the EN, the shrinkage parameter $%
\lambda $ is usually estimated either as the minimizer of 10-fold cross
validated mean-square predictive error, or by minimizing a generalized
information criterion $\mathrm{GIC}(\lambda )$ (Fan \&\ Tang, 2013\nocite%
{FanTang13}). Sparse shrinkage methods also include Bayesian variable
selection linear models that are equipped with indicator (0,1) parameters
that set slope coefficients to zero with positive prior probability (e.g.,
Bottolo \&\ Richardson, 2012\nocite{BottoloRichardson10}).

However, from a Bayesian perspective, there are at least three good reasons
why shrinkage methods like ridge regression may be preferred over the sparse
shrinkage methods. First, it may be argued that no continuous quantity like
a slope coefficient $\beta _{k}$ is ever precisely zero (Draper, 1999\nocite%
{Draper99}). Similarly, for a Bayesian variable selection model, the
posterior average of $\boldsymbol{\beta }$ is surely non-zero, after
averaging over the indicator parameters (Polson \&\ Scott 2012\nocite%
{PolsonScott12}). Then point-null hypotheses of the form $\beta _{k}=0$ seem
unrealistic, and the posterior average of $\boldsymbol{\beta }$ obtained
under either of these methods may be practically indistinguishable from the
posterior average of $\boldsymbol{\beta }$ obtained under a pure shrinkage
method. Second, a pure shrinkage method can provide computational gains over
sparse shrinking and Bayesian variable selection (model averaging), in terms
of speed and simplicity (Polson \&\ Scott 2012\nocite{PolsonScott12}).
Moreover, shrinkage regression models, in part because they never entirely
eliminate (zero-out)\ predictors, tend to have lower prediction error
compared to sparse shrinkage regression methods (e.g., Tibshirani, 1996%
\nocite{Tibshirani96}).

In the following sections, we describe and illustrate the marginal maximum
likelihood estimation methodology, for each of the Bayesian RR, PRR, and
GRR\ models. In order to provide full context for these models, Section 2.1
provides a brief review of the classical inferential theory of the Bayesian
linear model (e.g., Denison, et al. 2002\nocite{DenisonHolmesMallickSmith02}%
; O'Hagan \&\ Forster, 2004\nocite{OhaganForster04}). Then in Section 2.2,
we define the Bayesian RR, PRR, and GRR models. There we show that for each
of these Bayesian models, the posterior mean and covariance estimates of $%
\boldsymbol{\beta }_{\boldsymbol{\lambda }}$, and auxiliary statistics such
as posterior intervals and SN criteria, can be efficiently calculated the
help of the s.v.d. of $\mathbf{X}$, without needing to evaluate any
computationally-expensive matrix operations such as inverses, determinants,
and large-scale matrix multiplications. In practice, this is an important
feature when either $p$ or $n$ is large. Then in Section 2.3, we show that
for each of the three Bayesian ridge models, the marginal density $\pi (%
\mathcal{D}_{n}\,|\,\boldsymbol{\lambda })$ can be simplified into an
equation involving no matrix operations whatsoever, after taking a s.v.d.\
of $\mathbf{X}$. We will also show how the marginal maximum likelihood
estimate $\widehat{\boldsymbol{\lambda }}$ can be obtained for each of the
three Bayesian ridge models, under the Bayes empirical Bayes framework
(Deely \&\ Lindley, 1981\nocite{DeelyLindley81}). As mentioned, in this
study we will introduce fast iterative algorithms for estimating the
parameters of the Bayesian RR\ and PRR\ models. For the Bayesian GRR\ model,
we present a set of closed-form equations to provide a non-iterative and
automatic plug-in estimator of $\widehat{\boldsymbol{\lambda }}$, leading to
estimates of the coefficients and error variance. For each of the three
Bayesian ridge models, the estimate $\widehat{\boldsymbol{\lambda }}$
provides a basis to directly calculate the posterior mean and covariance
estimate of $(\boldsymbol{\beta },\sigma ^{2})$, as well as the auxiliary
statistics. In Section 3, we briefly review the literature to support the
idea that the Bayesian GRR model, including the Bayesian RR and PRR\ model,
has a large scope for applied statistics, because of the model's ability to
handle data with a very large number of covariates $p$ (e.g., hundreds or
thousands) and corresponding slope coefficient parameters. Ridge regression
can provide a flexible and approximate Bayesian nonparametric modeling (M%
\"{u}ller \&\ Quintana, 2004\nocite{MullerQuintana04}), even for linear
classification of a binary dependent variable.

Section 4 illustrates the marginal likelihood estimation methodology for the
three Bayesian ridge models, through the analysis of 10 real data sets. For
nearly all these data sets, the number of covariates ($p$) range from
several hundred to around fifteen thousand. Several of these data sets
involve more covariates than observations (i.e., $p>n$). We will also
compare the Bayesian RR, PRR, and GRR\ models under marginal likelihood
estimation, with other approaches. They include the Bayesian RR\ model with
ridge parameter $\lambda $ estimated either by (1)\ the extended HKB\
plug-in estimator (Cule \&\ DeIorio, 2013\nocite{CuleDeIorio13}); or (2)\ by
maximizing the original marginal likelihood equation that requires the
computation of matrix determinants; or (3)\ by maximizing the marginal
likelihood equation that does not require the computation of matrix
determinants but depends on the $n\times n$ orthogonal matrix $\mathbf{U}$
obtained from the s.v.d. of $\mathbf{X}$ (Neto et al., 2014\nocite%
{NetoJangFriendMargolin14}); or (4)\ by minimizing 10-fold cross-validated
mean-square predictive error; or (5)\ by minimizing the\textrm{\ }$\mathrm{%
GCV}(\lambda )$ criterion. They also include the ordinary Bayesian RR\
model, with $\boldsymbol{\beta },\sigma ^{2}\,|\,\lambda $ assigned a
conjugate multivariate-normal inverse-gamma prior distribution along with a
gamma prior distribution for $\lambda $. They also include the LASSO\ model
and the EN\ model (with parameter $\alpha =1/2$), each with shrinkage
parameter $\lambda $ estimated either by minimizing 10-fold cross-validated
mean-square predictive error, or by minimizing a\ Generalized Information
Criterion $\mathrm{GIC}(\lambda )$, or by minimizing a Bayesian information
criterion $\mathrm{BIC}(\lambda ),$ a special case of $\mathrm{GIC}(\lambda
) $. Here the $\mathrm{GIC}(\lambda )$ and $\mathrm{BIC}(\lambda )$ criteria
are defined in a way to provide model (or $\lambda $) selection that allows
the number of covariates $p$ to increase at an exponential rate with $n$, as 
$n\rightarrow \infty $ (see Fan \&\ Tang, 2013\nocite{FanTang13}). Across
all the different approaches mentioned here, we will compare computation
times for parameter estimation, as well as compare the predictive fit to the
data according to 10-fold cross-validated mean-squared predictive error.\
Also we will compare the different ridge regression models in terms of the
log marginal likelihood $\log \pi (\mathcal{D}_{n}\,|\smallskip \,\widehat{%
\boldsymbol{\lambda }})$ of the data conditionally on the estimate $\widehat{%
\boldsymbol{\lambda }}$. Section 4 also reports the results of a simulation
study that compares the performance of the 95\%, interquartile, and SN\
hypothesis testing criteria of the Bayesian RR, PRR, and GRR\ models
estimated under marginal maximum likelihood estimation, with the performance
of the LASSO and EN\ models. They are compared in terms of the ability to
differentiate between truly-significant from truly-insignificant predictors
(covariates), according to Receiver Operator Curve (ROC) analyses. Here, the
LASSO and EN\ models have shrinkage parameter $\lambda $ estimated either by
minimizing $\mathrm{GIC}(\lambda )$ or $\mathrm{BIC}(\lambda )$.

\section{Bayesian Generalized Ridge Regression}

\subsection{The Bayesian Linear Model}

Here we review inference methods with the Bayesian linear model, as given in
standard textbooks (e.g., Denison et al., 2002\nocite%
{DenisonHolmesMallickSmith02}; O'Hagan \&\ Forster, 2004\nocite%
{OhaganForster04})

Consider a given set of data, $\mathcal{D}_{n}=(\mathbf{X},\mathbf{y})$,
where $\mathbf{X}=(x_{ip})_{n\times p}$ and $\mathbf{y}=(y_{1},\ldots
,y_{n})^{\intercal }$. The Bayesian linear regression model, which assigns a
conjugate normal-inverse gamma (NIG)\ prior density to $(\boldsymbol{\beta }%
,\sigma ^{2})$, is defined by: 
\begin{subequations}
\label{lm}
\begin{eqnarray}
f(\mathbf{y}\mathcal{\,}|\,\mathbf{X},\boldsymbol{\beta },\sigma ^{2}) &=&%
\mathrm{n}_{n}(\mathbf{y}\,|\,\mathbf{X}\boldsymbol{\beta },\sigma ^{2}%
\mathbf{I}_{n})=\dprod\nolimits_{i=1}^{n}\mathrm{n}(y\,|\,\mathbf{x}%
_{i}^{\intercal }\boldsymbol{\beta },\sigma ^{2}),  \label{lm Like} \\
\pi (\boldsymbol{\beta },\sigma ^{2}) &=&\mathrm{n}_{p}(\boldsymbol{\beta \,}%
|\boldsymbol{\,}\mathbf{m},\sigma ^{2}\mathbf{V})\mathrm{ig}(\sigma ^{2}%
\boldsymbol{\,}|\,a,b),  \label{NIG prior0} \\
&=&\mathrm{nig}(\boldsymbol{\beta },\sigma ^{2}\boldsymbol{\,}|\boldsymbol{\,%
}\mathbf{m},\mathbf{V},a,b),  \label{lm NIG}
\end{eqnarray}%
where $\mathrm{n}_{n}(\cdot \,|\,\boldsymbol{\mu },\mathbf{\Sigma })$ and $%
\mathrm{n}(\cdot \,|\,\mu ,\sigma ^{2})$ denote the probability density
functions (p.d.f.s) of the $n$-variate normal distribution and the
univariate normal distribution (resp.), and where $\mathrm{ig}(\cdot 
\boldsymbol{\,}|\,a,b)$ denotes the density function of the inverse-gamma
distribution with shape $a$ and rate $b$ (and scale $1/b$). Also, $\mathrm{%
nig}(\boldsymbol{\beta },\sigma ^{2}\,|\,\mathbf{m},\mathbf{V},a,b)$ denotes
the p.d.f.\ of the NIG\ distribution, which as shown in (\ref{NIG prior0})-(%
\ref{lm NIG}) is defined by a product of the multivariate normal p.d.f. and
the inverse-gamma p.d.f. (Lindley \&\ Smith, 1972\nocite{LindleySmith72}).

Since the joint prior distribution of $(\boldsymbol{\beta },\sigma ^{2})$
has a NIG distribution, it then follows that, marginally, the prior
distribution of $\boldsymbol{\beta }$ is a Student distribution with mean $%
\mathbb{E}\left( \boldsymbol{\beta }\right) =\mathbf{m}$, covariance matrix $%
\mathbb{V}\left( \boldsymbol{\beta }\right) =(b/(a-1))\mathbf{V}$, and
degrees of freedom $2a$. Also, the prior distribution of $\sigma ^{2}$ is an
inverse-gamma distribution with mean $\mathbb{E}\left( \sigma ^{2}\right)
=b/(a-1)$ and variance $\mathbb{V}\left( \sigma ^{2}\right) =b^{2}/\left\{
(a-1)^{2}(a-2)\right\} $.

The prior predictive distribution of an observable dependent response, $y$,
conditionally on chosen covariates $\mathbf{x}$, is a Student distribution
with p.d.f.: 
\end{subequations}
\begin{equation}
\mathrm{st}(y\,|\,\mathbf{x}^{\intercal }\mathbf{m},b(1+\mathbf{x}%
^{\intercal }\mathbf{Vx}),a)=\diint \mathrm{n}(y\,|\,\mathbf{x}^{\intercal }%
\boldsymbol{\beta },\sigma ^{2})\mathrm{nig}(\boldsymbol{\beta \,},\sigma
^{2}\boldsymbol{\,}|\boldsymbol{\,}\mathbf{m},\mathbf{V},\,a,b)\mathrm{d}%
\boldsymbol{\beta }\mathrm{d}\sigma ^{2},  \label{prior pred y}
\end{equation}%
with mean $\mathbb{E}(Y\boldsymbol{\,}|\,\mathbf{x})=\mathbf{x}^{\intercal }%
\mathbf{m}$, variance $\mathbb{V}(Y\boldsymbol{\,}|\,\mathbf{x})=(b/(a-2))(1+%
\mathbf{x}^{\intercal }\mathbf{Vx})$, and degrees of freedom $a$. More
generally, the prior predictive distribution of any given vector of
observations $\mathbf{y}=(y_{1},\ldots ,y_{n})^{\intercal }$, conditionally
on any given design matrix $\mathbf{X}=(x_{ip})_{n\times p}$, is a
multivariate Student distribution with p.d.f.:%
\begin{equation}
\mathrm{st}(\mathbf{y}\,|\,\mathbf{Xm},b(\mathbf{I}_{p}+\mathbf{XVX}%
^{\intercal }),a)=\diint \mathrm{n}(y\,|\,\mathbf{X}\boldsymbol{\beta }%
,\sigma ^{2}\mathbf{I}_{n})\mathrm{nig}(\boldsymbol{\beta \,},\sigma ^{2}%
\boldsymbol{\,}|\boldsymbol{\,}\mathbf{m},\mathbf{V},\,a,b)\mathrm{d}%
\boldsymbol{\beta }\mathrm{d}\sigma ^{2},  \label{prior pred bold y}
\end{equation}%
along with mean $\mathbb{E}(Y_{1},\ldots ,Y_{n}\boldsymbol{\,}|\,\mathbf{X})=%
\mathbf{Xm}$ and covariance matrix $\mathbb{V}(Y_{1},\ldots ,Y_{n}%
\boldsymbol{\,}|\,\mathbf{X})=(b/(a-2))(\mathbf{I}_{p}+\mathbf{XVX}%
^{\intercal })$, where the diagonal elements of $\mathbb{V}_{n}(Y_{1},\ldots
,Y_{n}\boldsymbol{\,}|\,\mathbf{X})$ give the prior predictive variances of $%
(Y_{1},\ldots ,Y_{n})$, respectively.

A set of data $\mathcal{D}_{n}=(\mathbf{X},\mathbf{y})$ updates the NIG
prior density (\ref{lm NIG}) to a posterior density $\pi (\boldsymbol{\beta }%
,\sigma ^{2}\,|\,\mathcal{D}_{n})$. Given the conjugacy of the NIG\ prior
p.d.f. (\ref{lm NIG}) with the normal likelihood p.d.f. (\ref{lm Like}), the
posterior p.d.f. (distribution)\ is also a NIG, with p.d.f. given by: 
\begin{subequations}
\label{NIG post}
\begin{eqnarray}
\pi (\boldsymbol{\beta },\sigma ^{2}\,|\,\mathcal{D}_{n}) &=&\dfrac{\mathrm{n%
}_{n}(\mathbf{y}\,|\,\mathbf{X}\boldsymbol{\beta },\sigma ^{2}\mathbf{I}_{n})%
\mathrm{nig}(\boldsymbol{\beta },\sigma ^{2}\boldsymbol{\,}|\boldsymbol{\,}%
\mathbf{m},\mathbf{V},\,a,b)}{\diint \mathrm{n}_{n}(\mathbf{y}\,|\,\mathbf{X}%
\boldsymbol{\beta },\sigma ^{2}\mathbf{I}_{n})\mathrm{nig}(\boldsymbol{\beta
\,},\sigma ^{2}\boldsymbol{\,}|\boldsymbol{\,}\mathbf{m},\mathbf{V},\,a,b)%
\mathrm{d}\boldsymbol{\beta }\mathrm{d}\sigma ^{2}}  \label{NIG post 1} \\
&=&\mathrm{n}_{p}(\boldsymbol{\beta \,}|\boldsymbol{\,}\overline{\boldsymbol{%
\beta }},\sigma ^{2}\overline{\mathbf{V}})\mathrm{ig}(\sigma ^{2}\boldsymbol{%
\,}|\,\overline{a},\overline{b}) \\
&=&\mathrm{nig}(\boldsymbol{\beta },\sigma ^{2}\boldsymbol{\,}|\boldsymbol{\,%
}\overline{\boldsymbol{\beta }},\overline{\mathbf{V}},\overline{a},\overline{%
b}),
\end{eqnarray}%
with data-updated parameters: 
\end{subequations}
\begin{subequations}
\label{Updates}
\begin{eqnarray}
\overline{\mathbf{V}} &=&(\mathbf{V}^{-1}+\mathbf{X}^{\intercal }\mathbf{X)}%
^{-1},  \label{Update V} \\
\overline{\boldsymbol{\beta }} &=&\overline{\mathbf{V}}\left( \mathbf{V}^{-1}%
\mathbf{m}+\mathbf{X}^{\intercal }\mathbf{y}\right) , \\
\overline{a} &=&\,a+n/2, \\
\overline{b} &=&\,b+(\mathbf{y}^{\intercal }\mathbf{y}-\overline{\boldsymbol{%
\beta }}^{\intercal }\overline{\mathbf{V}}^{-1}\overline{\boldsymbol{\beta }}%
)/2.
\end{eqnarray}%
Above, the second term of the $\overline{b}$ equation is equal to half the
sum of squared residuals of the fit, using the posterior mean $\overline{%
\boldsymbol{\beta }}$ of $\boldsymbol{\beta }$. Since the joint posterior
distribution of $(\boldsymbol{\beta },\sigma ^{2})$ has a NIG distribution
with p.d.f. $\mathrm{nig}(\boldsymbol{\beta },\sigma ^{2}\boldsymbol{\,}|%
\boldsymbol{\,}\overline{\boldsymbol{\beta }},\overline{\mathbf{V}},%
\overline{a},\overline{b})$, it then follows that the marginal posterior
distribution of $\boldsymbol{\beta }$ is a Student distribution with mean $%
\mathbb{E}\left( \boldsymbol{\beta }\,|\,\mathcal{D}_{n}\right) =\overline{%
\boldsymbol{\beta }}$, covariance $\mathbb{V}\left( \boldsymbol{\beta }\,|\,%
\mathcal{D}_{n}\right) =(\overline{b}/(\overline{a}-1))\overline{\mathbf{V}}$%
, and degrees of freedom $2a+n$; and that the marginal posterior
distribution of $\sigma ^{2}$ is an inverse-gamma distribution with
probability density function $\mathrm{ig}(\sigma ^{2}\boldsymbol{\,}|\,%
\overline{a},\overline{b})$, mean $\mathbb{E}\left( \sigma ^{2}\,|\,\mathcal{%
D}_{n}\right) =\overline{\sigma }^{2}=\overline{b}/(\overline{a}-1)$ and
variance $\mathbb{V}\left( \sigma ^{2}\,|\,\mathcal{D}_{n}\right) =\overline{%
b}^{2}/\left\{ (\overline{a}-1)^{2}(\overline{a}-2)\right\} $.

If $\mathbf{V}^{-1}\rightarrow \mathbf{0}$, $a\rightarrow 0$, and $%
b\rightarrow 0$ in the NIG\ prior p.d.f. $\mathrm{nig}(\boldsymbol{\beta }%
,\sigma ^{2}\boldsymbol{\,}|\boldsymbol{\,}\mathbf{0},\mathbf{V},a,b)$, then
the posterior mean of $\boldsymbol{\beta }$ becomes the OLS\ estimator, with 
$\overline{\boldsymbol{\beta }}=\widehat{\boldsymbol{\beta }}=\overline{%
\mathbf{V}}\mathbf{X}^{\intercal }\mathbf{y}=(\mathbf{X}^{\intercal }\mathbf{%
X)}^{-1}\mathbf{X}^{\intercal }\mathbf{y}$, along with posterior covariance
matrix $\widehat{\sigma }^{2}\overline{\mathbf{V}}=\widehat{\sigma }^{2}(%
\mathbf{X}^{\intercal }\mathbf{X)}^{-1}$, where $\widehat{\sigma }^{2}=(%
\mathbf{y}^{\intercal }\mathbf{y}-\widehat{\boldsymbol{\beta }}^{\intercal }%
\overline{\mathbf{V}}^{-1}\widehat{\boldsymbol{\beta }})/(n-p)$ gives the
OLS estimate of $\sigma ^{2}$. Also, the sampling distribution of $\widehat{%
\boldsymbol{\beta }}\,|\,\widehat{\sigma }^{2}$ is a multivariate normal
distribution with density $\mathrm{n}_{p}(\boldsymbol{\beta \,}|\boldsymbol{%
\,}\widehat{\boldsymbol{\beta }},\widehat{\sigma }^{2}(\mathbf{X}^{\intercal
}\mathbf{X)}^{-1})$. The square roots of the diagonal elements $\mathrm{diag}%
(\widehat{\sigma }^{2}(\mathbf{X}^{\intercal }\mathbf{X)}^{-1})$ provide the
standard errors of $\widehat{\boldsymbol{\beta }}$.

The normalizing constant (denominator) of the posterior density (\ref{NIG
post 1}) is the marginal likelihood of the linear model, defined by: 
\end{subequations}
\begin{equation}
\pi (\mathcal{D}_{n})=\dfrac{|\overline{\mathbf{V}}|^{1/2}b^{a}\Gamma (%
\overline{a})}{|\mathbf{V}|^{1/2}\overline{b}^{\overline{a}}\Gamma (a)\pi
^{n/2}},  \label{margLike}
\end{equation}%
and corresponds to log marginal likelihood: 
\begin{subequations}
\label{logpdLine}
\begin{eqnarray}
\log \pi (\mathcal{D}_{n}) &=&\log |\overline{\mathbf{V}}|^{1/2}-\log |%
\mathbf{V}|^{1/2}+a\log b-\overline{a}\log \overline{b} \\
&&+\log \Gamma (\overline{a})-\log \Gamma (a)-\tfrac{n}{2}\log \pi \\
&=&\dsum\nolimits_{i=1}^{n}\log \pi \left( y_{i}\,|\,(y_{i-1},\mathbf{x}%
_{i-1}),\ldots ,(y_{1},\mathbf{x}_{1})\right) .
\end{eqnarray}

Suppose that we compare two different linear models, $\mathcal{M}_{1}$ and $%
\mathcal{M}_{2}$, assigned NIG\ prior distributions with densities $\mathrm{%
nig}(\boldsymbol{\beta },\sigma ^{2}\boldsymbol{\,}|\boldsymbol{\,}\mathbf{m}%
_{1},\mathbf{V}_{1},a_{1},b_{1})$ and $\mathrm{nig}(\boldsymbol{\beta }%
,\sigma ^{2}\boldsymbol{\,}|\boldsymbol{\,}\mathbf{m}_{2},\mathbf{V}%
_{2},a_{2},b_{2})$, respectively. Then the Bayes factor, defined by a ratio
of posterior odds to prior odds, given by 
\end{subequations}
\begin{equation}
B_{12}=\frac{\pi (\mathcal{D}_{n}\,|\,\mathcal{M}_{1})}{\pi (\mathcal{D}%
_{n}\,|\,\mathcal{M}_{2})}=\left. \dfrac{\pi (\mathcal{M}_{1}\,|\,\mathcal{D}%
_{n})}{\pi (\mathcal{M}_{2}\,|\,\mathcal{D}_{n})}\right/ \dfrac{\pi (%
\mathcal{M}_{1})}{\pi (\mathcal{M}_{2})},  \label{BF}
\end{equation}%
provides a measure of whether the data $\mathcal{D}_{n}$ have increased or
decreased the odds on model $\mathcal{M}_{1}$ relative to model $\mathcal{M}%
_{2}$ (Kass \&\ Raftery, 1995\nocite{KassRaftery95}).\ The natural log of
the Bayes factor, $\log B_{12}$, measures the weight of evidence of model $%
\mathcal{M}_{1}$ versus model $\mathcal{M}_{2}$ (Good, 1950\nocite{Good50}).
A Bayes factor $B_{12}$ ($\log B_{12}$, resp.) that takes on a value less
than 1 (less than 0, resp.) indicates higher support for model $\mathcal{M}%
_{2}$ over model $\mathcal{M}_{1}$; a value between 1 to 3 (between $0$ to $%
1.1$, resp.) indicates a slightly higher odds for model $\mathcal{M}_{1}$
versus model $\mathcal{M}_{2}$ that is "non-significant"; a value between 3
to 10 (between $1.1$ to $2.\allowbreak 3$, resp.) indicates substantially
higher odds for model $\mathcal{M}_{1}$; a value between 10 to 30 (between $%
2.\allowbreak 3$ to $\allowbreak 3.\allowbreak 4$, resp.) indicates a
strongly-higher odds in favor of model $\mathcal{M}_{1}$; a value between 30
to 100 (between $3.\allowbreak 4$ to $\allowbreak 4.\allowbreak 6$, resp.)
indicates very strongly-higher odds in favor of model $\mathcal{M}_{1}$; and
a value greater than $100$ ($4.\allowbreak 6$, resp.) indicates a
decisively-higher odds in favor of model $\mathcal{M}_{1}$ (Jeffreys, 1961%
\nocite{Jeffreys61}).

The posterior predictive distribution of an observable dependent response, $%
y $, given a chosen covariate vector $\mathbf{x}$, is given by the Student
distribution with p.d.f.:%
\begin{equation}
\mathrm{st}(y\,|\,\mathbf{x}^{\intercal }\overline{\boldsymbol{\beta }},%
\overline{b}(1+\mathbf{x}^{\intercal }\overline{\mathbf{V}}\mathbf{x}),%
\overline{a})=\dint \dint \mathrm{n}(y\,|\,\mathbf{x}^{\intercal }%
\boldsymbol{\beta },\sigma ^{2})\mathrm{nig}(\boldsymbol{\beta \,},\sigma
^{2}\boldsymbol{\,}|\boldsymbol{\,}\overline{\boldsymbol{\beta }},\overline{%
\mathbf{V}},\,\overline{a},\overline{b})\mathrm{d}\boldsymbol{\beta }\mathrm{%
d}\sigma ^{2},  \label{post pred y}
\end{equation}%
along with mean $\mathbb{E}_{n}(Y\boldsymbol{\,}|\,\mathbf{x})=\mathbf{x}%
^{\intercal }\overline{\boldsymbol{\beta }}$, variance $\mathbb{V}_{n}(Y%
\boldsymbol{\,}|\,\mathbf{x})=(\overline{b}/(\overline{a}-2))(1+\mathbf{x}%
^{\intercal }\overline{\mathbf{V}}\mathbf{x})$, and degrees of freedom $%
\overline{a}$. By extension, the posterior predictive distribution of any
given vector of dependent observations $\mathbf{y}=(y_{1},\ldots
,y_{n})^{\intercal }$, conditionally on any chosen design matrix $\mathbf{X}%
=(x_{ip})_{n\times p}$, is given by the multivariate Student distribution
with p.d.f.:%
\begin{equation}
\mathrm{st}(\mathbf{y}\,|\,\mathbf{X}\overline{\boldsymbol{\beta }},%
\overline{b}(\mathbf{I}_{p}+\mathbf{X\overline{\mathbf{V}}X}^{\intercal }),%
\overline{a})=\diint \mathrm{n}(y\,|\,\mathbf{X}\boldsymbol{\beta },\sigma
^{2}\mathbf{I}_{n})\mathrm{nig}(\boldsymbol{\beta \,},\sigma ^{2}\boldsymbol{%
\,}|\boldsymbol{\,}\overline{\boldsymbol{\beta }},\overline{\mathbf{V}},\,%
\overline{a},\overline{b})\mathrm{d}\boldsymbol{\beta }\mathrm{d}\sigma ^{2},
\label{post pred bold y}
\end{equation}%
along with mean vector $\mathbb{E}_{n}(Y_{1},\ldots ,Y_{n}\boldsymbol{\,}|\,%
\mathbf{X})=\mathbf{X}\overline{\boldsymbol{\beta }}$ and covariance matrix $%
\mathbb{V}_{n}(Y_{1},\ldots ,Y_{n}\boldsymbol{\,}|\,\mathbf{X})=(b/(a-2))(%
\mathbf{I}_{p}+\mathbf{X\overline{\mathbf{V}}X}^{\intercal })$.\ The
diagonal elements of $\mathbb{V}_{n}(Y_{1},\ldots ,Y_{n}\boldsymbol{\,\,}|\,%
\mathbf{X})$ give the posterior predictive variances of $(Y_{1},\ldots
,Y_{n})$, respectively.

The usual auxiliary statistics can be calculated from a Bayesian normal\
linear model that is fit to data $\mathcal{D}_{n}=(\mathbf{X},\mathbf{y})$.
They include:

\begin{itemize}
\item R-squared, $R^{2}=1-\{||\mathbf{y}-\mathbf{X}\overline{\boldsymbol{%
\beta }}||^{2}/||\mathbf{y}-\overline{y}\mathbf{I}_{n}||^{2}\}$, with $%
\overline{y}=\frac{1}{n}\tsum\nolimits_{i=1}^{n}y_{i}$;

\item The $n\times n$ hat matrix%
\begin{equation}
\mathbf{H}=\mathbf{X}(\mathbf{V}^{-1}+\mathbf{X}^{\intercal }\mathbf{X})^{-1}%
\mathbf{X}^{\intercal }=\mathbf{X}\overline{\mathbf{V}}\mathbf{X}^{\intercal
},  \label{HatMat}
\end{equation}%
where $\widehat{\mathbf{y}}=\mathbb{E}(Y_{1},\ldots ,Y_{n}\boldsymbol{\,}|\,%
\mathbf{X})=\mathbf{X}\overline{\boldsymbol{\beta }}=\mathbf{Hy}$, with
diagonal elements $\mathrm{diag}(\mathbf{H)=}$ $(h_{11},\ldots ,h_{nn}%
\mathbf{)}^{\intercal }$ measuring the influence or leverage of the
dependent observations $(y_{1},\ldots ,y_{n}\mathbf{)}$ on the estimates $(%
\widehat{y}_{1},\ldots ,\widehat{y}_{n}\mathbf{)}$, respectively (for
related definitions under the OLS context, see Hoaglin \&\ Welsch 1978\nocite%
{HoaglinWelsch78});

\item Standardized residuals, given by $r_{i}=\{y_{i}-\mathbb{E}_{n}(Y_{i}%
\boldsymbol{\,}|\,\mathbf{x}_{i})\}/\{(1-h_{ii})^{1/2}\mathbb{V}%
_{n}^{1/2}(Y_{i}\boldsymbol{\,}|\,\mathbf{x}_{i})\}$, or approximated by $%
r_{i}\approx \{y_{i}-\mathbb{E}_{n}(Y_{i}\boldsymbol{\,}|\,\mathbf{x}_{i})\}/%
\overline{\sigma }$, for $i=1,\ldots ,n$;

\item The change in coefficient estimates after removing $y_{i}$, given by $%
\overline{\boldsymbol{\beta }}-\overline{\boldsymbol{\beta }}_{(i)}=%
\overline{\mathbf{V}}\mathbf{x}_{i}(y_{i}-\mathbb{E}_{n}(Y_{i}\boldsymbol{\,}%
|\,\mathbf{x}_{i}))/(1-h_{ii})$ , assuming prior mean $\mathbf{m}=\mathbf{0}$
for $\boldsymbol{\beta }$ (Hoaglin \&\ Welsch 1978\nocite{HoaglinWelsch78});

\item The effective degrees of freedom of the model, given by the trace $%
\mathrm{df}=\mathrm{tr}(\mathbf{H)=}\tsum\nolimits_{i=1}^{n}h_{ii}$; while $%
\mathrm{tr}(\mathbf{HH}^{\intercal })$ is the degrees of freedom for the
variance $\sigma ^{2}$, and $\mathrm{tr}(2\mathbf{H}-\mathbf{HH}^{\intercal
} $) is the degrees of freedom for regression error (Hastie \&\ Tibshirani,
1990\nocite{HastieTibshirani90}).
\end{itemize}

In conclusion, as shown in equation (\ref{Update V}), the computation of the
NIG\ posterior mean $\boldsymbol{\,}\overline{\boldsymbol{\beta }}$ and
covariance matrix $\overline{\mathbf{V}}$ depends on a matrix inversion; as
shown in (\ref{margLike}), the computation of the marginal likelihood $\pi (%
\mathcal{D}_{n})$ depends on the computation of matrix determinants $|%
\overline{\mathbf{V}}|^{1/2}$ and $|\mathbf{V}|^{1/2}$; and as shown in (\ref%
{post pred y}) and (\ref{HatMat}), the computation of the posterior
predictive p.d.f. of $\mathbf{y}$ given $\mathbf{X}$, and of the hat matrix $%
\mathbf{H}$, depends on a matrix multiplication $\mathbf{X\overline{\mathbf{V%
}}X}^{\intercal }$. The matrix inverse, determinants, and multiplication are
computationally expensive when either $p$ or $n$ is large, especially when $%
\mathbf{X}^{\intercal }\mathbf{X}$ is non-diagonal. However, as we show in
the next two subsections, under either the Bayesian RR, PRR, or GRR\ models,
and after taking a singular value decomposition (s.v.d.) of $\mathbf{X}$,
the equations of the posterior covariance matrix, marginal likelihood, and
the posterior predictive p.d.f., and hat matrix, can each be simplified to a
form where they do not require the direct computation of any matrix
inverses, determinants, or large-scale matrix multiplications. This fact
helps speed up the computation of these quantities, considerably.

\subsection{Bayesian Ridge Regression}

The ridge regression (RR)\ model can be characterized as a Bayesian linear
model that assigns a normal $\mathrm{n}_{p}(\boldsymbol{\beta \,}|%
\boldsymbol{\,}\mathbf{0},\sigma ^{2}\lambda ^{-1}\mathbf{I}_{p})$ prior
distribution for $\boldsymbol{\beta }$, conditionally on $\sigma ^{2}$. If a
normal inverse-gamma $\mathrm{nig}(\boldsymbol{\beta },\sigma ^{2}%
\boldsymbol{\,}|\boldsymbol{\,}\mathbf{0},\lambda ^{-1}\mathbf{I}_{p},a,b)$
prior is assigned to $(\boldsymbol{\beta },\sigma ^{2})$, then all the same
inferential procedures of the Bayesian normal linear model apply to the
ridge regression model, as described in the previous subsection.

After taking a s.v.d.\ of $\mathbf{X}$, we can gain further insight about
the properties of the estimators under OLS, Bayesian RR, power ridge
regression (PRR), and generalized ridge regression (GRR) models.\ We can
also gain insight about more computationally-efficient strategies for
computing the quantities necessary for performing prior and posterior
inferences with each of the three ridge models.

Recall that the s.v.d.\ of a design matrix $\mathbf{X}$ is given by $\mathbf{%
X}=\mathbf{UDW}^{\intercal }$, where $\mathbf{U}$ and $\mathbf{W}$ are
orthogonal matrices of dimensions $n\times q$ and $p\times q$, respectively,
with $q=\min (n,p)$ and $\mathbf{Z}=\mathbf{UD=XW}$. Here, $\mathbf{D}=%
\mathrm{diag}(d_{1},\ldots ,d_{q})$ is the diagonal matrix of singular
values $d_{1}>\cdots >d_{q}>0$, with $(d_{1}^{2},\ldots ,d_{q}^{2})$ giving
at most the first $q\leq p$ non-zero eigenvalues $(d_{1}^{2},\ldots
,d_{p}^{2})^{\intercal }$ of $\mathbf{X}^{\intercal }\mathbf{X}$, and giving
the diagonal elements of $\mathbf{Z}^{\intercal }\mathbf{Z}$. Also, the
columns of $\mathbf{XW}$ give the $q$ principal components of $\mathbf{X}$,
so that the column-wise sum of squares over the rows yields returns the
eigenvalues $(d_{1}^{2},\ldots ,d_{q}^{2})$.

The canonical\ normal linear model, for the orthogonalized data $(\mathbf{Z},%
\mathbf{y})$, is defined by the multivariate normal likelihood density:%
\begin{equation}
\mathrm{n}_{n}(\mathbf{y}\,|\,\mathbf{XW}\boldsymbol{\alpha }=\mathbf{Z}%
\boldsymbol{\alpha }=\mathbf{X}\boldsymbol{\beta },\sigma ^{2}\mathbf{I}%
_{n}),  \label{Canonical Normal linear}
\end{equation}%
The OLS estimate of the canonical regression coefficients $\boldsymbol{%
\alpha }=(\alpha _{1},\ldots ,\alpha _{q})^{\intercal }$ is given by:%
\begin{equation}
\widehat{\boldsymbol{\alpha }}=\left( \widehat{\alpha }_{1},\ldots ,\widehat{%
\alpha }_{q}\right) ^{\intercal }=\mathbf{D}^{-1}\mathbf{U}^{\intercal }%
\mathbf{y}=(\mathbf{Z}^{\intercal }\mathbf{Z})^{-1}\mathbf{Z}^{\intercal }%
\mathbf{y}=\mathbf{\mathbf{D}}^{-2}\mathbf{Z}^{\intercal }\mathbf{y}=\mathrm{%
diag}(d_{1}^{2},\ldots ,d_{q}^{2})^{-1}\mathbf{Z}^{\intercal }\mathbf{y}.
\label{OLS canonical}
\end{equation}%
Then for the normal linear model for the original data $\mathcal{D}_{n}=(%
\mathbf{X},\mathbf{y})$, with likelihood defined by multivariate normal
likelihood p.d.f.,%
\begin{equation}
\mathrm{n}_{n}(\mathbf{y}\,|\,\mathbf{X}\boldsymbol{\beta },\sigma ^{2}%
\mathbf{I}_{n}),  \label{Normal linear}
\end{equation}%
the OLS estimate of the slope coefficients $\boldsymbol{\beta }$ is given by 
$\widehat{\boldsymbol{\beta }}=\mathbf{W}\widehat{\boldsymbol{\alpha }}$.
Also, for either the canonical linear model defined on the orthogonalized
data $(\mathbf{Z},\mathbf{y})$, or for the normal linear model defined for
the original data $\mathcal{D}_{n}=(\mathbf{X},\mathbf{y})$, the OLS
estimate of the error variance parameter $\sigma ^{2}$ satisfies the
equality:%
\begin{equation*}
\widehat{\sigma }^{2}=\tfrac{1}{n-p}||\mathbf{y}-\mathbf{Z}\widehat{%
\boldsymbol{\beta }}||^{2}=\newline
\tfrac{1}{n-p}||\mathbf{y}-\mathbf{X}\widehat{\boldsymbol{\beta }}||^{2}.
\end{equation*}

Suppose that an NIG prior distribution is assigned to the parameters $(%
\boldsymbol{\alpha },\sigma ^{2})$ of the canonical normal linear model,
with this prior having p.d.f.: 
\begin{subequations}
\label{GRR NIG Prior canonical}
\begin{eqnarray}
\pi (\boldsymbol{\alpha },\sigma ^{2}) &=&\mathrm{nig}(\boldsymbol{\alpha }%
,\sigma ^{2}\boldsymbol{\,}|\boldsymbol{\,}\mathbf{0},\mathrm{diag}(\lambda
_{1},\ldots ,\lambda _{q})^{-1},a,b)  \label{GRR NIG prior canonical 1} \\
&=&\mathrm{n}(\boldsymbol{\alpha }\,|\boldsymbol{\,}\mathbf{0},\sigma ^{2}%
\mathrm{diag}(\lambda _{1},\ldots ,\lambda _{q})^{-1})\mathrm{ig}(\sigma ^{2}%
\boldsymbol{\,}|\boldsymbol{\,}a,b),  \label{GRR NIG prior canonical 2} \\
&=&\mathrm{nig}(\boldsymbol{\alpha },\sigma ^{2}\boldsymbol{\,}|\boldsymbol{%
\,}\mathbf{0},\mathbf{V}_{\boldsymbol{\lambda }}^{(\boldsymbol{\alpha }%
)},a,b)  \label{GRR NIG prior canonical 3}
\end{eqnarray}%
In (\ref{GRR NIG prior canonical 2}), the multivariate normal conditional
prior p.d.f. 
\end{subequations}
\begin{equation}
\pi (\boldsymbol{\alpha }\,|\boldsymbol{\,}\sigma ^{2})=\mathrm{n}(%
\boldsymbol{\alpha }\,|\,\mathbf{0},\sigma ^{2}\mathbf{V}_{\boldsymbol{%
\lambda }}^{(\boldsymbol{\alpha })})=\mathrm{n}(\boldsymbol{\alpha }\,|\,%
\mathbf{0},\sigma ^{2}\mathrm{diag}(\lambda _{1},\ldots ,\lambda _{q})^{-1}),
\label{GRR prior canonical}
\end{equation}%
defines the GRR prior for the slope coefficients $\boldsymbol{\alpha }$ of
the canonical model, given $\sigma ^{2}$.

Based on standard results involving affine transformations, the
(conditional)\ multivariate normal prior distribution (p.d.f.) of (\ref{GRR
prior canonical}), for the slope coefficients $\boldsymbol{\alpha \,}|%
\boldsymbol{\,}\sigma ^{2}$ of the canonical model, implies that for the
original data space, the conditional prior distribution for $\boldsymbol{%
\beta }=\mathbf{W}\boldsymbol{\alpha }$ (given $\sigma ^{2}$) has
multivariate normal p.d.f.:%
\begin{equation}
\pi (\boldsymbol{\beta \,}|\boldsymbol{\,}\sigma ^{2})=\mathrm{n}(%
\boldsymbol{\beta \,}|\boldsymbol{\,}\mathbf{0},\sigma ^{2}\mathbf{V}_{%
\boldsymbol{\lambda }})=\mathrm{n}(\boldsymbol{\beta \,}|\boldsymbol{\,}%
\mathbf{0},\sigma ^{2}\mathbf{W}\mathrm{diag}(\lambda _{1},\ldots ,\lambda
_{q})^{-1}\mathbf{W}^{\intercal })=\mathrm{n}(\boldsymbol{\beta \,}|%
\boldsymbol{\,}\mathbf{0},\sigma ^{2}\mathbf{WV}_{\boldsymbol{\lambda }}^{(%
\boldsymbol{\alpha })}\mathbf{W}^{\intercal })\newline
.  \label{GRR prior}
\end{equation}%
The prior density (\ref{GRR prior}) gives the prior distribution (p.d.f.)\
for the generalized ridge regression (GRR)\ (Hoerl \&\ Kennard, 1970\nocite%
{HoerlKennard70}). Also, the NIG\ prior distribution (p.d.f.)\ for the
parameters $(\boldsymbol{\alpha },\sigma ^{2})$ of the canonical model,
given by (\ref{GRR NIG Prior canonical}), implies a NIG prior distribution
(p.d.f.)\ for the parameters $(\boldsymbol{\beta },\sigma ^{2})$ of the
normal linear model defined on the original data space, with prior p.d.f.
given by: 
\begin{subequations}
\label{NIG prior GRR}
\begin{eqnarray}
\pi (\boldsymbol{\beta },\sigma ^{2}) &=&\mathrm{nig}(\boldsymbol{\beta }%
,\sigma ^{2}\boldsymbol{\,}|\boldsymbol{\,}\mathbf{0},\mathbf{W}\mathrm{diag}%
(\lambda _{1},\ldots ,\lambda _{q})^{-1}\mathbf{W}^{\intercal },a,b) \\
&=&\mathrm{n}(\boldsymbol{\beta \,}|\boldsymbol{\,}\mathbf{0},\sigma ^{2}%
\mathbf{W}\mathrm{diag}(\lambda _{1},\ldots ,\lambda _{q})^{-1}\mathbf{W}%
^{\intercal })\newline
\mathrm{ig}(\sigma ^{2}\boldsymbol{\,}|\boldsymbol{\,}a,b)
\label{NIG ridge prior cond} \\
&=&\mathrm{nig}(\boldsymbol{\beta },\sigma ^{2}\boldsymbol{\,}|\boldsymbol{\,%
}\mathbf{0},\mathbf{W\mathbf{V}_{\boldsymbol{\lambda }}^{(\boldsymbol{\alpha 
})}W}^{\intercal },a,b) \\
&=&\mathrm{nig}(\boldsymbol{\beta },\sigma ^{2}\boldsymbol{\,}|\boldsymbol{\,%
}\mathbf{0},\mathbf{V}_{\boldsymbol{\lambda }},a,b).
\end{eqnarray}

It then follows that under the Bayesian GRR\ model, the prior predictive
distribution of an observable dependent response, $y$, conditionally on
chosen covariates $\mathbf{x}$, is a Student distribution with p.d.f.: 
\end{subequations}
\begin{subequations}
\label{GRR PriorPred}
\begin{eqnarray}
&&\mathrm{st}(y\,|\,\mathbf{0},b(1+\mathbf{x}^{\intercal }\mathbf{Vx}),a) \\
&=&\diint \mathrm{n}(y\,|\,\mathbf{x}^{\intercal }\boldsymbol{\beta },\sigma
^{2})\mathrm{nig}(\boldsymbol{\beta \,},\sigma ^{2}\boldsymbol{\,}|%
\boldsymbol{\,}\mathbf{0},\mathbf{W}\mathrm{diag}(\lambda _{1},\ldots
,\lambda _{q})^{-1}\mathbf{W}^{\intercal },\,a,b)\mathrm{d}\boldsymbol{\beta 
}\mathrm{d}\sigma ^{2},
\end{eqnarray}%
with mean $\mathbb{E}(Y\boldsymbol{\,}|\,\mathbf{x})=\mathbf{x}^{\intercal }%
\mathbf{m}$, variance $\mathbb{V}(Y\boldsymbol{\,}|\,\mathbf{x})=(b/(a-2))(1+%
\mathbf{x}^{\intercal }\mathbf{Vx})$, and degrees of freedom $a$. More
generally, the prior predictive distribution of any given vector of
observations $\mathbf{y}=(y_{1},\ldots ,y_{n})^{\intercal }$, conditionally
on any given design matrix $\mathbf{X}=(x_{ip})_{n\times p}$, is a
multivariate Student distribution with p.d.f.: 
\end{subequations}
\begin{eqnarray}
&&\mathrm{st}(\mathbf{y}\,|\,\mathbf{0},b(\mathbf{I}_{p}+\mathbf{XVX}%
^{\intercal }),a) \\
&=&\diint \mathrm{n}(y\,|\,\mathbf{X}\boldsymbol{\beta },\sigma ^{2}\mathbf{I%
}_{n})\mathrm{nig}(\boldsymbol{\beta \,},\sigma ^{2}\boldsymbol{\,}|%
\boldsymbol{\,}\mathbf{0},\mathbf{W}\mathrm{diag}(\lambda _{1},\ldots
,\lambda _{q})^{-1}\mathbf{W}^{\intercal },\,a,b)\mathrm{d}\boldsymbol{\beta 
}\mathrm{d}\sigma ^{2},
\end{eqnarray}%
along with mean $\mathbb{E}(Y_{1},\ldots ,Y_{n}\boldsymbol{\,}|\,\mathbf{X})=%
\mathbf{Xm}=\mathbf{0}$ (with $\mathbf{m}=\mathbf{0}$) and covariance matrix 
$\mathbb{V}(Y_{1},\ldots ,Y_{n}\boldsymbol{\,}|\,\mathbf{X})=(b/(a-2))(%
\mathbf{I}_{p}+\mathbf{XVX}^{\intercal })$, where the diagonal elements of $%
\mathbb{V}_{n}(Y_{1},\ldots ,Y_{n}\boldsymbol{\,}|\,\mathbf{X})$ give the
prior predictive variances of $(Y_{1},\ldots ,Y_{n})$, respectively.

A special case of the GRR\ prior (\ref{GRR prior}) is given by the ordinary
ridge regression (RR)\ prior, which assumes the equality constraint $\lambda
=\lambda _{1}=\cdots =\lambda _{q}$.\ In this case, the multivariate normal
prior p.d.f. (\ref{GRR prior}) for $\boldsymbol{\beta \,}|\boldsymbol{\,}%
\sigma ^{2}$ is given by: 
\begin{subequations}
\label{ridge prior special case}
\begin{eqnarray}
\pi (\boldsymbol{\beta \,}|\boldsymbol{\,}\sigma ^{2}) &=&\mathrm{n}(%
\boldsymbol{\beta \,}|\boldsymbol{\,}\mathbf{0},\sigma ^{2}\mathbf{V}_{%
\boldsymbol{\lambda }})=\mathrm{n}(\boldsymbol{\beta \,}|\boldsymbol{\,}%
\mathbf{0},\sigma ^{2}\mathbf{W}\mathrm{diag}(\lambda _{1},\ldots ,\lambda
_{q})^{-1}\mathbf{W}^{\intercal }) \\
&=&\mathrm{n}(\boldsymbol{\beta \,}|\boldsymbol{\,}\mathbf{0},\sigma ^{2}%
\mathbf{W\mathbf{V}_{\boldsymbol{\lambda }}^{(\boldsymbol{\alpha })}W}%
^{\intercal })=\mathrm{n}(\boldsymbol{\beta \,}|\boldsymbol{\,}\mathbf{0}%
,\sigma ^{2}\lambda ^{-1}\mathbf{I}_{p}),
\end{eqnarray}%
and corresponds to multivariate normal prior p.d.f.: 
\end{subequations}
\begin{equation}
\pi (\boldsymbol{\alpha }\,|\boldsymbol{\,}\sigma ^{2})=\mathrm{n}(%
\boldsymbol{\alpha \,}|\boldsymbol{\,}\mathbf{0},\sigma ^{2}\mathbf{\mathbf{V%
}_{\boldsymbol{\lambda }}^{(\boldsymbol{\alpha })}})=\mathrm{n}(\boldsymbol{%
\alpha }\,|\boldsymbol{\,}\mathbf{0},\sigma ^{2}\lambda ^{-1}\mathbf{I}_{q})
\label{RR prior canonical}
\end{equation}%
for the canonical normal linear model (\ref{Canonical Normal linear}).

Another special case of the GRR\ prior (\ref{GRR prior}) includes the power
ridge regression (PRR)\ prior (Frank \&\ Friedman, 1993\nocite%
{FrankFriedman93}), which assumes that the prior p.d.f. (\ref{GRR prior})
for $\boldsymbol{\beta \,}|\boldsymbol{\,}\sigma ^{2}$\ is: 
\begin{subequations}
\label{Power ridge prior special case}
\begin{eqnarray}
\pi (\boldsymbol{\beta \,}|\boldsymbol{\,}\sigma ^{2}) &=&\mathrm{n}(%
\boldsymbol{\beta \,}|\boldsymbol{\,}\mathbf{0},\sigma ^{2}\mathbf{W}\mathrm{%
diag}(\lambda _{1},\ldots ,\lambda _{q})^{-1}\mathbf{W}^{\intercal })\newline
\\
&=&\mathrm{n}(\boldsymbol{\beta \,}|\boldsymbol{\,}\mathbf{0},\sigma ^{2}%
\mathbf{W}\mathrm{diag}(d_{1}^{2\delta }/\lambda ,\ldots ,d_{q}^{2\delta
}/\lambda )\mathbf{W}^{\intercal }) \\
&=&\mathrm{n}(\boldsymbol{\beta \,}|\boldsymbol{\,}\mathbf{0},\sigma ^{2}%
\mathbf{W\mathbf{V}_{\boldsymbol{\lambda }}^{(\boldsymbol{\alpha })}W}%
^{\intercal }) \\
&=&\mathrm{n}(\boldsymbol{\beta \,}|\boldsymbol{\,}\mathbf{0},\sigma
^{2}\lambda ^{-1}(\mathbf{X}^{\intercal }\mathbf{X})^{\delta }),
\end{eqnarray}%
for parameters $\delta $ and $\lambda >0$, so that $\lambda
_{k}=d_{k}^{2\delta }/\lambda $, for $k=1,\ldots ,q$. In (\ref{Power ridge
prior special case}), the conditional prior p.d.f. for $\boldsymbol{\beta \,}%
|\boldsymbol{\,}\sigma ^{2}$, for the normal linear model (\ref{Normal
linear}), implies for the canonical linear model (\ref{Canonical Normal
linear}), a multivariate normal prior distribution (p.d.f.)\ for $%
\boldsymbol{\alpha \,}|\boldsymbol{\,}\sigma ^{2}$, defined by: 
\end{subequations}
\begin{equation}
\pi (\boldsymbol{\alpha }\,|\boldsymbol{\,}\sigma ^{2})=\mathrm{n}(%
\boldsymbol{\alpha }\,|\,\mathbf{0},\sigma ^{2}\mathbf{V}_{\boldsymbol{%
\lambda }}^{(\boldsymbol{\alpha })})=\mathrm{n}(\boldsymbol{\alpha }\,|%
\boldsymbol{\,}\mathbf{0},\mathrm{diag}(d_{1}^{2\delta }/\lambda ,\ldots
,d_{q}^{2\delta }/\lambda )).
\end{equation}%
As mentioned, if $\delta =0$, then the PR prior reduces to the ordinary
ridge prior $\mathrm{n}(\boldsymbol{\beta \,}|\boldsymbol{\,}\mathbf{0}%
,\sigma ^{2}/\lambda \mathbf{I}_{p})$, so that $\mathrm{diag}(d_{1}^{2\delta
},\ldots ,d_{q}^{2\delta })=\mathbf{I}_{q}$, thereby expressing no
preference for the eigenvalues. A prior parameter $\delta >0$ expresses
greater preference for larger eigenvalues, and defines a prior that gives
rise to (approximately) principle components regression and penalized least
squares regression. A prior choice $\delta <0$ expresses greater preference
for smaller eigenvalues for the coefficients. In particular, the choice $%
\delta =-1$ yields Zellner's (1986\nocite{Zellner86})\ $g$-prior, in which
case the conditional prior (\ref{Power ridge prior special case}) for $%
\boldsymbol{\beta \,}|\boldsymbol{\,}\sigma ^{2}$ becomes: 
\begin{subequations}
\label{gprior}
\begin{eqnarray}
\pi (\boldsymbol{\beta \,}|\boldsymbol{\,}\sigma ^{2}) &=&\mathrm{n}(%
\boldsymbol{\beta \,}|\boldsymbol{\,}\mathbf{0},\sigma ^{2}\mathbf{W}\mathrm{%
diag}(\lambda _{1},\ldots ,\lambda _{q})^{-1}\mathbf{W}^{\intercal }) \\
&=&\mathrm{n}(\boldsymbol{\beta \,}|\boldsymbol{\,}\mathbf{0},\sigma ^{2}%
\mathbf{W}\mathrm{diag}(d_{1}^{-2}/\lambda ,\ldots ,d_{q}^{-2}/\lambda )%
\mathbf{W}^{\intercal }) \\
&=&\mathrm{n}(\boldsymbol{\beta \,}|\boldsymbol{\,}\mathbf{0},\sigma ^{2}%
\mathbf{W\mathbf{V}_{\boldsymbol{\lambda }}^{(\boldsymbol{\alpha })}W}%
^{\intercal }) \\
&=&\mathrm{n}(\boldsymbol{\beta \,}|\boldsymbol{\,}\mathbf{0},\sigma
^{2}\lambda ^{-1}(\mathbf{X}^{\intercal }\mathbf{X})^{-1}).
\end{eqnarray}%
Also, if $\mathbf{X}^{\intercal }\mathbf{X}$ is singular and $\delta \neq 0$%
, then the PRR\ prior (\ref{Power ridge prior special case}) for the
conditional random variable $\boldsymbol{\beta \,}|\boldsymbol{\,}\sigma
^{2} $ becomes a singular normal distribution.

According to standard inference procedures with the Bayesian linear model,
under the NIG\ prior p.d.f. (\ref{GRR NIG Prior canonical}) for the
parameters $(\boldsymbol{\alpha },\sigma ^{2})$ of the canonical linear
model, including GRR\ prior p.d.f. (\ref{GRR prior canonical}) for $%
\boldsymbol{\alpha \,}|\boldsymbol{\,}\sigma ^{2}$, the posterior
distribution of $(\boldsymbol{\alpha },\sigma ^{2})$ is a NIG distribution
with p.d.f.: 
\end{subequations}
\begin{subequations}
\label{Post CanonRidge}
\begin{eqnarray}
\pi (\boldsymbol{\alpha },\sigma ^{2}\boldsymbol{\,}|\boldsymbol{\,}\mathcal{%
D}_{n}) &=&\mathrm{nig}(\boldsymbol{\alpha },\sigma ^{2}\boldsymbol{\,}|%
\boldsymbol{\,}\overline{\boldsymbol{\alpha }}_{\boldsymbol{\lambda }},%
\mathrm{diag}(\lambda _{1}+d_{1}^{2},\ldots ,\lambda _{q}+d_{q}^{2})^{-1},%
\overline{a},\overline{b}_{\boldsymbol{\lambda }}) \\
&=&\mathrm{n}_{p}(\boldsymbol{\alpha \,}|\boldsymbol{\,}\overline{%
\boldsymbol{\alpha }}_{\boldsymbol{\lambda }},\sigma ^{2}\mathrm{diag}%
(\lambda _{1}+d_{1}^{2},\ldots ,\lambda _{q}+d_{q}^{2})^{-1})\mathrm{ig}%
(\sigma ^{2}\boldsymbol{\,}|\boldsymbol{\,}\overline{a},\overline{b}_{%
\boldsymbol{\lambda }})  \label{Post CanonRidge 2} \\
&=&\mathrm{nig}(\boldsymbol{\alpha },\sigma ^{2}\boldsymbol{\,}|\boldsymbol{%
\,}\overline{\boldsymbol{\alpha }}_{\boldsymbol{\lambda }},(\mathrm{diag}%
(\lambda _{1},\ldots ,\lambda _{q})+\mathbf{Z}^{\intercal }\mathbf{Z})^{-1},%
\overline{a},\overline{b}_{\boldsymbol{\lambda }}) \\
&=&\mathrm{nig}(\boldsymbol{\alpha },\sigma ^{2}\boldsymbol{\,}|\boldsymbol{%
\,}\overline{\boldsymbol{\alpha }}_{\boldsymbol{\lambda }},\overline{\mathbf{%
V}}_{\boldsymbol{\lambda }}^{(\boldsymbol{\alpha })},\overline{a},\overline{b%
}_{\boldsymbol{\lambda }}),
\end{eqnarray}%
where 
\end{subequations}
\begin{equation}
\overline{\boldsymbol{\alpha }}_{\boldsymbol{\lambda }}=\overline{\mathbf{V}}%
_{\boldsymbol{\lambda }}^{(\boldsymbol{\alpha })}\mathbf{Z}^{\intercal }%
\mathbf{y}=\left( \overline{\alpha }_{\boldsymbol{\lambda }1},\ldots ,%
\overline{\alpha }_{\boldsymbol{\lambda }p}\right) ^{\intercal }=\newline
\left( \dfrac{d_{1}^{2}/\lambda _{1}}{1+d_{1}^{2}/\lambda _{1}}\widehat{%
\alpha }_{1},\ldots ,\dfrac{d_{q}^{2}/\lambda _{q}}{1+d_{q}^{2}/\lambda _{q}}%
\widehat{\alpha }_{q}\right) ^{\intercal },  \label{ridge alphabar}
\end{equation}%
and 
\begin{subequations}
\label{Ridge IG shape rate}
\begin{eqnarray}
\overline{a} &=&a+n/2, \\
\overline{b}_{\boldsymbol{\lambda }} &=&\,b+(\mathbf{y}^{\intercal }\mathbf{y%
}-\overline{\boldsymbol{\alpha }}_{\boldsymbol{\lambda }}^{\intercal }%
\overline{\mathbf{V}}_{(\boldsymbol{\alpha })\boldsymbol{\lambda }}^{-1}%
\overline{\boldsymbol{\alpha }}_{\boldsymbol{\lambda }})/2 \\
&=&b+\left( \mathbf{y}^{\intercal }\mathbf{y}-\tsum\nolimits_{k=1}^{q}%
\overline{\alpha }_{\boldsymbol{\lambda }k}^{2}(\lambda
_{k}+d_{k}^{2})\right) /2 \\
&=&\,b+\dfrac{1}{2}\left( \mathbf{y}^{\intercal }\mathbf{y}%
-\tsum\limits_{k=1}^{q}\dfrac{\widehat{\alpha }_{k}^{2}d_{k}^{4}}{\lambda
_{k}+d_{k}^{2}}\right) .
\end{eqnarray}%
It then follows that $\boldsymbol{\beta }=\mathbf{W}\boldsymbol{\alpha }$
has posterior distribution with p.d.f.: 
\end{subequations}
\begin{eqnarray}
\pi (\boldsymbol{\beta },\sigma ^{2}\boldsymbol{\,}|\boldsymbol{\,}\mathcal{D%
}_{n}) &=&\mathrm{nig}(\boldsymbol{\beta },\sigma ^{2}\boldsymbol{\,}|%
\boldsymbol{\,\,}\overline{\boldsymbol{\beta }}_{\boldsymbol{\lambda }},%
\mathbf{W}\mathrm{diag}(\lambda _{1}+d_{1}^{2},\ldots ,\lambda
_{q}+d_{q}^{2})^{-1}\mathbf{W}^{\intercal },\overline{a},\overline{b}_{%
\boldsymbol{\lambda }}) \\
&=&\mathrm{nig}(\boldsymbol{\beta },\sigma ^{2}\boldsymbol{\,}|\boldsymbol{\,%
}\overline{\boldsymbol{\beta }}_{\boldsymbol{\lambda }}=\mathbf{W}\overline{%
\boldsymbol{\alpha }}_{\boldsymbol{\lambda }},\mathbf{W\overline{\mathbf{V}}%
_{\boldsymbol{\lambda }}^{(\boldsymbol{\alpha })}W}^{\intercal },\overline{a}%
,\overline{b}_{\boldsymbol{\lambda }}) \\
&=&\mathrm{nig}(\boldsymbol{\beta \,}|\boldsymbol{\,}\overline{\boldsymbol{%
\beta }}_{\boldsymbol{\lambda }},\mathbf{\overline{\mathbf{V}}}_{\boldsymbol{%
\lambda }},\overline{a},\overline{b}_{\boldsymbol{\lambda }}), \\
&=&\mathrm{n}_{p}(\boldsymbol{\beta \,}|\boldsymbol{\,}\overline{\boldsymbol{%
\beta }}_{\boldsymbol{\lambda }},\sigma ^{2}\mathbf{\overline{\mathbf{V}}}_{%
\boldsymbol{\lambda }})\mathrm{ig}(\sigma ^{2}\boldsymbol{\,}|\boldsymbol{\,}%
\overline{a},\overline{b}_{\boldsymbol{\lambda }}), \\
&=&\mathrm{nig}(\boldsymbol{\beta },\sigma ^{2}\boldsymbol{\,}|\boldsymbol{\,%
}\overline{\boldsymbol{\beta }}_{\boldsymbol{\lambda }},\mathbf{\overline{%
\mathbf{V}}}_{\boldsymbol{\lambda }},\overline{a},\overline{b}_{\boldsymbol{%
\lambda }})
\end{eqnarray}%
with marginal posterior means and variances that are readily calculated by:%
\begin{eqnarray}
\mathbb{E}\left( \boldsymbol{\beta }_{\boldsymbol{\lambda }}\,|\,\mathcal{D}%
_{n}\right) &=&\overline{\boldsymbol{\beta }}_{\boldsymbol{\lambda }}=%
\mathbf{W}\overline{\boldsymbol{\alpha }}_{\boldsymbol{\lambda }},
\label{post mean beta} \\
\mathbb{V}\left( \boldsymbol{\beta }_{\boldsymbol{\lambda }}\,|\,\mathcal{D}%
_{n}\right) &=&(\overline{b}_{\boldsymbol{\lambda }}/(\overline{a}-1))%
\overline{\mathbf{V}}_{\boldsymbol{\lambda }},  \label{post var beta} \\
\mathbb{E}\left( \sigma ^{2}\,|\,\mathcal{D}_{n}\right) &=&\overline{\sigma }%
_{\boldsymbol{\lambda }}^{2}=\overline{b}_{\boldsymbol{\lambda }}/(\overline{%
a}-1), \\
\mathbb{V}\left( \sigma ^{2}\,|\,\mathcal{D}_{n}\right) &=&\overline{b}_{%
\boldsymbol{\lambda }}^{2}/\left\{ (\overline{a}-1)^{2}(\overline{a}%
-2)\right\} ,
\end{eqnarray}%
respectively. Above, $\overline{b}_{\boldsymbol{\lambda }}$ satisfies the
equality:%
\begin{equation}
\overline{b}_{\boldsymbol{\lambda }}=\,b+(\mathbf{y}^{\intercal }\mathbf{y}-%
\overline{\boldsymbol{\beta }}_{\boldsymbol{\lambda }}^{\intercal }\overline{%
\mathbf{V}}_{\boldsymbol{\lambda }}^{-1}\overline{\boldsymbol{\beta }}_{%
\boldsymbol{\lambda }})/2=\,b+\dfrac{1}{2}\left( \mathbf{y}^{\intercal }%
\mathbf{y}-\tsum\limits_{k=1}^{q}\dfrac{\widehat{\alpha }_{k}^{2}d_{k}^{4}}{%
\lambda _{k}+d_{k}^{2}}\right) ,  \label{Post CanonRidge b}
\end{equation}%
which is computationally-efficient since the latter term does not require a
large matrix multiplication.

While the equations (\ref{Post CanonRidge})-(\ref{Post CanonRidge b}) for
the posterior distribution of $(\boldsymbol{\alpha },\sigma ^{2})$\ and of $(%
\boldsymbol{\beta },\sigma ^{2})$ were stated for the Bayesian GRR\ model,
they easily extend to the Bayesian PRR\ model, and to the Bayesian RR\
model, using characterizations of these latter two models, stated earlier.
Specifically, for the Bayesian RR\ model, posterior equations (\ref{Post
CanonRidge})-(\ref{Post CanonRidge b}) hold after assuming $\lambda
_{k}=\lambda $, for $k=1,\ldots ,q$ and for parameter $\lambda $. For the
Bayesian PRR\ model, posterior equations (\ref{Post CanonRidge})-(\ref{Post
CanonRidge b}) hold after assuming $\lambda _{k}=d_{k}^{2\delta }/\lambda $,
for $k=1,\ldots ,q$ and for parameters $\left( \delta ,\lambda \right) $.
For the $g$-prior, given by a PRR\ prior with $\delta =-1$, the conditional
posterior distribution of $\boldsymbol{\beta \,}|\boldsymbol{\,}\sigma ^{2}$
has a $p$-variate normal p.d.f.:%
\begin{equation}
\pi (\boldsymbol{\beta \,}|\boldsymbol{\,}\sigma ^{2},\mathcal{D}_{n})=%
\mathrm{n}\left( \boldsymbol{\beta \,}\left\vert \allowbreak \left( \frac{1}{%
\lambda +1}\right) \widehat{\boldsymbol{\beta }},\left( \frac{1}{\lambda +1}%
\right) \sigma ^{2}(\mathbf{X}^{\intercal }\mathbf{X})^{-1}\right. \right) ,
\end{equation}%
so that under the $g$-prior, the conditional posterior mean and variance of $%
\boldsymbol{\beta }$ is a shrunken version of the OLS estimate $\widehat{%
\boldsymbol{\beta }}$.

Now we remark on the bias and mean-squared error properties of the OLS
estimator $\widehat{\boldsymbol{\beta }}$ and the GRR estimator $\overline{%
\boldsymbol{\beta }}_{\boldsymbol{\lambda }}$. First, it is well-known that
the OLS\ estimator is unbiased, that is, its bias $\mathrm{b}(\widehat{%
\boldsymbol{\beta }}\,|\,\boldsymbol{\beta })$ is given by:%
\begin{equation}
\mathrm{b}(\widehat{\boldsymbol{\beta }}\,|\,\boldsymbol{\beta })=\mathbb{E}%
_{(\mathbf{X},\mathbf{y})}\widehat{\boldsymbol{\beta }}-\boldsymbol{\beta }%
=\dint \widehat{\boldsymbol{\beta }}(\mathbf{X},\mathbf{y})\mathrm{d}F(%
\mathbf{X},\mathbf{y}\,|\,\boldsymbol{\beta },\sigma ^{2})-\boldsymbol{\beta 
}=\mathbf{0}
\end{equation}%
for any given true population parameter $\boldsymbol{\beta }$, with
expectation $\mathbb{E}_{(\mathbf{X},\mathbf{y})}$ taken over a given
population distribution $F(\mathbf{X},\mathbf{y}\,|\,\boldsymbol{\beta }%
,\sigma ^{2})$ of $(\mathbf{X}=(x_{ik})_{n\times p},\mathbf{y}%
=(y_{i})_{p\times 1})$, holding $n>0$ and $p>0$ fixed, and with $\widehat{%
\boldsymbol{\beta }}(\mathbf{X},\mathbf{y})$ denoting an OLS\ estimate
obtained from a sample of $(\mathbf{X},\mathbf{y})$ from that population.
The mean-squared error (MSE) of the OLS estimator, which is also the
variance of the estimator since it is unbiased, is given by: 
\begin{subequations}
\begin{eqnarray}
\mathrm{MSE}(\widehat{\boldsymbol{\beta }}\,|\,\boldsymbol{\beta }) &=&%
\mathbb{E}_{(\mathbf{X},\mathbf{y})}||\widehat{\boldsymbol{\beta }}-%
\boldsymbol{\beta }||^{2}=\mathrm{b}(\widehat{\boldsymbol{\beta }}\,|\,%
\boldsymbol{\beta })^{\top }\mathrm{b}(\widehat{\boldsymbol{\beta }}\,|\,%
\boldsymbol{\beta })+\mathrm{tr}\mathbb{V}_{(\mathbf{X},\mathbf{y})}(%
\widehat{\boldsymbol{\beta }}_{(\mathbf{X},\mathbf{y})}) \\
&=&\mathrm{tr}\mathbb{V}_{(\mathbf{X},\mathbf{y})}(\widehat{\boldsymbol{%
\beta }}_{(\mathbf{X},\mathbf{y})})=\sigma ^{2}\dsum\limits_{k=1}^{p}\dfrac{1%
}{d_{k}^{2}},
\end{eqnarray}%
where $||\cdot ||$ denotes the $L^{2}$ norm, and $\mathrm{tr}\mathbb{V}_{(%
\mathbf{X},\mathbf{y})}$ denotes the trace (tr)\ of the sampling covariance
matrix. Therefore, if $\mathbf{X}^{\mathbf{\intercal }}\mathbf{X}$ is
ill-conditioned with one or more of the eigenvalues $d_{k}^{2}$ near or at
zero, then the corresponding inverses $1/d_{k}^{2}$, and hence the $\mathrm{%
MSE}(\widehat{\boldsymbol{\beta }})$ of the OLS estimator diverges to
infinity.

Under the Bayesian GRR model, the bias for the estimator $\overline{%
\boldsymbol{\beta }}_{\boldsymbol{\lambda }}$ is given by: 
\end{subequations}
\begin{equation}
\mathrm{b}(\overline{\boldsymbol{\beta }}_{\boldsymbol{\lambda }}\,|\,%
\boldsymbol{\beta })=\mathbb{E}_{(\mathbf{X},\mathbf{y})}\overline{%
\boldsymbol{\beta }}_{\boldsymbol{\lambda }}-\boldsymbol{\beta }=\mathbf{W}%
\mathrm{diag}(\lambda _{1},\ldots ,\lambda _{q})\mathbf{W}^{\intercal }%
\boldsymbol{\beta }=\mathbf{W}\mathrm{diag}(\boldsymbol{\lambda })\mathbf{W}%
^{\intercal }\boldsymbol{\beta }
\end{equation}%
(e.g., Singh, 2010\nocite{Singh10}). In general, the bias is non-zero if $%
\boldsymbol{\lambda }$ is non-zero. The mean-squared error of the GRR model
is given by: 
\begin{subequations}
\begin{eqnarray}
\mathrm{MSE}(\overline{\boldsymbol{\beta }}_{\boldsymbol{\lambda }}\,|\,%
\boldsymbol{\beta }) &=&\mathbb{E}_{(\mathbf{X},\mathbf{y})}||\overline{%
\boldsymbol{\beta }}_{\boldsymbol{\lambda }}\{\mathbf{X},\mathbf{y}\}-%
\boldsymbol{\beta }||^{2}=\mathrm{b}(\overline{\boldsymbol{\beta }}_{%
\boldsymbol{\lambda }}\,|\,\boldsymbol{\beta })^{\top }\mathrm{b}(\overline{%
\boldsymbol{\beta }}_{\boldsymbol{\lambda }}\,|\,\boldsymbol{\beta })+%
\mathrm{tr}\mathbb{V}_{(\mathbf{X},\mathbf{y})}(\overline{\boldsymbol{\beta }%
}_{\boldsymbol{\lambda }}) \\
&=&\dsum\limits_{k=1}^{p}\dfrac{d_{k}^{2}\sigma ^{2}+\lambda _{k}^{2}\alpha
_{k}^{2}}{\left( d_{k}^{2}+\lambda _{k}\right) ^{2}}
\end{eqnarray}%
(Goldstein \&\ Smith, 1974\nocite{GoldsteinSmith74}, p.288), which as shown,
is a function of $\boldsymbol{\lambda }=(\lambda _{1},\ldots ,\lambda _{q})$
and is always decreasing at $\boldsymbol{\lambda }=\boldsymbol{0}$. This
implies the existence of a ranges of values of $\boldsymbol{\lambda }$ such
that the $\mathrm{MSE}(\overline{\boldsymbol{\beta }}_{\boldsymbol{\lambda }%
}\,|\,\boldsymbol{\beta })$\ of the ridge estimator is always lower that the 
$\mathrm{MSE}(\widehat{\boldsymbol{\beta }}\,|\,\boldsymbol{\beta })$\ of
the OLS\ estimator, for any linear model. This is true especially when $%
\mathbf{X}^{\mathbf{\intercal }}\mathbf{X}$ is ill-conditioned or singular.
Also the preceding discussion about the bias and the MSE for the estimator\ $%
\overline{\boldsymbol{\beta }}_{\boldsymbol{\lambda }}$ under the Bayesian
GRR\ model also extend to the Bayesian RR\ model under the assumption $%
\lambda _{k}=\lambda $, for $k=1,\ldots ,q$; and extend to the Bayesian PRR
model under the assumption $\lambda _{k}=d_{k}^{2\delta }/\lambda $, for $%
k=1,\ldots ,q$.

As a consequence of taking the s.v.d. of $\mathbf{X}$, the posterior
quantities do not require computationally-expensive matrix operations on
large non-diagonal matrices, for either the Bayesian GRR model, or special
cases of this model including the Bayesian RR model (where $\lambda
_{k}=\lambda $, $k=1,\ldots ,q$), and the Bayesian PRR model (where $\lambda
_{k}=d_{k}^{2\delta }/\lambda $, $k=1,\ldots ,q$). For example, for each of
these three ridge models, the marginal posterior mean of $\boldsymbol{\beta }%
_{\boldsymbol{\lambda }}$ can be efficiently computed by $\overline{%
\boldsymbol{\beta }}_{\boldsymbol{\lambda }}=\mathbf{W}\overline{\boldsymbol{%
\alpha }}_{\boldsymbol{\lambda }}$, as shown by equation (\ref{post mean
beta}); and the computation for the posterior rate $\overline{b}_{%
\boldsymbol{\lambda }}$ of the error variance parameter $\sigma ^{2}$ can be
computed without needing to evaluate the matrix product $\overline{%
\boldsymbol{\beta }}_{\boldsymbol{\lambda }}^{\intercal }\overline{\mathbf{V}%
}_{\boldsymbol{\lambda }}^{-1}\overline{\boldsymbol{\beta }}_{\boldsymbol{%
\lambda }}$, as shown by equation (\ref{Post CanonRidge b}).

As a consequence of taking the s.v.d. of $\mathbf{X}$, we can also calculate
other posterior quantities without needing to perform other
computationally-expensive matrix operations, for either of the three ridge
models. These operations include matrix inversion to obtain the $p\times p$
posterior covariance matrix $\mathbf{\overline{\mathbf{V}}}_{\boldsymbol{%
\lambda }}$; and include large-scale matrix products such as $\mathbf{x}%
^{\intercal }\mathbf{\overline{\mathbf{V}}}_{\boldsymbol{\lambda }}\mathbf{x}
$ or $\mathbf{X\overline{\mathbf{V}}_{\boldsymbol{\lambda }}X}^{\intercal }$
that are needed to make posterior predictive inferences. Such operations are
computationally expensive when either $p$ or $n$ is large.

To elaborate, we now describe the computationally efficient strategies for
the posterior quantities for the Bayesian GRR, RR, and PRR models. In the
remainder of this subsection we will explain these strategies in terms of
the Bayesian GRR model having NIG\ prior (\ref{GRR NIG Prior canonical}),
with the understanding that the Bayesian RR model assumes a specialized NIG
prior under the constraints $\lambda =\lambda _{1}=\cdots =\lambda _{q}$;
while the Bayesian PRR model assumes a specialized prior under the
constraints $\lambda _{k}=d_{k}^{2\delta }/\lambda $ for $k=1,\ldots ,q$.

The marginal posterior variances of $\boldsymbol{\beta }_{\boldsymbol{%
\lambda }}$, extracted from the marginal posterior covariance matrix in
equation (\ref{post var beta}), can be more simply computed as: 
\end{subequations}
\begin{equation}
\QDOVERD( ) {\overline{b}_{\boldsymbol{\lambda }}}{\overline{a}-1}\mathrm{%
diag}(\overline{\mathbf{V}}_{\boldsymbol{\lambda }})=%
\begin{pmatrix}
\widetilde{v}_{\boldsymbol{\lambda }1} \\ 
\vdots \\ 
\widetilde{v}_{\boldsymbol{\lambda }k} \\ 
\vdots \\ 
\widetilde{v}_{\boldsymbol{\lambda }p}%
\end{pmatrix}%
=\dfrac{\overline{b}_{\boldsymbol{\lambda }}}{(\overline{a}-1)}%
\begin{pmatrix}
\tsum\nolimits_{k=1}^{q}w_{1k}^{2}/(\lambda _{k}+d_{k}^{2}) \\ 
\vdots \\ 
\tsum\nolimits_{k=1}^{q}w_{kk}^{2}/(\lambda _{k}+d_{k}^{2}) \\ 
\vdots \\ 
\tsum\nolimits_{k=1}^{q}w_{pk}^{2}/(\lambda _{k}+d_{k}^{2})%
\end{pmatrix}%
,  \label{post var beta simple}
\end{equation}%
where the $w_{1k}$s are the elements of the matrix $\mathbf{W}$ obtained
from the s.v.d. of $\mathbf{X}$. The above computational strategy avoids the
computationally-demanding matrix inversion $\overline{\mathbf{V}}_{%
\boldsymbol{\lambda }}=(\mathbf{V}^{-1}+\mathbf{X}^{\intercal }\mathbf{X)}%
^{-1}$.

Also, while the posterior predictive distribution of an observable response, 
$y$, given a chosen covariate vector $\mathbf{x}$, is a Student distribution
with p.d.f.: 
\begin{equation}
\mathrm{st}(y\,|\,\mathbf{x}^{\intercal }\overline{\boldsymbol{\beta }}_{%
\boldsymbol{\lambda }},\overline{b}(1+\mathbf{x}^{\intercal }\mathbf{%
\overline{\mathbf{V}}}_{\boldsymbol{\lambda }}\mathbf{x}),\overline{a}%
)=\diint \mathrm{n}(y\,|\,\mathbf{x}^{\intercal }\boldsymbol{\beta },\sigma
^{2}\mathbf{I}_{n})\mathrm{nig}(\boldsymbol{\beta \,},\sigma ^{2}\boldsymbol{%
\,}|\boldsymbol{\,}\overline{\boldsymbol{\beta }}_{\boldsymbol{\lambda }},%
\overline{\mathbf{V}}_{\boldsymbol{\lambda }},\,\overline{a},\overline{b}_{%
\boldsymbol{\lambda }})\mathrm{d}\boldsymbol{\beta }\mathrm{d}\sigma ^{2},
\label{PostPred}
\end{equation}%
the variance $\mathbb{V}_{n}(Y\boldsymbol{\,}|\,\mathbf{x})$ of this
distribution can be efficiently computed as:%
\begin{equation}
\mathbb{V}_{n}(Y\boldsymbol{\,}|\,\mathbf{x})=\QDOVERD( ) {\overline{b}_{%
\boldsymbol{\lambda }}}{\overline{a}-2}(1+\mathbf{x}^{\intercal }\mathbf{%
\overline{\mathbf{V}}}_{\boldsymbol{\lambda }}\mathbf{x})=\QDOVERD( ) {%
\overline{b}_{\boldsymbol{\lambda }}}{\overline{a}-2}\left(
1+\dsum\limits_{l=1}^{q}\dfrac{\left(
\tsum\nolimits_{k=1}^{p}x_{k}w_{kl}\right) ^{2}}{\lambda _{l}+d_{l}^{2}}%
\right) ,
\end{equation}%
while avoiding the need to directly evaluate the computationally-expensive
matrix product $\mathbf{x}^{\intercal }\mathbf{\overline{\mathbf{V}}}_{%
\boldsymbol{\lambda }}\mathbf{x}$. Similarly, the posterior predictive
distribution of any chosen vector of dependent response $\mathbf{y}%
=(y_{1},\ldots ,y_{n})^{\intercal }$, conditionally on any chosen design
matrix $\mathbf{X}=(x_{ip})_{n\times p}$, is a multivariate Student
distribution with p.d.f.:%
\begin{equation}
\mathrm{st}(\mathbf{y}\,|\,\mathbf{X}\overline{\boldsymbol{\beta }}_{%
\boldsymbol{\lambda }},\overline{b}(\mathbf{I}_{p}+\mathbf{X\overline{%
\mathbf{V}}_{\boldsymbol{\lambda }}X}^{\intercal }),\overline{a})=\diint 
\mathrm{n}(y\,|\,\mathbf{X}\boldsymbol{\beta },\sigma ^{2}\mathbf{I}_{n})%
\mathrm{nig}(\boldsymbol{\beta \,},\sigma ^{2}\boldsymbol{\,}|\boldsymbol{\,}%
\overline{\boldsymbol{\beta }}_{\boldsymbol{\lambda }},\overline{\mathbf{V}}%
_{\boldsymbol{\lambda }},\,\overline{a},\overline{b}_{\boldsymbol{\lambda }})%
\mathrm{d}\boldsymbol{\beta }\mathrm{d}\sigma ^{2}.
\end{equation}%
Here, the diagonal elements of the covariance matrix $\mathbb{V}_{n}(\mathbf{%
y}\boldsymbol{\,}|\,\mathbf{X})=(b/(a-2))(\mathbf{I}_{p}+\mathbf{X\overline{%
\mathbf{V}}_{\boldsymbol{\lambda }}X}^{\intercal })$, which give the
marginal posterior predictive variances of $(Y_{n},\ldots ,Y_{n})\,|\,%
\mathbf{X}$, can be efficiently computed as:%
\begin{equation}
\mathrm{diag}\{\mathbb{V}_{n}(\mathbf{y}\boldsymbol{\,}|\,\mathbf{X}%
)\}=\QDOVERD( ) {\overline{b}_{\boldsymbol{\lambda }}}{\overline{a}-2}(%
\mathbf{I}_{p}+\mathbf{X\overline{\mathbf{V}}_{\boldsymbol{\lambda }}X}%
^{\intercal })=\QDOVERD( ) {\overline{b}_{\boldsymbol{\lambda }}}{\overline{a%
}-2}%
\begin{pmatrix}
1+\dsum\limits_{l=1}^{q}\dfrac{\left(
\tsum\nolimits_{k=1}^{p}x_{1k}w_{kl}\right) ^{2}}{\lambda _{l}+d_{l}^{2}} \\ 
\vdots \\ 
1+\dsum\limits_{l=1}^{q}\dfrac{\left(
\tsum\nolimits_{k=1}^{p}x_{nk}w_{kl}\right) ^{2}}{\lambda _{l}+d_{l}^{2}}%
\end{pmatrix}%
,
\end{equation}%
without needing to directly evaluate the computationally-expensive matrix
product $\mathbf{X\overline{\mathbf{V}}_{\boldsymbol{\lambda }}X}^{\intercal
}$.

Also, if $\mathbf{y}=(y_{1},\ldots ,y_{n})^{\intercal }$ and $\mathbf{X}%
=(x_{ip})_{n\times p}$ give the observed dependent responses and covariates
of the data $\mathcal{D}=(\mathbf{X},\mathbf{y})$, then the diagonal
elements of the hat matrix $\mathbf{H}=\mathbf{X\overline{\mathbf{V}}_{%
\boldsymbol{\lambda }}X}^{\intercal }$ from (\ref{HatMat}) are readily
obtained as:%
\begin{equation}
\mathrm{diag}(\mathbf{H)}=%
\begin{pmatrix}
h_{11} \\ 
\vdots \\ 
h_{nn}%
\end{pmatrix}%
=\QDOVERD( ) {\overline{a}-2}{\overline{b}_{\boldsymbol{\lambda }}}\mathrm{%
diag}\{\mathbb{V}_{n}(\mathbf{y}\boldsymbol{\,}|\,\mathbf{X})\}-1.
\end{equation}%
The Bayesian GRR model has degrees of freedom (effective number of model
parameters)\ that can be efficiently computed by:%
\begin{equation}
\mathrm{df}=\mathrm{tr}(\mathbf{H)=}\dsum\limits_{i=1}^{n}h_{ii}=\dsum%
\limits_{k=1}^{q}\dfrac{d_{k}^{2}}{d_{k}^{2}+\lambda _{k}},
\end{equation}%
with degrees of freedom for variance efficiently computed by $%
\tsum\nolimits_{k=1}^{q}\left. d_{k}^{4}\right/ (d_{k}^{2}+\lambda _{k})^{2}$%
.

Finally, for either the Bayesian GRR, RR, and PRR regression models, various
auxiliary statistics can be used to test for the (predictive)\ significance
of each covariate $X_{k}$, for $k=1,\ldots ,p$. Since the posterior
probability that $H_{0}:\beta _{k}=0$ is always zero (for $k=1,\ldots ,p$),
a common practice is to view the null hypothesis $H_{0}:\beta _{k}=0$ as an
approximation to the null hypothesis $H_{0}:\beta _{k}\in \lbrack -t,t]$ for
some small constant $t>0$ (Berger 1993\nocite{Berger93}). Then we may reject
the null hypothesis when zero is outside the 95\%\ marginal posterior
interval of $\beta _{k}$, 
\begin{equation}
\left( \mathrm{St}_{2a+n}^{-1}(.025\,|\,\overline{\beta }_{\boldsymbol{%
\lambda }k}/\widetilde{v}_{\boldsymbol{\lambda }k}^{1/2}),\mathrm{St}%
_{2a+n}^{-1}(.975\,|\,\overline{\beta }_{\boldsymbol{\lambda }k}/\widetilde{v%
}_{\boldsymbol{\lambda }k}^{1/2})\right) ,
\end{equation}%
or outside the interquartile (50\%)\ marginal posterior interval of $\beta
_{k}$, 
\begin{equation}
\left( \mathrm{St}_{2a+n}^{-1}(.25\,|\,\overline{\beta }_{\boldsymbol{%
\lambda }k}/\widetilde{v}_{\boldsymbol{\lambda }k}^{1/2}),\mathrm{St}%
_{2a+n}^{-1}(.75\,|\,\overline{\beta }_{\boldsymbol{\lambda }k}/\widetilde{v}%
_{\boldsymbol{\lambda }k}^{1/2})\right) ,
\end{equation}%
or when the scaled neighborhood (SN)\ criterion of $\beta _{k}$, given by%
\begin{equation}
\text{\textrm{SN}}_{k}=\mathrm{St}_{2a+n}(\{\widetilde{v}_{\boldsymbol{%
\lambda }k}^{1/2}-\overline{\beta }_{\boldsymbol{\lambda }k}\}/\widetilde{v}%
_{\boldsymbol{\lambda }k}^{1/2})-\,\mathrm{St}_{2a+n}(\{-\widetilde{v}_{%
\boldsymbol{\lambda }k}^{1/2}-\overline{\beta }_{\boldsymbol{\lambda }k}\}/%
\widetilde{v}_{\boldsymbol{\lambda }k}^{1/2}),
\end{equation}%
is less than $1/2$ (Li\ \&\ Lin, 2010\nocite{LiLin10}). Above, \textrm{St}$%
_{2a+n}(\cdot )$ denotes the standard Student cumulative distribution
function (c.d.f.) with mean zero, variance 1, and $2a+n$ degrees of freedom,
with \textrm{St}$_{2a+n}^{-1}(u)$ giving a quantile at $u\in \lbrack 0,1]$.
Also, the SN\ criterion is defined by the marginal posterior probability
that $\beta _{k}$ lies within the interval $[-\widetilde{v}_{k}^{1/2},%
\widetilde{v}_{k}^{1/2}]$ of $\beta _{k}$ values within one marginal
posterior standard deviation of $0$.

\subsection{MML Estimation of $\protect\lambda $ Based on the S.V.D.\label%
{Estimation of lambda}}

Consider the marginal likelihood (MML) estimation of the parameter $%
\boldsymbol{\lambda }$. For this, we will consider the "Bayes Empirical
Bayes" statistical framework (Deely \&\ Lindley, 1981\nocite{DeelyLindley81}%
), which defines the posterior density of $(\boldsymbol{\beta },\sigma ^{2})$
for the Bayesian GRR model as: 
\begin{subequations}
\label{BEB}
\begin{eqnarray}
\pi _{u}(\boldsymbol{\beta },\sigma ^{2}\,|\,\mathcal{D}_{n}) &=&\dint 
\mathrm{nig}(\boldsymbol{\beta },\sigma ^{2}\boldsymbol{\,}|\boldsymbol{\,}%
\overline{\boldsymbol{\beta }}_{\boldsymbol{\lambda }},\overline{\mathbf{V}}%
_{\boldsymbol{\lambda }},\overline{a},\overline{b}_{\boldsymbol{\lambda }%
})\pi (\boldsymbol{\lambda }\,|\,\mathcal{D}_{n})\mathrm{d}\boldsymbol{%
\lambda }, \\
\pi (\boldsymbol{\lambda }\,|\,\mathcal{D}_{n}) &\propto &\pi (\mathcal{D}%
_{n}\,|\,\boldsymbol{\lambda })\pi (\boldsymbol{\lambda }), \\
\pi (\boldsymbol{\lambda }) &=&\tprod\nolimits_{k=1}^{q}\mathrm{u}(\lambda
_{k}\boldsymbol{\,}|\,0,\lambda _{\max }),
\end{eqnarray}%
with uniform prior densities $\mathrm{u}(\lambda _{k}\boldsymbol{\,}%
|\,0,\lambda _{\max })$ assuming a suitably large $\lambda _{\max }$ such as 
$10^{10}$, and assuming a non-informative prior inverse-gamma density for $%
\sigma ^{2}$, approximated by $\mathrm{ig}(\sigma ^{2}\boldsymbol{\,}%
|\,a=0^{+},b=0^{+})$, with $0^{+}=\lim_{t\downarrow 0}t$. Then in the above
setup (\ref{BEB}), the MML point-estimate $\widehat{\boldsymbol{\lambda }}$
is not only the choice of $\boldsymbol{\lambda }$ which maximizes the
marginal likelihood $\pi (\mathcal{D}_{n}\,|\,\boldsymbol{\lambda })$, and
the log-marginal likelihood $\log \pi (\mathcal{D}_{n}\,|\,\boldsymbol{%
\lambda })$, but is also the mode of the posterior density $\pi (\boldsymbol{%
\lambda }\,|\,\mathcal{D}_{n})$; and gives rise to the posterior mode
point-estimate $\mathrm{nig}(\boldsymbol{\beta },\sigma ^{2}\boldsymbol{\,}|%
\boldsymbol{\,}\overline{\boldsymbol{\beta }}_{\widehat{\boldsymbol{\lambda }%
}},\overline{\mathbf{V}}_{\widehat{\boldsymbol{\lambda }}},\overline{a},%
\overline{b}_{\widehat{\boldsymbol{\lambda }}})$ for the posterior density
(distribution) of $(\boldsymbol{\beta },\sigma ^{2})$.

For the Bayesian RR model, with NIG\ prior density given by equation (\ref%
{NIG prior GRR}) (equation (\ref{GRR NIG Prior canonical}) for the canonical
space), assuming the constraint $\lambda =\lambda _{1}=\cdots =\lambda _{q}$%
, it can be shown using simple algebra that: 
\end{subequations}
\begin{equation}
\log \dfrac{|\overline{\mathbf{V}}_{\boldsymbol{\lambda }}|^{1/2}}{|\mathbf{V%
}_{\boldsymbol{\lambda }}|^{1/2}}=\log \dfrac{|\overline{\mathbf{V}}_{%
\boldsymbol{\lambda }}^{(\boldsymbol{\alpha })}|^{1/2}}{|\mathbf{V}_{%
\boldsymbol{\lambda }}^{(\boldsymbol{\alpha })}|^{1/2}}=\dfrac{1}{2}\left\{
q\log \lambda -\dsum\limits_{k=1}^{q}\log \left( \lambda +d_{k}^{2}\right)
\right\} ,
\end{equation}%
and%
\begin{equation}
\overline{b}_{\boldsymbol{\lambda }}=b+(\mathbf{y}^{\intercal }\mathbf{y}-%
\overline{\boldsymbol{\beta }}_{\boldsymbol{\lambda }}^{\intercal }\overline{%
\mathbf{V}}_{\boldsymbol{\lambda }}^{-1}\overline{\boldsymbol{\beta }}_{%
\boldsymbol{\lambda }})/2=b+\dfrac{1}{2}\left( \mathbf{y}^{\intercal }%
\mathbf{y}-\dsum\limits_{k=1}^{q}\dfrac{\widehat{\alpha }_{k}^{2}d_{k}^{4}}{%
\lambda +d_{k}^{2}}\right) .
\end{equation}

\begin{center}
---------------------------------------------------------------------------------------------------------------------

Figure 1 \ \ \ in \ \texttt{http://www.uic.edu/\symbol{126}%
georgek/HomePage/figuresRidge.pdf}

---------------------------------------------------------------------------------------------------------------------
\end{center}

Then for this model, the log-marginal likelihood $\log \pi (\mathcal{D}_{n}%
\boldsymbol{\,}|\,\boldsymbol{\lambda })$ can be expressed as: 
\begin{subequations}
\begin{eqnarray}
\log \pi (\mathcal{D}_{n}\boldsymbol{\,}|\,\boldsymbol{\lambda }) &=&\log |%
\overline{\mathbf{V}}_{\boldsymbol{\lambda }}|^{1/2}-\log |\mathbf{V}_{%
\boldsymbol{\lambda }}|^{1/2}+a\log b-\overline{a}\log \overline{b}_{%
\boldsymbol{\lambda }} \\
&&+\log \Gamma (\overline{a})-\log \Gamma (a)-\tfrac{n}{2}\log \pi \\
&=&\tfrac{q}{2}\log \lambda -\tfrac{1}{2}\tsum\limits_{k=1}^{q}\log \left(
\lambda +d_{k}^{2}\right) +a\log b \\
&&-\overline{a}\log \,\left\{ b+\dfrac{1}{2}\left( \mathbf{y}^{\intercal }%
\mathbf{y}-\dsum\limits_{k=1}^{q}\dfrac{\widehat{\alpha }_{k}^{2}d_{k}^{4}}{%
\lambda +d_{k}^{2}}\right) \right\} \\
&&+\log \Gamma (\overline{a})-\log \Gamma (a)-\tfrac{n}{2}\log \pi .
\end{eqnarray}%
After retaining only the terms that depend on $\lambda $, we find that the
marginal maximum likelihood estimate $\widehat{\lambda }$ is the choice $%
\lambda \in \lbrack 0,\infty )$ which maximizes: 
\end{subequations}
\begin{equation}
q\log \lambda -\dsum\limits_{k=1}^{q}\log \left( \lambda +d_{k}^{2}\right)
-n\log \left( \mathbf{y}^{\intercal }\mathbf{y}-\dsum\limits_{k=1}^{q}\dfrac{%
\widehat{\alpha }_{k}^{2}d_{k}^{4}}{\lambda +d_{k}^{2}}\right) .
\label{RR maximize}
\end{equation}

Given data $\mathcal{D}_{n}$, the marginal maximum likelihood estimate $%
\widehat{\lambda }$ for the ridge model can be quickly obtained by a
two-step algorithm. The first step evaluates $\log \pi (\mathcal{D}%
_{n}\,|\,\lambda )$ for successive values $\lambda =\tfrac{1}{4}k$, $%
k=0,1,2,\ldots $, until a value $k^{\ast }$ is found such that $\log \pi (%
\mathcal{D}_{n}\,|\,\lambda =\tfrac{1}{4}k^{\ast })<\log \pi (\mathcal{D}%
_{n}\,|\,\lambda =\tfrac{1}{4}(k^{\ast }-1))$. The second step obtains the
estimate $\widehat{\lambda }$ as the the value $\lambda \in \lbrack \max \{0,%
\tfrac{1}{4}(k^{\ast }-1)\},\tfrac{1}{4}(k^{\ast }+1)]$ which minimizes $%
-\log \pi (\mathcal{D}_{n}\,|\,\lambda )$. The minimization step can be
performed by using the \texttt{fminbnd()} function of MATLAB (Natick, MA).

For the Bayesian RR model, Figure 1 presents the relationship between $%
\lambda $, the log marginal likelihood $\log \pi (\mathcal{D}_{n}\boldsymbol{%
\,}|\,\boldsymbol{\lambda })$, mean-squared error of the RR\ estimate $%
\mathrm{MSE}(\overline{\boldsymbol{\beta }}_{\boldsymbol{\lambda }}\,|\,%
\boldsymbol{\beta })$, and the generalized cross-validation criterion $%
\mathrm{GCV}(\lambda )$. This relationship is based on fitting the Bayesian
RR\ model to a data set that was simulated according to $n=100$
observations, $p=10$ covariates, with each dependent observation sampled as $%
Y_{i}\,|\,\mathbf{x}_{i}\sim \mathrm{N}(\mathbf{x}_{i}^{\mathbf{\intercal }}%
\boldsymbol{\beta },s^{2}=1)$, covariates $\mathbf{x}_{i}=\left(
x_{i1},\ldots ,x_{ip}\right) ^{\intercal }$ drawn from a $p$-variate normal
distribution with variances equal 1 and inter-variable correlations of $1/2$%
. Among other things, the figure shows that, as a function of $\lambda $,
the maximization of the log-marginal likelihood roughly corresponds to a
minimization of the mean-squared error. Also, this figure suggests that the
function $\log \pi (\mathcal{D}_{n}\,|\,\lambda )$ is log-concave, and this
will be further illustrated on real data sets in Section 4.

For the Bayesian PRR model, which in terms of the NIG\ prior (\ref{NIG prior
GRR}) (equation (\ref{GRR NIG Prior canonical}) for the canonical space)
assumes that $\boldsymbol{\lambda }$ is a function of two parameters $%
(\lambda ,\delta )$ and the eigenvalues $(d_{k}^{2})_{k=1}^{q}$ of $\mathbf{X%
}^{\mathbf{\intercal }}\mathbf{X}$, the direct evaluation of the matrix
determinants can be avoided because:%
\begin{equation}
\log \dfrac{|\overline{\mathbf{V}}_{\boldsymbol{\lambda }}|^{1/2}}{|\mathbf{V%
}_{\boldsymbol{\lambda }}|^{1/2}}=\log \dfrac{|\overline{\mathbf{V}}_{%
\boldsymbol{\lambda }}^{(\boldsymbol{\alpha })}|^{1/2}}{|\mathbf{V}_{%
\boldsymbol{\lambda }}^{(\boldsymbol{\alpha })}|^{1/2}}=\dfrac{1}{2}%
\dsum\limits_{k=1}^{q}\left\{ \log (\lambda /d_{k}^{2\delta })-\log (\lambda
/d_{k}^{2\delta }+d_{k}^{2})\right\} ,
\end{equation}%
and because:%
\begin{equation}
\overline{b}_{\boldsymbol{\lambda }}=b+(\mathbf{y}^{\intercal }\mathbf{y}-%
\overline{\boldsymbol{\beta }}_{\boldsymbol{\lambda }}^{\intercal }\overline{%
\mathbf{V}}_{\boldsymbol{\lambda }}^{-1}\overline{\boldsymbol{\beta }}_{%
\boldsymbol{\lambda }})/2=b+\dfrac{1}{2}\left( \mathbf{y}^{\intercal }%
\mathbf{y}-\dsum\limits_{k=1}^{q}\frac{(d_{k}^{4\delta +4}\widehat{\alpha }%
_{k}^{2})\left( \lambda /d_{k}^{2\delta }+d_{k}^{2}\right) }{\left( \lambda
+d_{k}^{2\delta +2}\right) ^{2}}\right) .
\end{equation}%
Then for this model, we can write the log marginal likelihood function as:%
\begin{eqnarray}
\log \pi (\mathcal{D}_{n}\boldsymbol{\,}|\,\boldsymbol{\lambda }) &=&\log
\pi (\mathcal{D}_{n}\boldsymbol{\,}|\,\lambda ,\delta ) \\
&=&\log |\overline{\mathbf{V}}_{\boldsymbol{\lambda }}|^{1/2}-\log |\mathbf{V%
}_{\boldsymbol{\lambda }}|^{1/2}+a\log b-\overline{a}\log \overline{b}_{%
\boldsymbol{\lambda }} \\
&&+\log \Gamma (\overline{a})-\log \Gamma (a)-\tfrac{n}{2}\log \pi \\
&=&\tfrac{1}{2}\dsum\limits_{k=1}^{q}\left\{ \log (\lambda /d_{k}^{2\delta
})-\log (\lambda /d_{k}^{2\delta }+d_{k}^{2})\right\} +a\log b \\
&&-\overline{a}\log \,\left\{ \dfrac{1}{2}\left( \mathbf{y}^{\intercal }%
\mathbf{y}-\dsum\limits_{k=1}^{q}\frac{(d_{k}^{4\delta +4}\widehat{\alpha }%
_{k}^{2})\left( \lambda /d_{k}^{2\delta }+d_{k}^{2}\right) }{\left( \lambda
+d_{k}^{2\delta +2}\right) ^{2}}\right) \right\} \\
&&+\log \Gamma (a+n/2)-\log \Gamma (a)-\tfrac{n}{2}\log \pi .
\end{eqnarray}%
After retaining only the terms that depend on $(\lambda ,\delta )$, it then
follows that the marginal maximum likelihood estimate $(\widehat{\lambda },%
\widehat{\delta })$ is the choice of parameters $(\lambda ,\delta )$ that
maximizes:%
\begin{equation}
\dsum\limits_{k=1}^{q}\left\{ \log (\lambda /d_{k}^{2\delta })-\log (\lambda
/d_{k}^{2\delta }+d_{k}^{2})\right\} -n\log \,\left( \mathbf{y}^{\intercal }%
\mathbf{y}-\dsum\limits_{k=1}^{q}\frac{(d_{k}^{4\delta +4}\widehat{\alpha }%
_{k}^{2})\left( \lambda /d_{k}^{2\delta }+d_{k}^{2}\right) }{\left( \lambda
+d_{k}^{2\delta +2}\right) ^{2}}\right) .  \label{PRR maximize}
\end{equation}

The estimate the marginal maximum likelihood estimate $(\widehat{\lambda },%
\widehat{\delta })$ can be obtained by a two stage algorithm. After
initiating with $\widehat{\delta }=0$, the first stage of the algorithm
evaluates $\log \pi (\mathcal{D}_{n}\,|\,\lambda ,\widehat{\delta })$
searches through successive values $\lambda =\tfrac{1}{4}k$, $k=0,1,2,\ldots 
$, until a value $k^{\ast }$ is found such that $\log \pi (\mathcal{D}%
_{n}\,|\,\lambda =\tfrac{1}{4}k^{\ast },\widehat{\delta })<\log \pi (%
\mathcal{D}_{n}\,|\,\lambda =\tfrac{1}{4}(k^{\ast }-1),\widehat{\delta })$,
and then obtains the estimate $\widehat{\lambda }$ as the the value $\lambda
\in \lbrack \max \{0,\tfrac{1}{4}(k^{\ast }-1)\},\tfrac{1}{4}(k^{\ast }+1)]$
which minimizes $-\log \pi (\mathcal{D}_{n}\,|\,\lambda ,\widehat{\delta })$
using the \texttt{fminbnd()} MATLAB\ function. The second stage of the
algorithm updates the estimate $\widehat{\delta }$ as the minimizer of $%
-\log \pi (\mathcal{D}_{n}\,|\,\widehat{\lambda },\delta )$ using \texttt{%
fminbnd()} again. The two stages are repeated until the last update $(%
\widehat{\lambda },\widehat{\delta })$ produces an increase in $\log \pi (%
\mathcal{D}_{n}\,|\,\widehat{\lambda },\delta )$ that is less than $10^{-4}$%
. We will find in Section 4 that this marginal maximum likelihood estimation
algorithm is reasonably fast for a large range of data sets, but it is
necessarily slower than the estimation algorithm for the Bayesian RR model.

For the Bayesian PRR model, Figure 2 presents the relationship between the
parameters $\lambda $ and $\delta $, the log marginal likelihood $\log \pi (%
\mathcal{D}_{n}\boldsymbol{\,}|\,\lambda ,\delta )$, and the mean-squared
error of the PRR\ estimate $\mathrm{MSE}(\overline{\boldsymbol{\beta }}_{%
\boldsymbol{\lambda }}\,|\,\boldsymbol{\beta })$. This displayed
relationship is based on fitting the Bayesian PRR\ model to the same
simulated data set that was analyze by the Bayesian RR\ model, as mentioned
earlier.

\begin{center}
---------------------------------------------------------------------------------------------------------------------

Figure 2 \ \ \ in \ \texttt{http://www.uic.edu/\symbol{126}%
georgek/HomePage/figuresRidge.pdf}

---------------------------------------------------------------------------------------------------------------------
\end{center}

For the Bayesian GRR\ model, with NIG\ prior density given by equation (\ref%
{NIG prior GRR}) (equation (\ref{GRR NIG Prior canonical}) for the canonical
space), the log marginal likelihood of the data can be written as:%
\begin{eqnarray}
\log \pi (\mathcal{D}_{n}\boldsymbol{\,}|\,\boldsymbol{\lambda }) &=&\log |%
\overline{\mathbf{V}}_{\boldsymbol{\lambda }}|^{1/2}-\log |\mathbf{V}_{%
\boldsymbol{\lambda }}|^{1/2}+a\log b-\overline{a}\log \overline{b}_{%
\boldsymbol{\lambda }} \\
&&+\log \Gamma (\overline{a})-\log \Gamma (a)-\tfrac{n}{2}\log \pi \\
&=&\tfrac{1}{2}\dsum\limits_{k=1}^{q}\left\{ \log (\lambda _{k})-\log
(\lambda _{k}+d_{k})\right\} +a\log b \\
&&-\overline{a}\log \,\left\{ b+\dfrac{1}{2}\left( \mathbf{y}^{\intercal }%
\mathbf{y}-\dsum\limits_{k=1}^{q}\dfrac{\widehat{\alpha }_{k}^{2}d_{k}^{4}}{%
\lambda _{k}+d_{k}^{2}}\right) \right\} \\
&&+\log \Gamma (a+n/2)-\log \Gamma (a)-\tfrac{n}{2}\log \pi .
\end{eqnarray}%
For the general ridge model, the direct evaluation of the matrix
determinants $|\overline{\mathbf{V}}|$ and $|\mathbf{V}|$ can be avoided
because:%
\begin{equation}
\log \dfrac{|\overline{\mathbf{V}}_{\boldsymbol{\lambda }}|^{1/2}}{|\mathbf{V%
}_{\boldsymbol{\lambda }}|^{1/2}}=\log \dfrac{|\overline{\mathbf{V}}_{%
\boldsymbol{\lambda }}^{(\boldsymbol{\alpha })}|^{1/2}}{|\overline{\mathbf{V}%
}_{\boldsymbol{\lambda }}^{(\boldsymbol{\alpha })}|^{1/2}}=\dfrac{1}{2}%
\left\{ \dsum\limits_{k=1}^{q}\log (\lambda _{k})-\dsum\limits_{k=1}^{q}\log
(\lambda _{k}+d_{k})\right\} ,
\end{equation}%
and%
\begin{equation}
\overline{b}_{\boldsymbol{\lambda }}=b+(\mathbf{y}^{\intercal }\mathbf{y}-%
\overline{\boldsymbol{\beta }}_{\boldsymbol{\lambda }}^{\intercal }\overline{%
\mathbf{V}}_{\boldsymbol{\lambda }}^{-1}\overline{\boldsymbol{\beta }}_{%
\boldsymbol{\lambda }})/2=b+\dfrac{1}{2}\left( \mathbf{y}^{\intercal }%
\mathbf{y}-\dsum\limits_{k=1}^{q}\dfrac{\widehat{\alpha }_{k}^{2}d_{k}^{4}}{%
\lambda _{k}+d_{k}^{2}}\right) .
\end{equation}%
It then follows that the marginal maximum likelihood estimate of $\widehat{%
\boldsymbol{\lambda }}$ is the maximizer of%
\begin{equation}
\dsum\limits_{k=1}^{q}\left\{ \log (\lambda _{k})-\log (\lambda
_{k}+d_{k})\right\} -n\log \,\left( \mathbf{y}^{\intercal }\mathbf{y}%
-\dsum\limits_{k=1}^{q}\dfrac{\widehat{\alpha }_{k}^{2}d_{k}^{4}}{\lambda
_{k}+d_{k}^{2}}\right) .
\end{equation}%
We can immediately see from the equation above that the estimate $\widehat{%
\boldsymbol{\lambda }}$ is obtained by finding the estimate $\widehat{%
\lambda }_{k}$ that maximizes the functions:%
\begin{equation}
\phi (\lambda _{k})=\log (\lambda _{k})-\log (\lambda _{k}+d_{k})-n\log
\,\left( \mathbf{y}^{\intercal }\mathbf{y}-\dfrac{\widehat{\alpha }%
_{k}^{2}d_{k}^{4}}{\lambda _{k}+d_{k}^{2}}\right) ,
\end{equation}%
for all $k=1,\ldots ,q$, each of which have first derivative:%
\begin{equation}
\dfrac{\mathrm{d}}{\mathrm{d}\lambda _{k}}\phi (\lambda _{k})=-\frac{1}{2}%
\dfrac{d_{k}^{2}\left( d_{k}^{4}\widehat{\alpha }_{k}^{2}+n\lambda
_{k}d_{k}^{2}\widehat{\alpha }_{k}^{2}-(\mathbf{y}^{\intercal }\mathbf{y}%
)d_{k}^{2}-(\mathbf{y}^{\intercal }\mathbf{y})\lambda _{k}\right) }{\lambda
_{k}\left( d_{k}^{2}+\lambda _{k}\right) \left( -d_{k}^{4}\widehat{\alpha }%
_{k}^{2}+(\mathbf{y}^{\intercal }\mathbf{y})d_{k}^{2}+(\mathbf{y}^{\intercal
}\mathbf{y})\lambda _{k}\right) }.
\end{equation}%
Solving for $\lambda _{k}$ at $\dfrac{d}{d\lambda _{k}}\phi (\lambda _{k})=0$
yields:%
\begin{equation}
\widehat{\lambda }_{k}=\max \left( \dfrac{d_{k}^{2}\mathbf{y}^{\intercal }%
\mathbf{y}-d_{k}^{4}\widehat{\alpha }_{k}^{2}}{d_{k}^{2}n\widehat{\alpha }%
_{k}^{2}-\mathbf{y}^{\intercal }\mathbf{y}},\lambda _{\max }\mathbf{1}\left( 
\dfrac{d_{k}^{2}\mathbf{y}^{\intercal }\mathbf{y}-d_{k}^{4}\widehat{\alpha }%
_{k}^{2}}{d_{k}^{2}n\widehat{\alpha }_{k}^{2}-\mathbf{y}^{\intercal }\mathbf{%
y}}\leq 0\right) \right) ,\text{ for }k=1,\ldots ,q,  \label{GRR maximize}
\end{equation}%
to provide the marginal maximum likelihood estimate $\widehat{\boldsymbol{%
\lambda }}=(\widehat{\lambda }_{1},\ldots ,\widehat{\lambda }_{k},\ldots ,%
\widehat{\lambda }_{q})$ for the Bayesian GRR\ model.

\section{On the Ridge Models\ and Corresponding Estimation Approaches}

Sections 1-3 focused on Bayesian GRR, PRR, and RR linear regression models,
assuming a continuous-valued dependent variable. These ridge models and
corresponding estimation approaches may appear limited in scope. However, by
drawing from the existing literature, it may be argued that this scope can
be large for applied regression analysis. We list some of the arguments
below.

\subsection{Ridge Regression as (approximate)\ Bayesian Nonparametric
Modeling}

The ridge regression model allows for a fast least-squares estimation of
model parameters, even when the number of covariates $p$ is very large
(e.g., in the hundreds or thousands), and when $p>n$. This modeling
approach, where the number of coefficient parameters $p$ may be very large,
is similar in spirit to Bayesian nonparametric (BNP)\ modeling.\ This
modeling approach involves the specification of models with infinitely (or
massively) many parameters, for the purposes of providing flexible and
robust statistical inference (M\"{u}ller \&\ Quintana, 2004\nocite%
{MullerQuintana04}). So for the ridge regression model, if the number of
covariates $p$ is chosen to be an increasing function of $n$, then the model
meets this technical definition of the BNP\ model as $n\rightarrow \infty $.
Indeed, it was recently shown that, despite its simple prior structure, the
Bayesian RR model displayed good predictive performance for many data sets
(Griffin \&\ Brown, 2013\nocite{GriffinBrown13}).

For example, a flexible linear model where $p$ grows with $n$, may assume
the mean function:%
\begin{equation}
\mathbb{E}[Y\,|\,\mathbf{x}]=\mathbf{x}^{\intercal }\boldsymbol{\beta }%
=\tsum\limits_{k=1}^{L}\beta _{k}x_{k}+\tsum\limits_{i=1}^{n}\beta _{i}B_{i}(%
\mathbf{x}),  \label{SplineFunct}
\end{equation}%
where $B_{i}(\mathbf{x})$ is a multivariate spline, such as a cubic spline $%
B_{i}(\mathbf{x})=||\mathbf{x}-\mathbf{x}_{i}||^{3}$ with knots $\mathbf{x}%
_{i},$ for $i=1,\ldots ,n$ (M\"{u}ller \& Rios Insua, 1998\nocite%
{MullerRiosinsua98}; Denison et al., 2002\nocite{DenisonHolmesMallickSmith02}%
, p. 102). Here, $\tsum\nolimits_{i=1}^{n}\beta _{i}B_{i}(\mathbf{x})$ is
the linear combination of basis functions that captures departures of
linearity of the underlying regression function. The covariates of this mean
function (\ref{SplineFunct}) can be easily specified (before centering and
scaling all the $p=L+n$ covariates), before subjecting $\boldsymbol{\beta }$
to ridge shrinkage posterior estimation. Also, high dimensional (large $p$)
shrinkage linear regression models can be characterized by L\'{e}vy
processes (see Polson \&\ Scott, 2012\nocite{PolsonScott12}).

\subsection{Ridge Regression as a Student Process}

As the prior predictive density equation (\ref{GRR PriorPred}) shows, the
Bayesian GRR model assumes that $h(\mathbf{x})=\mathbf{x}^{\top }\boldsymbol{%
\beta }$ is a Student process with zero mean function $\mu (\mathbf{x})=%
\mathbb{E}(h(\mathbf{x}))=0$ and non-stationary covariance function%
\begin{equation}
\mathcal{C}(\mathbf{x},\mathbf{x}^{\prime })=\mathbb{E}(h(\mathbf{x})h(%
\mathbf{x}^{\prime }))=b(1+\mathbf{x}^{\intercal }\mathbf{W}\mathrm{diag}%
(\lambda _{1},\ldots ,\lambda _{q})^{-1}\mathbf{W}^{\intercal }\mathbf{x}%
^{\prime })(a-2)^{-1}.
\end{equation}%
Given $\sigma ^{2}$, $h(\mathbf{x})=\mathbf{x}^{\top }\boldsymbol{\beta }$
is a zero-mean Gaussian process (GP)\ under the "weight-space view," with
covariance function 
\begin{equation}
\mathcal{C}(\mathbf{x},\mathbf{x}^{\prime };\mathbf{V})=\mathbf{x}^{\top
}\sigma ^{2}\mathbf{W}\mathrm{diag}(\lambda _{1},\ldots ,\lambda _{q})^{-1}%
\mathbf{W}^{\intercal }\mathbf{x}^{\prime }
\end{equation}%
(Rasmussen \&\ Williams, 2006\nocite{RasmussenWilliams06}). Since the
Student distribution assigns more probability in the tails, compared to the
normal distribution, the Student process provides more robust inference than
the GP (Denison et al. 2002\nocite{DenisonHolmesMallickSmith02}, p. 29).
However, the difference between the two processes is minimal for reasonable
sample sizes ($n>100$), because the Student process tends to a GP as $%
\overline{a}\rightarrow \infty $, as shown by the posterior predictive (\ref%
{PostPred}).

Nevertheless, the GP, and the Student process, provide examples of flexible
BNP\ models (M\"{u}ller \&\ Quintana, 2004\nocite{MullerQuintana04}). While
the "weight-space view" may not provide a fully-flexible GP\ modeling
approach (Rasmussen \&\ Williams, 2006\nocite{RasmussenWilliams06}), an
important limitation of the GP approach is that it usually requires repeated
inversions of $n\times n$ matrices, in order to compute the covariance
function for different values of the covariance parameters. Such matrix
inversions are computationally demanding or even prohibitive when $n$ is
sufficiently large. A possible remedy for large-$n$ settings is to
approximate a pure GP\ model by using a lower-dimensional covariance
function matrix. Alternatively, as done in this paper, a flexible and more
interpretable GP can always be specified, in a far more
computationally-efficient manner, by adopting the weight-space view and
taking the number of covariates $p$ to be very large, with the covariates
possibly including spline terms.

\subsection{Ridge Regression for Classification and Bayesian Density
Regression}

As mentioned in Section 1, we assume that dependent variable observations
are zero-mean centered and continuous. This is done with no loss of
generality, as we argue now. First, binary regression, that is regression
involving a binary (two-class)\ dependent variable, is often of interest in
statistical practice. Logistic regression provides a standard binary
regression model. However, both frequentist and Bayesian estimation of the
logistic regression model requires iteration, and this estimation can become
too computationally expensive for data sets that are sufficiently large. An
alternative and computationally-fast (albeit less natural)\ approach is to
code the binary class observations as $\widetilde{Y}_{i}\in \{-1,1\}$ (for $%
i=1,\ldots ,n$), and then apply the cutoff of 0 to the predictions based on
a shrinkage linear regression model fit by penalized least-squares (Hastie
et al., 2009, Section 16.4\nocite{HastieTibsFriedman09}). In this spirit,
the ridge regression model can be fit to the dependent responses $\mathbf{y}=%
\widetilde{\mathbf{y}}-\overline{y}$ (with $\overline{y}$ the mean of $%
\widetilde{\mathbf{y}}=(\widetilde{y}_{1},\ldots ,\widetilde{y}%
_{n})^{\intercal }$). Then for a given covariate vector $\mathbf{x}$, the
posterior predictive probability of a positive class $Y_{i}^{(c)}=1$ can be
estimated by%
\begin{equation}
\widehat{\Pr }(Y+\overline{y}\geq 0\,|\,\mathbf{x})=\dint\limits_{0}^{\infty
}\mathrm{st}\left( y+\overline{y}\,|\,\mathbf{x}^{\intercal }\overline{%
\boldsymbol{\beta }}_{\lambda },\overline{b}_{\lambda }\left(
1+\dsum\limits_{l=1}^{q}\dfrac{\left(
\tsum\nolimits_{k=1}^{p}x_{k}w_{kl}\right) ^{2}}{\lambda _{l}+d_{l}^{2}}%
\right) ,\overline{a}\right) \mathrm{d}y,  \label{ClassPred}
\end{equation}%
and the posterior predictive probability of class $\widetilde{Y}=-1$ can be
estimated by $1-\widehat{\Pr }(Y+\overline{y}\geq 0\,|\,\mathbf{x})$, where
the Student density in (\ref{ClassPred}) is from (\ref{PostPred}).

When there are more than two classes, labeled as $c=1,\ldots ,C$, with
observations $c_{l}$ (for $l=1,\ldots ,n_{c}$), it is possible to take a
"one-versus-all" approach to linear classification (Rifkin \&\ Klatau, 2004%
\nocite{RifkinKlautau04}). That is, consider the vector of $n=n_{c}C$
centered dependent variable observations $\mathbf{y}=\mathbf{y}^{(c)}-%
\overline{y}$, where $\mathbf{y}^{(c)}=(\mathbf{y}_{1}^{\intercal },\ldots ,%
\mathbf{y}_{l}^{\intercal },\ldots ,\mathbf{y}_{n_{c}}^{\intercal
})^{\intercal }$, $\mathbf{y}_{l}^{\intercal }=(y_{l}^{(1)},\ldots
,y_{l}^{(C)})^{\intercal }$ and $y_{l}^{(c)}=\mathbf{1}(c_{l}=c)-\mathbf{1}%
(c_{l}\neq c)$ for $c=1,\ldots ,C$ and $l=1,\ldots ,n_{c}$, and with $%
\overline{y}$ the mean of $\mathbf{y}^{(c)}$. Each observation vector $%
\mathbf{y}_{l}^{\intercal }$, for $l=1,\ldots ,n_{c}$, corresponds to a
covariate matrix $\underline{\mathbf{X}}_{l}$ consisting of row vectors $%
\underline{\mathbf{x}}_{lc}^{\intercal }=(\mathbf{1}(c_{l}=1)\underline{%
\mathbf{x}}_{l}^{\intercal },\ldots ,\mathbf{1}(c_{l}=C)\underline{\mathbf{x}%
}_{l}^{\intercal })$, for $c=1,\ldots ,C$, respectively. Also let $\mathbf{X}
$ be the design matrix, after centering and scaling each of the columns of $%
\underline{\mathbf{X}}=(\underline{\mathbf{X}}_{1}^{\intercal },\ldots ,%
\underline{\mathbf{X}}_{n_{c}}^{\intercal })^{\intercal }$, with $\underline{%
\mathbf{X}}$ having column means $\boldsymbol{\mu }_{x}$ and column standard
deviations $\boldsymbol{\sigma }_{x}$. Then we may consider a linear model $%
\mathrm{n}_{n}(\mathbf{y}\,|\,\mathbf{X}\boldsymbol{\beta },\sigma ^{2}%
\mathbf{I}_{n})$, with $\boldsymbol{\beta }$ subject to ridge regression
shrinkage estimation. For a given covariate vector $\mathbf{x}$, we may
estimate the posterior predictive probabilities $\widehat{\Pr }(Y+\overline{y%
}\geq 0\,|\,\mathbf{z}_{c})$, for $c=1,\ldots ,C$, with $\mathbf{z}_{c}=(%
\mathbf{x}_{c}-\boldsymbol{\mu }_{x})\boldsymbol{\sigma }_{x}^{-1}$ and $%
\underline{\mathbf{x}}_{c}^{\intercal }=(\mathbf{1}(c=1)\underline{\mathbf{x}%
}^{\intercal },\ldots ,\mathbf{1}(c=C)\underline{\mathbf{x}}_{l}^{\intercal
})$. Then the optimal class prediction is given by the class $c\in
\{1,\ldots ,C\}$ with the highest $P_{c}=\widehat{\Pr }(Y+\overline{y}\geq
0\,|\,\mathbf{z}_{c})$.

Moreover, $P_{c0}-\xi $, for $c=0,1,\ldots ,C$, collectively provide a
corrected\ histogram probability density estimate for the $C$ classes, if $%
\xi $ is chosen so that $\tsum\nolimits_{c=1}^{C}P_{c0}-\xi =1$ (Glad et al.
2003\nocite{GladHjortUshakov03}). If so desired, the given histogram may be
smoothed, say, by linear interpolation. Then here, the linear model provides
a type of "Bayesian density regression" (e.g., see Karabatsos \&\ Walker,
2012\nocite{KarabatsosWalker12c}). Also, this histogram/density correction
method provides another way to estimate class probabilities from linear
classifiers (see Wu et al., 2004\nocite{WuLinWeng04}).

\section{Illustrations\label{Illustrations}}

\subsection{Illustrations on Real Data}

We illustrate the marginal maximum likelihood estimation algorithm on a
variety of data sets that range widely in terms of sample size ($n$)\ and
the number of covariates ($p$). The ten data sets are described as follows:

\begin{itemize}
\item The \texttt{Iris} data set (Fisher, 1936\nocite{Fisher36}), with
septal length as the dependent variable, and with covariates of septal
width, petal length, and petal width.

\item The \texttt{Teacher} data set, a time series data set collected to
investigate the effect of a new teacher curriculum, versus the old
curriculum, on the respective teaching abilities of undergraduate teacher
education students who attended one of four Chicago universities (see
Karabatsos \&\ Walker, 2015\nocite{KarabatsosWalker15}). The dependent
response is a test score on math teaching ability, obtained after completing
a math teaching course. The covariates include time (in years), indicator
(0,1)\ of new curriculum, and 347 indicators (0,1) of student, which were
included for the purposes of providing a robust posterior (mean) estimate of
the new curriculum effect (coefficient).

\item Two versions of the classic, \texttt{Diabetes} data set (Efron et al.
2004\nocite{EfronHastieJohnstoneTibshirani04}), each with dependent variable
defined by a measure of disease progression one year after baseline. This
first version of the data set, \texttt{DiabetesQ}, includes 65 covariates
defined by 10 covariates at baseline, their squares, and their two-way
interactions. The second version of the data set, \texttt{DiabetesS},
includes over 500 covariates, including the 65 quadratic covariates, plus
442 65-variate cubic splines with knots defined by all $n=442$ observed
values of the 65-dimensional covariate vectors (see Section 3).

\item The \texttt{Meaning} data set, which contains the ratings of 1,194
high school students from 9 public high schools of Chicago and New York. The
students individually provided ratings on an 18-item survey about the
meaningfulness of past reading experiences (Tatum \&\ Karabatsos, 2013\nocite%
{TatumKarabatsos13}). In total, the data set includes nearly 21,000 ratings,
and over 100 student, teacher, and school background covariates.

\item The \texttt{Blog} data set, with the dependent variable being the
number of comments on the blog after 24 hours. The sample size exceeds
52,000, and there are over 2,500 covariates, formed by taking the powers of
1, 2, and 3 of each of the 840 original covariates. This large data set
file, named blogData\_train.csv, was obtained from
https://archive.ics.uci.edu/ml/datasets/BlogFeedback.

\item The \texttt{Wheat} and \texttt{Yarn} data sets, which have been used
to illustrate the weaknesses of RR in past research. The \texttt{Yarn} data
set contains 28 samples of Positive Emission Tomography (PET) yarns, with
dependent variable being is density of yarn, and covariates (predictors)\
being the Near Infrared (NIR) spectrum of 268 wavelengths. This data set was
obtained from the pls package (Wehrens \&\ Mevik, 2007\nocite{WehrensMevik07}%
) of the R software (2014\nocite{Rsoftware15}). We consider this data set,
since the ordinary ridge model was found to have relatively worse
cross-validated predictive error, compared to linear models assigned other
more sophisticated shrinkage priors (Griffin \&\ Brown, 2013\nocite%
{GriffinBrown13}). This is apparently true because the RR model can shrink
over-shrink the OLS\ coefficients $\widehat{\boldsymbol{\alpha }}=\left( 
\widehat{\alpha }_{1},\ldots ,\widehat{\alpha }_{q}\right) ^{\intercal }$
for components that have relatively small eigenvalues (Polson \&\ Scott, 2012%
\nocite{PolsonScott12}; Griffin \&\ Brown, 2013\nocite{GriffinBrown13}). A
similar point was made in a previous study (Fearn, 1983\nocite{Fearn83})
that provided the \texttt{Wheat} data set, consisting of $24$ observations
of protein content, regressed on the reflectance of NIR radiation by the
wheat samples at six different wavelengths in the range 1680-2310 nm.

\item The \texttt{Lymphoma} data set, containing nearly 80 observations on
over 7,000 covariates, with dependent variable indicating $(1,-1)$ the
presence of diffuse large b-cell lymphoma versus Follicular Lymphoma (FL)
morphology (Shipp et al. 2012\nocite{ShippRossEtAl02}). This large data set
is available through a web link associated with the cited article.

\item The \texttt{Cancer} data set (Petricoin, et al. 2002\nocite%
{Petricoin_EtAl02}), containing observations of a binary dependent variable
indicating $(1,-1)$ either the presence or absence of ovarian cancer, and
containing data on over 15,000 covariates. This large data set is available
through a web link associated with the cited article.
\end{itemize}

\noindent For each data set, before fitting the ridge regression model, the
observations of each of the variables were mean-centered, and each of the $p$
covariates were also scaled to have variance 1, as mentioned in Section 1.
For each of the \texttt{Iris}, \texttt{Teacher}, \texttt{Diabetes}, \texttt{%
Meaning}, \texttt{Blog}, \texttt{Wheat}, and \texttt{Yarn} data sets, the
dependent responses were also scaled to have variance 1.

For each of the 10 data sets, Table \ref{Table Times} presents computation
times for various tasks of data analysis, in seconds. All reported
computation times are based on a 64-bit laptop computer with 2.8 GHz Intel
Core i7 processor and 16 megabytes RAM, with no parallel computing
techniques used. Fairly similar computation times were obtained from an
older 64-bit laptop with a 2.4 GHz Intel i5 processor and 4 megabytes RAM.

Table \ref{Table Times} reports the time needed: to load the data into
MATLAB (times reported under column "data"); to compute the singular value
decomposition (s.v.d.) of the given design matrix $\mathbf{X}$ (column
"s.v.d."); and to compute OLS\ estimates $\widehat{\boldsymbol{\beta }}$
provided a non-singular $\mathbf{X}$ (column "OLS", with "Sing" indicating a
singular $\mathbf{X}$).

\noindent 
\begin{table}[H] \centering%
\begin{tabular}{lccccccc}
&  &  & \multicolumn{5}{c}{\textbf{Computation Time (seconds)}} \\ 
\cline{4-8}
\textbf{Data Set} & $n$ & $p$ & \textbf{data} & \textbf{s.v.d.} & \textbf{OLS%
} & \textbf{RR} & \textbf{RRN} \\ \hline
\texttt{Iris} & $150$ & $3$ & $0.01$ & $.001$ & $0.0003$ & $0.003$ & $0.01$
\\ 
\texttt{Teacher} & $347$ & $349$ & $1.35$ & $.04$ & Sing & $0.003$ & $0.17$
\\ 
\texttt{DiabetesQ} & $442$ & $65$ & $0.93$ & $.01$ & $0.002$ & $0.008$ & $%
0.07$ \\ 
\texttt{DiabetesS} & $442$ & $507$ & $1.\,\allowbreak 00$ & $.03$ & Sing & $%
0.054$ & $0.189$ \\ 
\texttt{Meaning} & $20,994$ & $113$ & $8.20$ & $0.05$ & Sing & $0.019$ & $%
29266$ \\ 
\texttt{Blog} & $52,397$ & $2,520$ & $3.77$ & $2.05$ & Sing & $0.074$ & $>8$h
\\ 
\texttt{Wheat} & $24$ & $6$ & $0.52$ & $0.001$ & $0.002$ & $0.001$ & $0.06$
\\ 
\texttt{Yarn} & $28$ & $268$ & $1.01$ & $0.001$ & Sing & $0.002$ & $0.004$
\\ 
\texttt{Lymphoma} & $77$ & $7,129$ & $1.82$ & $0.01$ & Sing & $0.002$ & $%
0.01 $ \\ 
\texttt{Cancer} & $253$ & $15,154$ & $11.34$ & $0.10$ & Sing & $0.002$ & $%
0.34$ \\ \hline
& \multicolumn{1}{l}{} &  &  &  &  &  &  \\ 
\textbf{Data Set} & $n$ & $p$ & \textbf{RRf} & \textbf{HKB} & \textbf{GCV} & 
\textbf{Ridge10} &  \\ \cline{1-7}
\texttt{Iris} & $150$ & $3$ & $0.001$ & $0.00002$ & $14.56$ & $0.71$ &  \\ 
\texttt{Teacher} & $347$ & $349$ & $>8$h & $0.05$ & $43.80$ & $6.16$ &  \\ 
\texttt{DiabetesQ} & $442$ & $65$ & $0.02$ & $0.00003$ & $23.80$ & $1.57$ & 
\\ 
\texttt{DiabetesS} & $442$ & $507$ & $0.05$ & $0.18$ & $58.30$ & $11.35$ & 
\\ 
\texttt{Meaning} & $20,994$ & $113$ & $>8$h & $0.002$ & $644.48$ & $44.46$ & 
\\ 
\texttt{Blog} & $52,397$ & $2,520$ & $>8$h & $0.03$ & $9074.19$ & $851.96$ & 
\\ 
\texttt{Wheat} & $24$ & $6$ & $0.22$ & $0.0003$ & $13.62$ & $0.21$ &  \\ 
\texttt{Yarn} & $28$ & $268$ & $0.03$ & $0.001$ & $16.99$ & $4.83$ &  \\ 
\texttt{Lymphoma} & $77$ & $7,129$ & $>8$h & $0.002$ & $208.30$ & $123.33$ & 
\\ 
\texttt{Cancer} & $253$ & $15,154$ & $>8$h & $0.02$ & $2245.26$ & $300.74$ & 
\\ \hline
\end{tabular}%
\caption{Computation times, for s.v.d.: singular value decomposition; OLS:
Ordinary Least-Squares Estimate; RR: MML estimation for Ridge Regression
model, using algorithm of this paper, based on the s.v.d.; RRN: Ridge
regression using Neto et al. (2014) marginal likelihood equation based on
the s.v.d.; RRf: Ridge regression using full marginal likelihood equation,
using direct calculation of matrix determinants; HKB: Ridge regression using
HKB plug-in estimator of lambda; GCV: Ridge regression using minimum GCV
estimator of lambda; Ridge10: Ridge regression using estimator of lambda
based on minimizing 10-fold cross-validated mean square predictive error;
Sing: OLS estimate non-existent because the cross-product of the design
matrix X was singular. > 8h means that computations took over 8 hours to complete.}%
\label{Table Times}%
\end{table}%

Table \ref{Table Times} also reports the time needed to obtain the ridge
marginal maximum likelihood (MML) estimate $\widehat{\lambda }$ for the
Bayesian RR model, for various methods. Methods include the MML\ estimation
algorithm introduced in Section 2.3, which maximizes equation (\ref{RR
maximize}) based on the s.v.d. of $\mathbf{X}$ (column "RR"); the MML
estimation algorithm that directly uses the $n\times n$ orthogonal matrix $%
\mathbf{U}$ obtained from the s.v.d. $\mathbf{X}=\mathbf{UDW}^{\intercal }$
to compute the marginal likelihood over trial values of $\lambda $ (Neto et
al., 2014\nocite{NetoJangFriendMargolin14}) (column "RRN"); and the MML
estimation algorithm that computes the full marginal likelihood equation (%
\ref{margLike}) directly, including the direct computation of the matrix
determinants, and without taking a s.v.d. of $\mathbf{X}$ (column "RRf").
The table also reports the computation time to obtain the ridge estimate $%
\widehat{\lambda }$, based on other estimators.\ The other estimators
include the HKB\ plug-in estimator $\widehat{\lambda }_{\text{HKB}}$ (column
"HKB"); the estimator of $\lambda $ based on finding the value $\lambda $
that minimizes the $\mathrm{GCV}(\lambda )$\ criterion over a grid of trial
lambda values ranging from 0 to 500 separated by .005 (column "GCV"); and
the estimator of $\lambda $ based on minimizing the 10-fold cross-validated
mean square predictive error over a grid of 100 trial lambda values
equally-spaced on a log scale that range from 0 a value $\lambda _{\max }$.
Here, $\lambda _{\max }$ is chosen large enough to produce a coefficient
vector estimate $\widehat{\boldsymbol{\beta }}_{\lambda _{\max }}$ that is
virtually a zero $\mathbf{0}$ vector, a default of the \texttt{lasso()}
procedure of MATLAB (Friedman et al. 2010\nocite{FriedmanHastieTibshirani10}%
). In general, for a trial value $\boldsymbol{\lambda }$ of the shrinkage
parameter of a linear model (e.g., the Bayesian RR model has scalar
shrinkage parameter $\boldsymbol{\lambda }=\lambda $), the 10-fold
cross-validated mean-squared predictive error of the trial value $%
\boldsymbol{\lambda }$ is given by:%
\begin{equation}
\mathrm{CV}_{10}(\boldsymbol{\lambda })=\dfrac{1}{n}\dsum\limits_{k=1}^{10}%
\dsum\limits_{i\in C_{k}}\left( y_{i}-\mathbf{x}_{i}^{\intercal }\widehat{%
\boldsymbol{\beta }}_{\boldsymbol{\lambda }}^{(-k)}\right) ^{2},  \label{CV}
\end{equation}%
based on a random partition of the observation indices $i=1,\ldots ,n$ into
subsets $C_{k}$, for $k=1,\ldots ,10$. The mean square statistic $\mathrm{CV}%
_{10}(\boldsymbol{\lambda })$ has standard error:%
\begin{eqnarray}
\mathrm{SE}\left[ \mathrm{CV}_{10}(\boldsymbol{\lambda })\right] &=&\dfrac{1%
}{\sqrt{10}}\left\{ \dfrac{1}{9}\dsum\limits_{k=1}^{10}\left( \mathrm{CV}%
_{10}^{(k)}(\boldsymbol{\lambda })-\overline{\mathrm{CV}_{10}^{(k)}(%
\boldsymbol{\lambda })}\right) ^{2}\right\} ^{1/2}, \\
\text{with }\mathrm{CV}_{10}^{(k)}(\boldsymbol{\lambda }) &=&\dfrac{1}{n_{k}}%
\dsum\limits_{i\in C_{k}}\left( y_{i}-\mathbf{x}_{i}^{\intercal }\widehat{%
\boldsymbol{\beta }}_{\boldsymbol{\lambda }}^{(-k)}\right) ^{2}.
\end{eqnarray}

As shown in Table \ref{Table Times} (column "RR"), for each of the 10 data
sets, the MML\ estimation algorithm that maximizes equation (\ref{RR
maximize}) (in Section 2.3, using the s.v.d. of $\mathbf{X}$) computed the
MML\ estimate $\widehat{\lambda }$ in less than one-tenth of a second. For
all 10 data sets, this MML estimation algorithm was faster than the MML\
estimation algorithm that instead uses the $n\times n$ orthogonal matrix $%
\mathbf{U}$ of the s.v.d. to compute the marginal likelihood equation (of
Neto et al., 2014\nocite{NetoJangFriendMargolin14}) over trial values of $%
\lambda $; and faster than the MML\ estimation algorithm that instead
computes the full marginal likelihood equation (\ref{margLike}) directly,
including the computation of the matrix determinants. For each of the 
\texttt{Meaning} and \texttt{Blog} data sets, the MML\ estimation algorithm
that computed the marginal likelihood using the matrix $\mathbf{U}$ (Neto et
al., 2014\nocite{NetoJangFriendMargolin14}) needed over 8 hours to obtain
the ridge MML\ estimate $\widehat{\lambda }$. Also, for 5 of the 10 data
sets, the MML estimation algorithm that directly computes the full marginal
likelihood equation (\ref{margLike}) led to a MML\ estimation algorithm that
needed over 8 hours to obtain the ridge MML\ estimate $\widehat{\lambda }$.
Finally, as might be expected, the MML\ estimation algorithm that maximizes
equation (\ref{RR maximize}) required more computation time than the
automatic plug-in HKB\ estimator of the ridge parameter $\widehat{\lambda }_{%
\text{HKB}}$, and required less computation time than the ridge estimator of 
$\lambda $ that is based on finding the value of $\lambda $ that minimizes
the $\mathrm{GCV}(\lambda )$ criterion.

\begin{table}[H] \centering%
\begin{tabular}{lccccccc}
\hline
&  &  & \multicolumn{5}{c}{\textbf{Computation Time (seconds)}} \\ 
\cline{4-8}\cline{8-8}
\textbf{Data Set} & $n$ & $p$ & RR & PRR & GRR & FB & L10CV \\ \hline
\texttt{Iris} & $150$ & $3$ & $0.003$ & $0.05$ & $0.00001$ & $11.55$ & $0.97$
\\ 
\texttt{Teacher} & $347$ & $349$ & $0.003$ & $0.44$ & $0.00002$ & $153.31$ & 
$10.94$ \\ 
\texttt{DiabetesQ} & $442$ & $65$ & $0.008$ & $0.44$ & $0.00001$ & $18.29$ & 
$61.93$ \\ 
\texttt{DiabetesS} & $442$ & $507$ & $0.054$ & $2.75$ & $0.00002$ & $331.19$
& $780.14$ \\ 
\texttt{Meaning} & $20,994$ & $113$ & $0.019$ & $3.86$ & $0.00003$ & $38.44$
& $679.07$ \\ 
\texttt{Blog} & $52,397$ & $2,520$ & $0.074$ & $26.70$ & $0.00003$ & $%
1373.39 $ & $11259.2$ \\ 
\texttt{Wheat} & $24$ & $6$ & $0.001$ & $0.02$ & $0.00001$ & $10.77$ & $%
13.96 $ \\ 
\texttt{Yarn} & $28$ & $268$ & $0.002$ & $0.02$ & $0.00001$ & $13.39$ & $%
5.80 $ \\ 
\texttt{Lymphoma} & $77$ & $7,129$ & $0.002$ & $0.30$ & $0.00002$ & $44.35$
& $69.73$ \\ 
\texttt{Cancer} & $253$ & $15,154$ & $0.002$ & $0.04$ & $0.00002$ & $299.30$
& $272.81$ \\ \hline
& \multicolumn{1}{l}{} &  &  &  &  &  &  \\ 
\textbf{Data Set} & $n$ & $p$ & EN10CV & LGIC & EGIC & LBIC & EBIC \\ \hline
\texttt{Iris} & $150$ & $3$ & $1.14$ & $0.14$ & $0.15$ & $0.17$ & $0.17$ \\ 
\texttt{Teacher} & $347$ & $349$ & $9.87$ & $1.56$ & $1.66$ & $1.63$ & $1.62$
\\ 
\texttt{DiabetesQ} & $442$ & $65$ & $49.15$ & $5.19$ & $4.27$ & $4.93$ & $%
4.34$ \\ 
\texttt{DiabetesS} & $442$ & $507$ & $557.19$ & $72.50$ & $50.78$ & $72.45$
& $50.67$ \\ 
\texttt{Meaning} & $20,994$ & $113$ & $703.43$ & $78.93$ & $81.59$ & $92.55$
& $90.67$ \\ 
\texttt{Blog} & $52,397$ & $2,520$ & $16890.3$ & $1290.1$ & $1510.8$ & $%
1458.6$ & $1683.6$ \\ 
\texttt{Wheat} & $24$ & $6$ & $7.95$ & $1.43$ & $0.85$ & $1.41$ & $0.86$ \\ 
\texttt{Yarn} & $28$ & $268$ & $6.41$ & $0.67$ & $0.75$ & $0.64$ & $0.75$ \\ 
\texttt{Lymphoma} & $77$ & $7,129$ & $68.58$ & $10.94$ & $11.30$ & $11.18$ & 
$11.08$ \\ 
\texttt{Cancer} & $253$ & $15,154$ & $265.21$ & $41.73$ & $41.20$ & $41.16$
& $40.53$ \\ \hline
\end{tabular}%
\caption{Computation time to estimate the shrinkage parameter (lambda), 
for RR: Ridge Regression via MML estimation;
PRR: Power Ridge Regression via MML estimation;
GRR: Generalized Ridge Regression via MML estimation; 
FB: Ridge Regression based on MCMC sampling; 
L10CV: LASSO via 10-fold cross-validation;
EN10CV: Elastic Net via 10-fold cross-validation;
LGIC: LASSO via GIC minimization;
EGIC: Elastic Net via GIC minimization;
LBIC: LASSO via BIC minimization;
EBIC: Elastic Net via BIC minimization.
Sing: OLS estimate non-existent because the cross-product of the design matrix X was singular.}%
\label{Table Times 2}%
\end{table}%

Table \ref{Table Times 2} compares the computation times of the MML\
estimation algorithm that finds the MML estimate $\widehat{\lambda }$ which
maximizes equation (\ref{RR maximize}) (in Section 2.3, using s.v.d. of $%
\mathbf{X}$), against the computation times of other models and related
estimation approaches, namely:

\begin{enumerate}
\item The Bayesian PRR\ model under the MML\ estimation algorithm that finds
the MML estimate $(\widehat{\lambda },\widehat{\delta })$ maximizes equation
(\ref{PRR maximize}) (Section 2.3, using s.v.d. of $\mathbf{X}$);

\item Bayesian GRR\ model that makes use of the automatic plug-in estimator $%
\widehat{\boldsymbol{\lambda }}$ of equation (\ref{GRR maximize}) for MML\
estimation (Section 2.3, using s.v.d. of $\mathbf{X}$);

\item The Bayesian RR\ model with full posterior distribution estimated
using 110,000 MCMC sampling iterations, with the model assuming normal data
likelihood density,\newline
$\mathrm{n}_{n}(\mathbf{y}\,|\,\mathbf{X}\boldsymbol{\beta },\sigma ^{2}%
\mathbf{I}_{n})$, and parameters $(\boldsymbol{\beta },\sigma ^{2},\lambda )$
assigned a conjugate normal inverse-gamma prior distribution with
probability density function:%
\begin{equation*}
\pi (\boldsymbol{\beta },\sigma ^{2},\lambda )=\mathrm{n}(\boldsymbol{\beta }%
\,|\,\mathbf{0},\sigma ^{2}\lambda ^{-1}\mathbf{I}_{p})\mathrm{ig}(\sigma
^{2}\,|\,a,b)\mathrm{ga}(\lambda \,|\,a_{\lambda },b_{\lambda })
\end{equation*}%
(Denison et al. 2002\nocite{DenisonHolmesMallickSmith02}), based on an
attempt to specify a non-informative prior $(\sigma ^{2},\lambda )\,$by
choosing each of the prior parameters $(a,b,a_{\lambda },b_{\lambda })$ to
be near zero;

\item Three approaches to the LASSO\ model with shrinkage parameter
estimated, respectively, by the choice of value $\lambda $ which minimizes
either the 10-fold cross-validated mean-squared predictive error $\mathrm{CV}%
_{10}(\lambda )$ of equation (\ref{CV}); or the Generalized Information
Criterion $\mathrm{GIC}(\lambda )$ (Fan \&\ Tang, 2013\nocite{FanTang13}),
given by:%
\begin{equation*}
\mathrm{GIC}(\lambda )=\frac{1}{n}\left\{ ||\mathbf{y}-\mathbf{X}\widehat{%
\boldsymbol{\beta }}_{\lambda }||^{2}+a_{n}|\mathcal{S}_{\lambda }|\right\}
\end{equation*}%
\noindent where $|\mathcal{S}_{\lambda }|$ is the number of non-zero terms
in $\widehat{\boldsymbol{\beta }}_{\lambda }$. The choice $a_{n}=\log \{\log
(n)\}\log (p)$ has been recommended in practice (Fan \&\ Tang, 2013, p. 539%
\nocite{FanTang13}). The modified version of the Bayesian information
criterion (BIC; Schwarz, 1978\nocite{Schwarz78}), $\mathrm{BIC}(\lambda )$,
is a special case of the $\mathrm{GIC}(\lambda )$ that assumes $a_{n}=\log
\{\log (n)\}\log (n)$, and provides consistent model selection in terms of
identifying the subset of the coefficients $\boldsymbol{\beta }$ that are
zero (Fan \&\ Tang, 2013, p. 539\nocite{FanTang13});

\item Three approaches to the EN\ model assuming $\alpha =1/2$ and with
shrinkage parameter estimated, respectively, by the choice of value $\lambda 
$ which minimizes either the 10-fold cross-validated mean-squared predictive
error $\mathrm{CV}_{10}(\lambda )$ of equation (\ref{CV}); the Generalized
Information Criterion $\mathrm{GIC}(\lambda )$ (Fan \&\ Tang, 2013\nocite%
{FanTang13}); or the modified BIC.\newline
\end{enumerate}

\noindent The approaches based on $\mathrm{CV}_{10}(\lambda )$ minimization
are each based on obtaining the estimate $\widehat{\lambda }$ as the value
of $\lambda $ that minimizes $\mathrm{CV}_{10}(\lambda )$ over a grid of 100 
$\lambda $ values ranging from $0$ to $\lambda _{\max }$, equally-spaced on
a log scale, such that $\lambda _{\max }$ is chosen as the minimum value of $%
\lambda $ where $\widehat{\boldsymbol{\beta }}_{\lambda }=\mathbf{0}$.\ This
is a default for the \texttt{lasso()} procedure of MATLAB (Friedman et al.
2010\nocite{FriedmanHastieTibshirani10}). The approaches based on $\mathrm{%
GIC}(\lambda )$ minimization (or modified $\mathrm{BIC}(\lambda )$
minimization) are each based on obtaining the estimate $\widehat{\lambda }$
as the value of $\lambda $ that minimizes\ $\mathrm{GIC}(\lambda )$ (or the
modified $\mathrm{BIC}(\lambda )$) over a grid of 200 $\lambda $ values
ranging from $0$ to $\lambda _{\max }$, equally-spaced on a log scale, such
that $\lambda _{\max }$ is chosen as the minimum value of $\lambda $ where $%
\widehat{\boldsymbol{\beta }}_{\lambda }=\mathbf{0}$, as suggested in a
prior study (Fan \&\ Tang, 2013\nocite{FanTang13}).

The MCMC\ approach to estimating the Bayesian\ RR\ model deserves some
elaboration. In each MCMC\ sampling iteration, a sample from the full
conditional posterior distribution of the coefficients $\boldsymbol{\beta }$
is obtained by first drawing a random sample of $\boldsymbol{\alpha }$ from
the full conditional posterior distribution, given by:%
\begin{equation}
\Pi \left( \boldsymbol{\alpha \,}|\,\mathcal{D}_{n},\sigma ^{2}\right)
=\tprod\nolimits_{k=1}^{p}\mathrm{n}(\alpha _{k}\boldsymbol{\,}|\boldsymbol{%
\,}\overline{\alpha }_{\boldsymbol{\lambda }k},\sigma ^{2}(\lambda
+d_{k}^{2})^{-1})
\end{equation}%
and then taking $\boldsymbol{\beta }=\mathbf{W}\boldsymbol{\alpha }$ as a
sample from the full conditional posterior distribution of $\boldsymbol{%
\beta }$ (Polson \&\ Scott, 2012\nocite{PolsonScott12}); and then a sample
of the error variance $\sigma ^{2}$ from its full conditional distribution
is obtained by drawing from an inverse gamma distribution with p.d.f. $%
\mathrm{ig}(\sigma ^{2}\boldsymbol{\,}|\boldsymbol{\,}\overline{a},\overline{%
b}_{\boldsymbol{\lambda }})$, according to equations (\ref{ridge alphabar})
and (\ref{Ridge IG shape rate}) assuming $\lambda =\lambda _{1}=\cdots
=\lambda _{r}$; whereas a sample of $\lambda $ from its full conditional
posterior distribution is obtained by drawing from a gamma distribution with
p.d.f. $\mathrm{ga}(a_{\lambda }+p/2,b_{\lambda }+(2\sigma
^{2})^{-1}\tsum\nolimits_{k=1}^{p}\beta _{k}^{2})$. A total of 110,000 MCMC
sampling iterations has been suggested as providing a converged estimate of
the posterior distribution of the model parameters $(\boldsymbol{\beta }%
,\sigma ^{2},\lambda )$ for the Bayesian RR model (Tsonias \&\ Tassiopoulos,
2014\nocite{TsionasTassiopoulos14}).

\noindent 
\begin{table}[H] \centering%
\begin{tabular}{lcccccccc}
& \multicolumn{8}{c}{$\lambda $ estimate (and $\delta $ for PRR; for GRR,
min $\lambda $ and max $\lambda $)} \\ \cline{2-9}
\textbf{Data Set} & RR & $\,\text{HKB}$ & $\text{GCV}$ & $\text{FB}$ & $%
\text{PRR}$ & $\delta _{\text{PRR}}$ & \multicolumn{2}{c}{GRR} \\ \hline
\texttt{Iris} & $0.17$ & $0.16$ & $0.07$ & $0.001$ & $0.03$ & $-0.53$ & $%
0.48 $ & $2.98$ \\ 
\texttt{Teacher} & $10^{-4}$ & $2\times 10^{4}$ & $0.001$ & $2.53$ & $0.001$
& $2.78$ & $181.93$ & $10^{10}$ \\ 
\texttt{DiabetesQ} & $67.70$ & $0.35$ & $85.95$ & $10^{-4}$ & $536.79$ & $%
0.33$ & $41.41$ & $10^{10}$ \\ 
\texttt{DiabetesS} & $446.05$ & $10^{7}$ & $103.89$ & $164.91$ & $2\times
10^{3}$ & $0.19$ & $2.21$ & $10^{10}$ \\ 
\texttt{Meaning} & $250.45$ & $0.55$ & $4.83$ & $10^{-4}$ & $0.86$ & $-0.65$
& $56.07$ & $10^{10}$ \\ 
\texttt{Blog} & $410.88$ & $10^{-10}$ & $4.81$ & $10^{-16}$ & $0.49$ & $%
-0.68 $ & $5\times 10^{3}$ & $10^{10}$ \\ 
\texttt{Wheat} & $0.003$ & $0.003$ & $0.01$ & $0.02$ & $0.005$ & $-0.35$ & $%
0.08$ & $10^{10}$ \\ 
\texttt{Yarn} & $10^{-4}$ & $9.79$ & $10^{-3}$ & $10^{3}$ & $4\times 10^{-5}$
& $0.11$ & $10^{10}$ & $10^{10}$ \\ 
\texttt{Lymphoma} & $10^{-4}$ & $8\times 10^{5}$ & $10^{-3}$ & $2\times
10^{19}$ & $4\times 10^{-5}$ & $0.29$ & $2\times 10^{4}$ & $10^{10}$ \\ 
\texttt{Cancer} & $10^{-4}$ & $4\times 10^{8}$ & $10^{-3}$ & $3\times
10^{19} $ & $4\times 10^{-5}$ & $0.11$ & $10^{10}$ & $10^{10}$ \\ \hline
&  &  &  &  &  &  &  &  \\ 
\textbf{Data Set} & $\text{L10CV}$ & $\text{E10CV}$ & $\text{LGIC}$ & $\text{%
EGIC}$ & $\text{LBIC}$ & $\text{EBIC}$ &  &  \\ \cline{1-7}
\texttt{Iris} & $10^{-4}$ & $0.0002$ & $10^{-4}$ & $0.0002$ & $0.02$ & $0.02$
&  &  \\ 
\texttt{Teacher} & $0.02$ & $0.04$ & $0.16$ & $0.33$ & $0.16$ & $0.33$ &  & 
\\ 
\texttt{DiabetesQ} & $0.04$ & $0.07$ & $0.12$ & $0.24$ & $0.12$ & $0.12$ & 
&  \\ 
\texttt{DiabetesS} & $0.04$ & $0.07$ & $0.12$ & $0.24$ & $0.24$ & $0.24$ & 
&  \\ 
\texttt{Meaning} & $10^{-5}$ & $4\times 10^{-5}$ & $0.01$ & $0.02$ & $0.02$
& $0.04$ &  &  \\ 
\texttt{Blog} & $10^{-3}$ & $0.0003$ & $0.02$ & $0.04$ & $0.02$ & $0.05$ & 
&  \\ 
\texttt{Wheat} & $.0003$ & $10^{-3}$ & $0.01$ & $0.01$ & $0.02$ & $0.01$ & 
&  \\ 
\texttt{Yarn} & $0.02$ & $10^{-3}$ & $0.29$ & $1.95$ & $0.07$ & $1.95$ &  & 
\\ 
\texttt{Lymphoma} & $0.04$ & $0.14$ & $0.56$ & $1.12$ & $0.56$ & $1.12$ &  & 
\\ 
\texttt{Cancer} & $0.01$ & $0.01$ & $0.18$ & $1.70$ & $0.18$ & $0.77$ &  & 
\\ \hline
\end{tabular}%
\caption{Lambda estimates,  
for RR: Ridge Regression via MML estimation;
HKB: Ridge Regression via the HKB estimator;
GCV: Ridge Regression via the minimum GCV estimator;
FB: Ridge Regression via MCMC sampling; 
PRR: Power Ridge Regression via MML estimation (including delta);
GRR: Generalized Ridge Regression via MML estimation; 
L10CV: LASSO via 10-fold cross-validation;
EN10CV: Elastic Net via 10-fold cross-validation;
LGIC: LASSO via GIC minimization;
EGIC: Elastic Net via GIC minimization;
LBIC: LASSO via BIC minimization;
EBIC: Elastic Net via BIC minimization.}\label{Table Lambda Estimates}%
\end{table}%
\noindent

We see from Table \ref{Table Times 2} that, predictably, the Bayesian GRR\
model with automatic plug-in MML estimator $\widehat{\boldsymbol{\lambda }}$
obtained the fastest computation times across all the data sets by far,
followed by the iterative Bayesian RR\ MML\ estimation procedure, and then
followed by the iterative Bayesian PRR\ MML\ estimation procedure. Also, the
MCMC\ estimation procedures for the Bayesian RR model, the cross-validation
estimation procedures, and the estimation procedures based either on
minimizing $\mathrm{GIC}(\lambda )$\ or minimizing $\mathrm{BIC}(\lambda )$,
were noticeably slower. Table \ref{Table Lambda Estimates} presents the
point-estimates of $\lambda $ (or $\boldsymbol{\lambda }$ or $(\lambda
,\delta )$) for the various estimation procedures.

\noindent 
\begin{table}[H] \centering%
\begin{tabular}{lccccc}
& \multicolumn{5}{c}{Log marginal likelihood $\log \pi (\mathcal{D}_{n}\,|\,%
\widehat{\boldsymbol{\lambda }})$} \\ \cline{2-6}
\textbf{Data Set} & RR & RR:HKB & RR:GCV & PRR & GRR \\ \hline
\texttt{Iris} & $-61.73^{\ast }$ & $-61.73^{\ast }$ & $-62.12^{\ast }$ & $%
-61.02^{\ast }$ & $-64.40$ \\ 
\texttt{Teacher} & $-396.88$ & $-409.60$ & $-398.34$ & $-385.07$ & $%
-333.14^{\ast }$ \\ 
\texttt{DiabetesQ} & $-389.63$ & $-494.49$ & $-390.06$ & $-388.88$ & $%
-373.33^{\ast }$ \\ 
\texttt{DiabetesS} & $-375.83$ & $-511.75$ & $-392.63$ & $-375.61$ & $%
-325.36^{\ast }$ \\ 
\texttt{Meaning} & $-21262.87$ & $-21480.97$ & $-21378.56$ & $-21232.10$ & $%
-21168.23^{\ast }$ \\ 
\texttt{Blog} & $-44738.55$ & $-48392.54$ & $-44898.27$ & $-44334.46$ & $%
-43931.02^{\ast }$ \\ 
\texttt{Wheat} & $-31.50^{\ast }$ & $-31.51^{\ast }$ & $-31.74^{\ast }$ & $%
-31.05^{\ast }$ & $-40.37$ \\ 
\texttt{Yarn} & $40.22$ & $-0.49$ & $38.72$ & $42.02^{\ast }$ & $-33.56$ \\ 
\texttt{Lymphoma} & $-73.32$ & $-107.32$ & $-74.79$ & $-71.83^{\ast }$ & $%
-69.49^{\ast }$ \\ 
\texttt{Cancer} & $52.74$ & $-298.40$ & $51.27$ & $55.00^{\ast }$ & $-61.43$
\\ \hline
\end{tabular}%
\caption{Log-marginal likelihood, for RR: the Ridge Regression model via MML estimation; 
RR: HKB: the Ridge Regression model via the HKB estimator; 
PRR: the Power Ridge Regression model via MML estimation; 
GRR: Generalized Ridge Regression via MML estimation.}\label{Table LML
results ridge}%
\end{table}%

\begin{center}
---------------------------------------------------------------------------------------------------------------------

Figure 3 \ \ \ in \ \texttt{http://www.uic.edu/\symbol{126}%
georgek/HomePage/figuresRidge.pdf}

---------------------------------------------------------------------------------------------------------------------
\end{center}

Focusing on the point-estimation procedures involving either the Bayesian
RR, PRR, or GRR models, Table \ref{Table LML results ridge} compares the
log-marginal likelihood between the procedures and models. For the Bayesian
RR model, we find that for all 10 data sets, the MML\ estimation procedure
led to a higher log-marginal likelihood, compared to the HKB estimator, and
estimation via GCV minimization. Also, the Bayesian GRR model under MML
estimation attained the highest marginal likelihood for 6 of the 10 data
sets, while the Bayesian PRR model under MML estimation attained the highest
marginal likelihood for all the remaining 4 data sets, compared to the
Bayesian RR model under either MML, HKB, or GCV estimation of the shrinkage
parameter ($\lambda $). For the \texttt{Iris} and \texttt{Wheat}\ data sets,
the Bayesian RR and PRR models and associated estimation procedures were
tied for first, whereas for the \texttt{Lymphoma} data set the Bayesian PRR
and GRR models and associated MML\ estimation procedures were tied for first
(within 1.1 units of the log-marginal likelihood; see Section 1).

\begin{center}
---------------------------------------------------------------------------------------------------------------------

Figure 4 \ \ \ in \ \texttt{http://www.uic.edu/\symbol{126}%
georgek/HomePage/figuresRidge.pdf}

---------------------------------------------------------------------------------------------------------------------
\end{center}

For each of the ten data sets, Figure 3 presents the log marginal likelihood 
$\log \pi (\mathcal{D}_{n}\,|\,\lambda )$ as a function of $\lambda $, along
with the estimate $\widehat{\lambda }$ shown by a vertical dashed line. This
figure shows the concavity of $\log \pi (\mathcal{D}_{n}\,|\,\lambda )$.
Figure 4 presents, for the Bayesian PRR model, the surface of the log
marginal likelihood $\log \pi (\mathcal{D}_{n}\,|\,\lambda ,\delta )$ as a
function of $(\lambda ,\delta )$, for each of the 10 data sets.

For further illustration, Figure 5 presents the results of the \texttt{%
DiabetesQ} data, under the Bayesian GRR model estimated under MML\
maximization. As reported in Table \ref{Table LML results ridge}, this model
attained the highest log-marginal likelihood for this data set. In this
figure, the left panel presents the coefficient estimates $\overline{%
\boldsymbol{\beta }}_{\widehat{\lambda }}$ for the 65 covariates. For each
of the covariates, respectively, the middle panel presents a box plot of the
marginal posterior distributions of the coefficients including the posterior
mean estimate, the 50\%\ marginal posterior interval, and the 95\% marginal
posterior credible interval; and the right panel presents the results of the
scaled neighborhood criterion. According to the 95\%\ marginal posterior
credible interval of the middle panel, the 15 following covariates were
significant predictors of disease progression: \texttt{age}, \texttt{sex},%
\texttt{\ bmi,} \texttt{map}, \texttt{tch}, \texttt{ltg}, \texttt{glu}, 
\texttt{age*sex}, \texttt{age*ltg}, \texttt{sex*tch}, \texttt{sex*glu}, 
\texttt{bmi*map}, \texttt{map*ltg}, \texttt{age\symbol{94}2}, and \texttt{sex%
\symbol{94}2}.

\begin{center}
---------------------------------------------------------------------------------------------------------------------

Figure 5 \ \ \ in \ \texttt{http://www.uic.edu/\symbol{126}%
georgek/HomePage/figuresRidge.pdf}

---------------------------------------------------------------------------------------------------------------------
\end{center}

Across the 10 data sets, Table \ref{Table CV MS compare} compares the
cross-validated mean-squared predictive error $\mathrm{CV}_{10}(\widehat{%
\boldsymbol{\lambda }})$ between the Bayesian RR, PRR, and GRR models, and
the LASSO\ and EN\ models, conditionally on the estimate $\widehat{%
\boldsymbol{\lambda }}$ (e.g., the estimate $\widehat{\lambda }$ or $(%
\widehat{\lambda },\widehat{\delta })$ or $\widehat{\boldsymbol{\lambda }}$,
as applicable) obtained from any one of the various estimation procedures
mentioned earlier. The Bayesian GRR model under MML\ estimation obtained the
best (lowest) $\mathrm{CV}_{10}(\widehat{\boldsymbol{\lambda }})$ criterion
for 5 of the 10 real data sets, whereas the Bayesian PRR model under MML
estimation obtained the best (lowest) $\mathrm{CV}_{10}(\widehat{\boldsymbol{%
\lambda }})$ criterion for 4 of the 10 data sets. For the \texttt{Iris}, 
\texttt{Meaning}, and \texttt{Blog} data sets, multiple models (and
associated estimation procedures) were tied for best, including the Bayesian
PRR and GRR\ models for the \texttt{Iris} and \texttt{Meaning}\ data sets.

Table \ref{Table CV relMS compare}, for the various models and associated
estimation procedures, compares the ratio of the best minimizing $\mathrm{CV}%
_{10}(\widehat{\boldsymbol{\lambda }})$ mean-square criterion to the $%
\mathrm{CV}_{10}(\widehat{\boldsymbol{\lambda }})$ mean-square criterion.\
This ratio is a measure of relative predictive efficiency, for each of the
10 data sets. From this perspective, the Bayesian PRR\ model under MML\
estimation has the best relative predictive efficiency on the average.

In closing, we return to the discussion of the \texttt{Wheat} and \texttt{%
Yarn} data sets, which posed problems for the RR model. This is because the
model can over-shrink the OLS\ coefficients $\widehat{\boldsymbol{\alpha }}%
=\left( \widehat{\alpha }_{1},\ldots ,\widehat{\alpha }_{q}\right)
^{\intercal }$ for the principal components in $\mathbf{XW}$ of $\mathbf{X}$
that have relatively small eigenvalues (Polson \&\ Scott, 2012\nocite%
{PolsonScott12}; Griffin \&\ Brown, 2013\nocite{GriffinBrown13}). However,
for these two data sets, the Bayesian PRR model performed best among all the
ridge models in terms of marginal likelihood according to Table \ref{Table
LML results ridge}. Table \ref{Table CV MS compare} shows that in terms of
10-fold cross-validated mean-square predictive error, the Bayesian PRR model
performed among the best, and decisively the best for the \texttt{Yarn} data
set, compared to all the ridge, LASSO\ and EN\ models and associated
estimation methods. In fact for the more challenging \texttt{Yarn} data set,
the Bayesian PRR model obtained a mean-squared predictive error of near
zero. Hence the Bayes PRR model can help solve the aforementioned problem
that data sets, like the \texttt{Wheat} and \texttt{Yarn}, pose to the RR\
model.

\begin{table}[H] \centering%
\begin{tabular}{lccccccc}
& \multicolumn{7}{c}{10-fold CV($\boldsymbol{\lambda }$)\ Mean Squared Error}
\\ \cline{2-8}
\textbf{Data Set} & OLS & RR & $\text{HKB}$ & $\text{GCV}$ & $\text{FB}$ & 
PRR & GRR \\ \hline
\texttt{Iris} & $0.15^{\ast }$ & $0.15^{\ast }$ & $0.15^{\ast }$ & $%
0.15^{\ast }$ & $0.15^{\ast }$ & $0.15^{\ast }$ & $0.15^{\ast }$ \\ 
\texttt{Teacher} & Sing & $0.92$ & $0.94$ & $0.89^{\ast }$ & $0.90$ & $0.92$
& $0.92$ \\ 
\texttt{DiabetesQ} & $0.58$ & $0.51$ & $0.58$ & $0.53$ & $0.58$ & $0.51$ & $%
0.49^{\ast }$ \\ 
\texttt{DiabetesS} & Sing & $0.54$ & $0.99$ & $0.50$ & $0.52$ & $0.57$ & $%
0.39^{\ast }$ \\ 
\texttt{Meaning} & Sing & $0.87^{\ast }$ & $0.87^{\ast }$ & $0.87^{\ast }$ & 
$0.87^{\ast }$ & $0.87^{\ast }$ & $0.87^{\ast }$ \\ 
\texttt{Blog} & Sing & $0.65$ & $0.65$ & $0.64^{\ast }$ & $6^{9}$ & $0.65$ & 
$0.64^{\ast }$ \\ 
\texttt{Wheat} & $0.03$ & $0.03$ & $0.02^{\ast }$ & $0.03$ & $0.02^{\ast }$
& $0.03$ & $0.06$ \\ 
\texttt{Yarn} & Sing & $0.27$ & $0.86$ & $0.22$ & $0.10$ & $.00002^{\ast }$
& $0.04$ \\ 
\texttt{Lymphoma} & Sing & $0.17$ & $0.67$ & $0.15^{\ast }$ & $0.70$ & $0.17$
& $0.16$ \\ 
\texttt{Cancer} & Sing & $0.03^{\ast }$ & $0.91$ & $0.03^{\ast }$ & $0.91$ & 
$0.03^{\ast }$ & $0.13$ \\ 
&  &  &  &  &  &  &  \\ \cline{2-7}
\textbf{Data Set} & $\text{L10CV}$ & E10CV & LGIC & EGIC & LBIC & EBIC &  \\ 
\cline{1-7}
\texttt{Iris} & $0.15^{\ast }$ & $0.15^{\ast }$ & $0.15^{\ast }$ & $%
0.15^{\ast }$ & $0.17$ & $0.16$ &  \\ 
\texttt{Teacher} & $0.95$ & $0.95$ & $0.99$ & $0.98$ & $0.98$ & $0.99$ &  \\ 
\texttt{DiabetesQ} & $0.51$ & $0.50$ & $0.55$ & $0.56$ & $0.55$ & $0.55$ & 
\\ 
\texttt{DiabetesS} & $0.51$ & $0.50$ & $0.55$ & $0.57$ & $0.56$ & $0.56$ & 
\\ 
\texttt{Meaning} & $0.87^{\ast }$ & $0.87^{\ast }$ & $0.88$ & $0.88$ & $0.90$
& $0.90$ &  \\ 
\texttt{Blog} & $0.64^{\ast }$ & $0.64^{\ast }$ & $0.65$ & $0.65$ & $0.66$ & 
$0.66$ &  \\ 
\texttt{Wheat} & $0.04$ & $0.04$ & $0.05$ & $0.13$ & $0.18$ & $0.13$ &  \\ 
\texttt{Yarn} & $0.001$ & $0.001$ & $0.12$ & $1.00$ & $0.01$ & $1.05$ &  \\ 
\texttt{Lymphoma} & $0.24$ & $0.25$ & $0.75$ & $0.76$ & $0.75$ & $0.76$ & 
\\ 
\texttt{Cancer} & $0.03^{\ast }$ & $0.03^{\ast }$ & $0.11$ & $0.93$ & $0.11$
& $0.32$ &  \\ \hline
\end{tabular}%
\caption{10-fold cross-validated (CV ) mean-squared error,
for OLS: Ordinary Least-Squares estimator (Sing. means nonexistent OLS estimate due to X being singular);
RR: Ridge Regression via MML estimation;
HKB: Ridge Regression via the HKB estimator;
GCV: Ridge Regression via the minimum GCV estimator;
FB: Ridge Regression via lambda estimate obtained from MCMC sampling; 
PRR: Power Ridge Regression via MML estimation;
GRR: Generalized Ridge Regression via MML estimation; 
L10CV: LASSO via 10-fold cross-validation;
EN10CV: Elastic Net via 10-fold cross-validation;
LGIC: LASSO via GIC minimization;
EGIC: Elastic Net via GIC minimization;
LBIC: LASSO via BIC minimization;
EBIC: Elastic Net via BIC minimization.}\label{Table CV MS compare}%
\end{table}%

\begin{table}[H] \centering%
\begin{tabular}{lccccccc}
& \multicolumn{7}{c}{Ratio of smallest 10-fold CV($\lambda $)\ MS\ error} \\ 
& \multicolumn{7}{c}{to model 10-fold CV($\lambda $) MS error} \\ \cline{2-8}
\textbf{Data Set} & OLS & RR & $\text{HKB}$ & $\text{GCV}$ & $\text{FB}$ & 
PRR & GRR \\ \hline
\texttt{Iris} & $1.00$ & $1.00$ & $1.00$ & $1.00$ & $1.00$ & $1.00$ & $1.00$
\\ 
\texttt{Teacher} & Sing & $0.97$ & $0.95$ & $1.00$ & $0.99$ & $0.97$ & $0.95$
\\ 
\texttt{DiabetesQ} & $0.84$ & $0.96$ & $0.84$ & $0.92$ & $0.84$ & $0.96$ & $%
1.00$ \\ 
\texttt{DiabetesS} & Sing & $0.72$ & $0.39$ & $0.78$ & $0.75$ & $0.68$ & $%
1.00$ \\ 
\texttt{Meaning} & Sing & $1.00$ & $1.00$ & $1.00$ & $1.00$ & $1.00$ & $0.98$
\\ 
\texttt{Blog} & Sing & $0.98$ & $0.98$ & $1.00$ & $0.00$ & $0.98$ & $1.00$
\\ 
\texttt{Wheat} & $0.67$ & $0.67$ & $1.00$ & $0.67$ & $1.00$ & $0.67$ & $0.33$
\\ 
\texttt{Yarn} & Sing & $0.00$ & $0.00$ & $0.00$ & $0.00$ & $1.00$ & $0.00$
\\ 
\texttt{Lymphoma} & Sing & $0.88$ & $0.22$ & $1.00$ & $0.21$ & $0.88$ & $%
0.94 $ \\ 
\texttt{Cancer} & Sing & $1.00$ & $0.03$ & $1.00$ & $0.03$ & $1.00$ & $0.23$
\\ 
&  &  &  &  &  &  &  \\ 
\textbf{Data Set} & $\text{L10CV}$ & E10CV & LGIC & EGIC & LBIC & EBIC &  \\ 
\cline{1-7}
\texttt{Iris} & $1.00$ & $1.00$ & $1.00$ & $1.00$ & $0.88$ & $0.94$ &  \\ 
\texttt{Teacher} & $0.94$ & $0.94$ & $0.90$ & $0.91$ & $0.91$ & $0.90$ &  \\ 
\texttt{DiabetesQ} & $0.96$ & $0.98$ & $0.89$ & $0.88$ & $0.89$ & $0.89$ & 
\\ 
\texttt{DiabetesS} & $0.76$ & $0.78$ & $0.71$ & $0.68$ & $0.70$ & $0.70$ & 
\\ 
\texttt{Meaning} & $1.00$ & $1.00$ & $0.99$ & $0.99$ & $0.97$ & $0.97$ &  \\ 
\texttt{Blog} & $1.00$ & $1.00$ & $0.98$ & $0.98$ & $0.97$ & $0.97$ &  \\ 
\texttt{Wheat} & $0.50$ & $0.50$ & $0.40$ & $0.15$ & $0.11$ & $0.15$ &  \\ 
\texttt{Yarn} & $0.02$ & $0.02$ & $0.00$ & $0.00$ & $0.00$ & $0.00$ &  \\ 
\texttt{Lymphoma} & $0.63$ & $0.60$ & $0.20$ & $0.20$ & $0.20$ & $0.20$ & 
\\ 
\texttt{Cancer} & $1.00$ & $1.00$ & $0.27$ & $0.03$ & $0.27$ & $0.09$ &  \\ 
\hline
\end{tabular}%
\caption{Relative predictive efficiency,
for OLS: Ordinary Least-Squares estimator (Sing. means nonexistent OLS estimate due to X being singular);
RR: Ridge Regression via MML estimation;
HKB: Ridge Regression via the HKB estimator;
GCV: Ridge Regression via the minimum GCV estimator;
FB: Ridge Regression via lambda estimate obtained from MCMC sampling; 
PRR: Power Ridge Regression via MML estimation (including delta);
GRR: Generalized Ridge Regression via MML estimation; 
L10CV: LASSO via 10-fold cross-validation;
EN10CV: Elastic Net via 10-fold cross-validation;
LGIC: LASSO via GIC minimization;
EGIC: Elastic Net via GIC minimization;
LBIC: LASSO via BIC minimization;
EBIC: Elastic Net via BIC minimization.}\label{Table CV relMS compare}%
\end{table}%

\subsection{Simulation\ Study}

Here, we report the results of a simulation study that investigates the
ability of various statistical procedures to detect "significant" covariates
having non-zero\ coefficients, and detect "non-significant" covariates
having zero-valued coefficients, using Receiver Operator Curve (ROC)\
analysis. Statistical procedures include, under the Bayesian RR, PRR, and
GRR models, the posterior interquartile criterion, the 95\% posterior
interval criterion, and the scaled neighborhood (SN)\ criterion. Procedures
also include, under the LASSO model and the EN\ model (with $\alpha =1/2$),
the identification of zero-valued (and non-zero valued) coefficient
estimates in $\widehat{\boldsymbol{\beta }}_{\widehat{\lambda }}$, with
shrinkage parameter estimate $\widehat{\lambda }$ obtained either by
minimizing $\mathrm{GIC}(\lambda )$ or by minimizing $\mathrm{BIC}(\lambda )$
over a grid of 200 $\lambda $ values ranging from $0$ to $\lambda _{\max }$,
equally-spaced on a log scale (with $\lambda _{\max }$ the minimum value of $%
\lambda $ leading to $\widehat{\boldsymbol{\beta }}_{\lambda }=\mathbf{0}$;
Fan \&\ Tang, 2013\nocite{FanTang13}).

The simulation study is based on a fully-crossed $10\times 2$ design, and 50
data sets were simulated for each of the 20 cells of the design. The $10$
"row"\ levels of the simulation design reflect, respectively, the 10 real
data sets mentioned in the previous subsection, in terms of the sample size (%
$n$), the number of covariates ($p$), the covariate matrix of the $p$
covariates, and in terms of the error variance $\sigma ^{2}$ defined as
identical to the posterior-mean error variance estimate $\overline{\sigma }%
^{2}$ of the Bayesian PRR model.\ The error variance estimates for the 10
data sets listed in Table \ref{Table CV relMS compare} are given
respectively by:%
\begin{equation*}
(.15,3\times 10^{-13},.48,.49,.87,.62,.03,3.8\times 10^{-8},4.1\times
10^{-10},1.3\times 10^{-9}).
\end{equation*}%
Each of the $2$ "column"\ levels of the simulation design are based on
simulating $\beta _{k}\sim \mathrm{N}(0,1)$ independently for $k=1,\ldots ,p$%
, where for the first design level, $25\%$ of the betas were randomly
selected to be set to zero, and for the second design level around $75\%$ of
betas were randomly selected to be set to zero. Note that for each of the $%
20 $ ($=10\times 2$) conditions of the design, the $p$ coefficients\ ($\beta
_{k}$s) are simulated once in this way, then for each of the 50
replications, covariate data are simulated by $n$ independent samples $%
\mathbf{x}_{i}\sim \mathrm{N}_{p}(\mathbf{0},\frac{1}{n}\mathbf{X}^{\mathbf{%
\intercal }}\mathbf{X}+.001\mathbf{I}_{p})$ for $i=1,\ldots ,n$. Then
dependent variable data were simulated by $n$ conditionally independent
samples $y_{i}\sim \mathrm{N}(\mathbf{x}_{i}\boldsymbol{\beta },\overline{%
\sigma }^{2})$ for $i=1,\ldots ,n$. Here, $\frac{1}{n}\mathbf{X}^{\mathbf{%
\intercal }}\mathbf{X}+.001\mathbf{I}_{p}$ is the covariance matrix of the
covariates, with a small diagonal constant added to ensure a positive
definite matrix, while $\overline{\sigma }^{2}$ is the relevant posterior
mean estimate of $\sigma ^{2}$ under the Bayesian PRR model. For the 6 of
the 20 conditions that simulate the \texttt{DiabetesS}, \texttt{Meaning},
and\ \texttt{Blog} data sets, the LASSO\ and EN\ models were not run, using
either $\mathrm{GIC}(\lambda )$ or modified $\mathrm{BIC}(\lambda )$
minimization. This is because in these cases these models (and associated
minimization procedures) were too computationally-slow to be fitted in a
reasonable time over the 50 data replications.

For the ROC analyses of the simulated data, sensitivity is defined as the
estimated probability that a given statistic correctly identifies a truly
zero-valued coefficient, and specificity is defined as the estimated
probability that the statistic correctly identifies a truly nonzero-valued
coefficient. In all ROC analyses, the definitions of sensitivity and
specificity account for the fact that a higher value of the standard Student
c.d.f. statistic $\mathrm{St}_{2a+n}(\overline{\beta }_{\widehat{\boldsymbol{%
\lambda }}k}/\widetilde{v}_{\boldsymbol{\lambda }k}^{1/2})$ ($k=1,\ldots ,p$%
) (relevant to the interquartile and 95\%\ posterior interval criteria), and
that a lower value of \textrm{SN}$_{k}$, each indicates a covariate has a
coefficient that is more significantly-different than zero. Also, for the
LASSO\ and EN\ models, estimated under either $\mathrm{GIC}(\lambda )$ or
modified $\mathrm{BIC}(\lambda )$ minimization, the sensitivity is defined
as the probability that the true-zero data-generating coefficients in $%
\boldsymbol{\beta }$ have corresponding coefficient estimates in $\widehat{%
\boldsymbol{\beta }}_{\widehat{\lambda }}$ that are zero. The sensitivity is
defined as the probability that the true-nonzero data-generating
coefficients in $\boldsymbol{\beta }$ have corresponding coefficient
estimates in $\widehat{\boldsymbol{\beta }}_{\widehat{\lambda }}$ that are
nonzero.

Table \ref{SimulROC} presents the ROC\ results of the simulation study, in
terms of means and standard deviations of various ROC-based\ statistics over
the $20$ simulation conditions. The Table shows that for each of the
statistics $\mathrm{St}_{2a+n}(\overline{\beta }_{\widehat{\boldsymbol{%
\lambda }}k}/\widetilde{v}_{\boldsymbol{\lambda }k}^{1/2})$ and \textrm{SN}$%
_{k}$, the Bayesian RR and PRR models had slightly higher estimated area
under the curve (AUC) on average than the Bayesian GRR model. The \textrm{SN}%
$_{k}$ criterion had highest sensitivity on average under the Bayesian RR
model compared to the PRR and GRR models, while the \textrm{SN}$_{k}$
criterion had highest specificity on average under the Bayesian GRR model
compared to the RR and PRR models. The interquartile posterior interval
criterion had highest sensitivity on average under the Bayesian GRR model
compared to the RR and PRR models, and had highest specificity on average
under the Bayesian RR model compared to the PRR and GRR models. The 95\%
posterior interval criterion had highest sensitivity on average under the
Bayesian GRR model compared to the RR and PRR models, and had highest
specificity on average under the Bayesian RR and PRR\ models compared to the
GRR model. Finally, the Bayesian RR, PRR, and GRR models, the posterior
interquartile criterion, the 95\% posterior interval criterion, and the
scaled neighborhood (SN)\ criterion, in nearly all instances, attained
higher sensitivity and specificity, on average, compared to the LASSO model
and EN\ models estimated under either $\mathrm{GIC}(\lambda )$ or modified $%
\mathrm{BIC}(\lambda )$ minimization.

\noindent 
\begin{table}[H] \centering%
\begin{tabular}{lcccc}
\hline
& \textbf{CI} & \textbf{SN} & \multicolumn{2}{c}{\textbf{SN}} \\ 
\textbf{Model} & \textbf{AUC} & \textbf{AUC} & \textbf{Sens} & \textbf{Spec}
\\ \hline
RR & $.65$ $(.17)$ & $.65$ $(.17)$ & $.80$ $(.14)$ & $.41$ $(.33)$ \\ 
PRR & $.64$ $(.16)$ & $.64$ $(.16)$ & $.67$ $(.20)$ & $.53$ $(.29)$ \\ 
GRR & $.58$ $(.13)$ & $.58$ $(.14)$ & $.29$ $(.12)$ & $.77$ $(.15)$ \\ 
& \multicolumn{1}{l}{} & \multicolumn{1}{l}{} & \multicolumn{1}{l}{} & 
\multicolumn{1}{l}{} \\ \hline
& \multicolumn{2}{c}{\textbf{IQR PI}} & \multicolumn{2}{c}{\textbf{95\% PI}}
\\ 
\textbf{Model} & \textbf{Sens} & \textbf{Spec} & \textbf{Sens} & \textbf{Spec%
} \\ \hline
RR & $.49$ $(.32)$ & $.68$ $(.17)$ & $.25$ $(.31)$ & $.96$ $(.07)$ \\ 
PRR & $.61$ $(.27)$ & $.56$ $(.18)$ & $.36$ $(.30)$ & $.85$ $(.20)$ \\ 
GRR & $.84$ $(.11)$ & $.20$ $(.09)$ & $.52$ $(.31)$ & $.58$ $(.24)$ \\ 
&  &  &  &  \\ \cline{1-4}
& \textbf{IQR PI} & \textbf{95\% PI} & \textbf{SN} &  \\ 
\textbf{Model} & \textbf{S+S} & \textbf{S+S} & \textbf{S+S} &  \\ \cline{1-4}
RR & $1.18$ $(.19)$ & $1.20$ $(.30)$ & $1.21$ $(.23)$ &  \\ 
PRR & $1.17$ $(.19)$ & $1.21$ $(.28)$ & $1.20$ $(.23)$ &  \\ 
GRR & $1.04$ $(.06)$ & $1.10$ $(.17)$ & $1.06$ $(.09)$ &  \\ \cline{1-4}
\textbf{Model} & \textbf{S+S} & \textbf{Sens} & \textbf{Spec} & 
\multicolumn{1}{l}{} \\ \cline{1-4}
LS:GIC & $.57$ $(.25)$ & $.30$ $(.26)$ & $.27$ $(.27)$ & \multicolumn{1}{l}{}
\\ 
LS:BIC & $.57$ $(.25)$ & $.30$ $(.26)$ & $.27$ $(.27)$ & \multicolumn{1}{l}{}
\\ 
EN:GIC & $.55$ $(.24)$ & $.26$ $(.25)$ & $.29$ $(.28)$ & \multicolumn{1}{l}{}
\\ 
EN:BIC & $.55$ $(.24)$ & $.26$ $(.25)$ & $.29$ $(.28)$ & \multicolumn{1}{l}{}
\\ \hline
\end{tabular}%
\caption{For ROC analyses of simulated data, the average (standard deviation in parentheses) 
of ROC statistics over all the simulation conditions.
ROC statistics include the Area Under the Curve (AUC), Sensitivity (Sens), Specificity (Spec),
and Sensitivity and Specificity (S+S) statistics.
These statistics are calcualted for the 95 percent marginal posterior credible interval criterion (CI),
for the interquartile range (or 50 percent) marginal posterior credible interval criterion (IQR),
Ridge Regression (RR), Power Ridge Regression (PRR), and
Generalized Ridge Regression (GRR) under MML estimation.
They were also calculated for LASSO estimated under either GIC minimization (LS: GIC) 
or BIC minimization (LS: BIC); and for the Elastic Net estimated under either GIC minimization (EN: GIC) or BIC minimization (EN: BIC).}%
\label{SimulROC}%
\end{table}%

In terms of sensitivity \textit{plus} specificity, the following results
were obtained. For the Bayesian RR, PRR, and GRR models, the posterior
interquartile criterion, the 95\% posterior interval criterion, and the SN\
criterion, in nearly all instances, attained higher sensitivity plus
specificity on average, compared to the LASSO model and EN\ models estimated
under either $\mathrm{GIC}(\lambda )$ or modified $\mathrm{BIC}(\lambda )$
minimization. Sensitivity plus specificity was best on average under the
Bayesian RR and PRR models, closely followed by the Bayesian GRR model.
There were no large differences in the posterior interquartile criterion,
the 95\% posterior interval criterion, and the SN\ criterion, within each of
the three ridge models. After considering the mean plus/minus 2 times the
standard deviation of sensitivity plus specificity over the simulation
conditions, no significant differences were found in sensitivity plus
specificity between the three ridge models, and between the posterior
interquartile, 95\% posterior interval, and the SN\ criteria within each of
the three ridge models.

\section{Conclusions}

While regression analysis is ubiquitous in many applied research areas,
these days the field of statistics continues to enter more into the "Big
data" era as computer storage continues to be cheaper. As a result,
researchers have received more liberty to express their interest in
performing regression analyses of ultra large (or big)\ data sets, either in
terms of the number of observations and/or in terms of the number of
covariates which may easily number at least in the hundreds or thousands.
However, as the number of covariates in the data set increases, the risk for
having a singular matrix $\mathbf{X}$ increases, either due to the presence
of highly-correlated covariates and/or due to the number of covariates
exceeding the sample size (i.e., $p>n$). Then for the linear model, the
ordinary least-squares (OLS)\ estimate is either poorly-conditioned or
non-existent, with high or infinite sampling variance, respectively. The
ridge regression (RR)\ model provides one potential solution to this
problem, through the use of a ridge parameter ($\lambda $) that controls the
variance through shrinkage estimation. This leads to estimates of the linear
regression coefficients with improved MSE\ and prediction, after introducing
some bias. However, in practice, estimation of the ridge parameter in the
RR\ model is often undertaken on the basis of cross-validation or MCMC
methods, which are computationally slow for big data sets. Moreover, the RR
model can exhibit poor predictive behavior when the principal components of $%
\mathbf{X}$, having small eigenvalues, are significantly correlated with the
dependent variable.

To address the aforementioned issues, we introduced for each of the Bayesian
RR, PRR\ and GRR models, very fast methods for finding the marginal maximum
likelihood (MML)\ estimate of the ridge parameter(s) on the basis of the
s.v.d. of $\mathbf{X}$. They in turn, for each model, provide fast
computations of posterior mean and variance of the coefficients and error
variance parameters, while avoiding the need to evaluate
computationally-expensive matrix operations such as matrix inverses,
determinants, and large matrix multiplications. For each of the three ridge
models, the marginal maximum likelihood of the ridge parameter(s) is
preferred according to the Bayes factor over all pairwise comparisons of all
possible ridge parameter values. Also for each model, the maximum of the
marginal likelihood corresponds to the posterior mode of the ridge parameter
under the Bayes Empirical Bayes statistical framework, when the parameter is
assigned a uniform prior. Moreover, while the MML\ estimation of the ridge
parameter(s)\ for the Bayesian RR model and for the PRR\ model can be
undertaken using fast iterative algorithms, we introduced a closed-form and
automatic plug-in estimator for the ridge parameters of the most general,
Bayesian GRR model.

Through the analysis of 10 real data sets, involving hundreds to several
thousands of observations and/or covariates, often where the number of
covariates exceeds the number of observations (i.e., $p>n$), we showed that
for each of the three Bayesian ridge models, the MML\ estimation methods
introduced in this paper rapidly obtained the MML\ estimate of the ridge
parameter(s). Also, the MML\ estimation methods were faster than other
methods to estimating the shrinkage parameter ($\lambda $), involving either
cross-validation, MCMC, or the LASSO model or EN model involving estimation
via $\mathrm{GIC}(\lambda )$ or modified $\mathrm{BIC}(\lambda )$
minimization. Furthermore, while for more than half of the 10 data sets, the
Bayesian GRR model attained higher marginal likelihood than the RR\ and PRR\
models, all three ridge models tended to outperform the LASSO\ and EN\
models in terms of 10-fold cross-validated mean-square predictive error.
Moreover, we showed that compared to the RR\ model, the Bayesian PRR model
can provide better and good predictive performance for data sets where the
principal components of $\mathbf{X}$ with small eigenvalues are
significantly correlated with the dependent variable. However, $\mathbf{X}$
can be transformed into a better-conditioned matrix, to allow for a more
appropriate analysis with the ordinary RR model (Hoerl, et al. 1985\nocite%
{HoerlKennardHoerl85}). Finally, a simulation study, which reflected key
characteristics of these 10 data sets, demonstrated that test statistics
used for the Bayesian RR, PRR, and GRR models tended to have higher
sensitivity and specificity for detecting significant and non-significant
covariates (predictors), compared to the LASSO\ and EN\ models estimated
under $\mathrm{GIC}(\lambda )$ or modified $\mathrm{BIC}(\lambda )$
minimization. The test statistics for the ridge models included the
posterior interquartile criterion, the 95\%\ posterior interval criterion,
and the SN\ criterion, each of which can be rapidly computed as a function
of the posterior means and variances of the coefficients of the linear model.

A\ free menu-driven software package for the Bayesian RR, PRR, and GRR
models, estimated under MML, can be downloaded from:%
\begin{equation*}
\text{\texttt{http://www.uic.edu/\symbol{126}%
georgek/HomePage/BayesRidgeSoftware.html}}
\end{equation*}%
This package includes most of the data sets that were studied in this paper
(others can be obtained from web pages cited earlier). The webpage also
contains example ridge\ analysis output of a large data set having many
covariates.

\section{Acknowledgements}

This research is supported in part by National Science Foundation grant
SES-1156372.\noindent

\newpage

\noindent {\Large References}

\begin{description}
\item Akaike, H. (1973). Information theory and the an extension of the
maximum likelihood principle. In B. Petrov \& F. Csaki (Eds.), \textit{%
Second International Symposium on Information Theory} (p. 267-281).
Budapest: Academiai Kiado.

\item Berger, J. (1993). \textit{Statistical Decision Theory and Bayesian
Analysis} (2nd ed.). New York: Springer.

\item Bernardo, J., \& Ju\'{a}rez, M. (2003). Intrinsic estimation. In J.
Bernardo, M. Bayarri, J. Berger, A. Dawid, D. Heckerman, A. Smith, \& M.
West (Eds.), \textit{Bayesian statistics} (p. 465-476). Oxford University
Press.

\item Bottolo, L., \& Richardson, S. (2010). Evolutionary stochastic search
for Bayesian model exploration. \textit{Bayesian Analysis}, \textit{5},
583-618.

\item Chen, M.-H., \& Ibrahim, J. (2003). Conjugate priors for generalized
linear models. \textit{Statistica Sinica}, \textit{13}, 461-476.

\item Cooley, W., \& Lohnes, P. (1971). \textit{Multivariate Data Analysis}.
New York: John Wiley and Sons.

\item Cowles, M., \& Carlin, B. (1996). Markov chain Monte Carlo convergence
diagnostics: A comparative review. \textit{Journal of the American
Statistical Association}, \textit{91}, 833-904.

\item Cule, E., \& De Iorio, M. (2012). \textit{A semi-automatic method to
guide the choice of ridge parameter in ridge regression}. Preprint
arXiv:1205.0686.

\item Cule, E., \& De Iorio, M. (2013). Ridge regression in prediction
problems: Automatic choice of the ridge parameter. \textit{Genetic
Epidemiology}, \textit{37}, 704-714.

\item Deely, J., \& Lindley, D. (1981). Bayes empirical Bayes. \textit{%
Journal of the American Statistical Association}, \textit{76}, 833-841.

\item Denison, D., Holmes, C., Mallick, B., \& Smith, A. (2002). \textit{%
Bayesian Methods for Nonlinear Classification and Regression}. New York:
John Wiley and Sons.

\item Draper, D. (1999). Bayesian Model Averaging: A Tutorial: Comment. 
\textit{Statistical Science}, \textit{14}, 405-409.

\item Efron, B., Hastie, T., Johnstone, I., \& Tibshirani, R. (2004). Least
angle regression. \textit{Annals of Statistics}, \textit{32}, 407-499.

\item Fan, Y., \& Tang, C.-Y. (2013). Tuning parameter selection in high
dimensional penalized likelihood. \textit{Journal of the Royal Statistical
Society: Series B}, \textit{75}, 531-552.

\item Fearn, T. (1983). A misuse of ridge regression in the calibration of a
near infrared reflectance instrument. \textit{Journal of the Royal
Statistical Society}, \textit{Series C}, \textit{32}, 73-79.

\item Fisher, R. (1936). The use of multiple measurements in taxonomic
problems. \textit{Annals of Eugenics}, \textit{7}, 179-188.

\item Flegal, J., \& Jones, G. (2011). Implementing Markov chain Monte
Carlo: Estimating with confidence. In S. Brooks, A. Gelman, G. Jones, \& X.
Meng (Eds.), \textit{Handbook of Markov Chain Monte Carlo} (p. 175-197).
Boca Raton, FL: CRC.

\item Frank, I., \& Friedman, J. (1993). A statistical view of some
chemometrics regression tools (with discussion). \textit{Technometrics}, 
\textit{35}, 109-148.

\item Friedman, J., Hastie, T., \& Tibshirani, R. (2010). Regularization
paths for generalized linear models via coordinate descent. \textit{Journal
of Statistical Software}, \textit{33}.

\item Glad, I., Hjort, N., \& Ushakov, N. (2003). Correction of density
estimators that are not densities. \textit{Scandinavian Journal of Statistics%
}, \textit{30}, 415-427.

\item Goldstein, M., \& Smith, A. (1974). Ridge-type estimators for
regression analysis. \textit{Journal of the Royal Statistical Society}, 
\textit{Series B}, \textit{36}, 284-291.

\item Golub, G., Heath, M., \& Wahba, G. (1979). Generalized
cross-validation as a method for choosing a good ridge parameter. \textit{%
Technometrics}, \textit{21}, 215-223.

\item Good, I. (1950). \textit{Probability and the Weighing of Evidence}.
London: Charles Griffin and Company.

\item Griffin, J., \& Brown, P. (2013). Some priors for sparse regression
modelling. \textit{Bayesian Analysis}, \textit{8}, 691-702.

\item Hastie, T., \& Tibshirani, R. (1990). \textit{Generalized Additive
Models}. London: Chapman and Hall.

\item Hastie, T., Tibshiriani, R., \& Friedman, J. (2009). \textit{The
Elements of Statistical Learning: Data Mining, Inference, and Prediction
(2nd ed.)}. New York: Springer-Verlag.

\item Hoaglin, D., \& Welsch, R. (1978). The hat matrix in regression and
ANOVA. \textit{The American Statistician}, \textit{32}, 17-22.

\item Hoerl, A., \& Kennard, R. (1970). Ridge regression: Biased estimation
for nonorthogonal problems. \textit{Technometrics}, \textit{12}, 55-67.

\item Hoerl, A., Kennard, R., \& Baldwin, K. (1975). Ridge regression: Some
simulations. \textit{Communications in Statistics: Theory and Methods}, 
\textit{4}, 105-123.

\item Hoerl, A., Kennard, R., \& Hoerl, R. (1985). Practical use of ridge
regression: A challenge met. \textit{Journal of the Royal Statistical Society%
}, \textit{Series C}, \textit{34}, 114-120.

\item Jeffreys, H. (1961). \textit{Theory of Probability, 3rd Edition}.
Oxford University Press.

\item Johnson, N., Kotz, S., \& Balakrishnan, N. (1994). \textit{Continuous
univariate distributions, vol. 1}. New York: Wiley.

\item Karabatsos, G. (2014). \textit{Fast Marginal Likelihood Estimation of
the Ridge Parameter in Ridge Regression (Tech. Rep.)}. ArXiv preprint
1409.2437.

\item Karabatsos, G., \& Walker, S. (2012). Adaptive-modal Bayesian
nonparametric regression. \textit{Electronic Journal of Statistics}, \textit{%
6}, 2038-2068.

\item Karabatsos, G., \& Walker, S. (2015, to appear). A Bayesian
Nonparametric Causal Model for Regression Discontinuity Designs. In P. M\"{u}%
ller \& R. Mitra (Eds.), \textit{Nonparametric Bayesian Methods in
Biostatistics and Bioinformatics}. New York. (ArXiv preprint 1311.4482):
Springer-Verlag.

\item Kass, R., \& Raftery, A. (1995). Bayes factors.\textit{\ Journal of
the American Statistical Association}, \textit{90}, 773-795.

\item Li, Q., \& Lin, N. (2010). The Bayesian elastic net. \textit{Bayesian
Analysis}, \textit{5}, 151-170.

\item Lindley, D., \& Smith, A. (1972). Bayes estimates for the linear model
(with discussion). \textit{Journal of the Royal Statistical Society}, 
\textit{Series B}, \textit{34}, 1-41.

\item M\"{u}ller, P., \& Insua, D. (1998). Issues in Bayesian analysis of
neural network models. \textit{Neural Computation}, \textit{10}, 749-770.

\item M\"{u}ller, P., \& Quintana, F. (2004). Nonparametric Bayesian data
analysis. \textit{Statistical Science}, \textit{19}, 95-110.

\item Neto, E., Jang, I., Friend, S., \& Margolin, A. (2014). The stream
algorithm: Computationally efficient ridge-regression via Bayesian model
averaging, and applications to pharmacogenomic prediction of cancer cell
line sensitivity. \textit{Pacific Symposium on Biocomputing}, \textit{19},
27-38.

\item O'Hagan, A., \& Forster, J. (2004). \textit{Kendall's Advanced Theory
of Statistics: Bayesian Inference (Vol. 2B)}. London: Arnold.

\item Petricoin, E., Ardekani, A., Hitt, B., Levine, P., Fusaro, V.,
Steinberg, S., Mills, G., Simone, C., Fishman, D., Kohn, E., \& Liotta, L.
(2002). Use of proteomic patterns in serum to identify ovarian cancer. 
\textit{The Lancet}, \textit{359}, 572-577.

\item Polson, N., \& Scott, J. (2012). Local shrinkage rules, L\'{e}vy
processes and regularized regression. \textit{Journal of the Royal
Statistical Society: Series B}, \textit{74}, 287-311.

\item R Development Core Team. (2015). \textit{R: A language and Environment
for Statistical Computing}. Vienna, Austria.

\item Rasmussen, C., \& Williams, C. (2006). \textit{Gaussian Processes for
Machine Learning}. Cambridge, MA: The MIT Press.

\item Rifkin, R., \& Klautau, A. (2004). In defense of one-vs-all
classification. \textit{Journal of Machine Learning Research}, \textit{5},
101-141.

\item Ripley, B. (1996). \textit{Pattern Recognition and Neural Networks}.
Cambridge University press.

\item Schwarz, G. (1978). Estimating the dimension of a model. \textit{%
Annals of Statistics}, \textit{6}, 461-464.

\item Searle, S. (1982). \textit{Matrix Algebra Useful for Statistics}. New
York: John Wiley and Sons.

\item Shao, J. (1993). Linear model selection by cross-validation. \textit{%
Journal of the American Statistical Association}, \textit{88}, 486-494.

\item Shipp, M., Ross, K., Tamayo, P., Weng, A., Kutok, J., Aguiar, R.,
Gaasenbeek, M., Angelo, M., Reich, M., Pinkus, G., Ray, T., Koval, M., Last,
K., Norton, A., Lister, T., Mesirov, J., Neuberg, D., Lander, E., Aster, J.,
\& Golub, T. (2002). Diffuse large b-cell lymphoma outcome prediction by
gene-expression profiling and supervised machine learning. \textit{Nature
Medicine}, \textit{8}, 68-74.

\item Singh, R. (2010). A survey of ridge regression for improvement over
ordinary least squares. \textit{IUP Journal of Computational Mathematics}, 
\textit{3}, 54-74.

\item Tatum, A., \& Karabatsos, G. (2013). \textit{A survey study of teens
and texts (Tech. Rep.)}. University of Illinois-Chicago.

\item Tibshirani, R. (1996). Regression shrinkage and selection via the
Lasso. \textit{Journal of the Royal Statistical Society}, \textit{Series B}, 
\textit{58}, 267-288.

\item Tsionas, E., \& Tassiopoulos, A. (2014). \textit{Bayesian implications
of ridge regression and Zellner's g-prior (Tech. Rep.)}. Athens University
of Economics and Business - Department of Economics. SSRN:
http://ssrn.com/abstract=2387582.

\item Walker, S., \& Page, C. (2001). Generalized ridge regression and a
generalization of the Cp statistic. \textit{Journal of Applied Statistics}, 
\textit{28}, 911-922.

\item Wehrens, R., \& Mevik, B.-H. (2007). \textit{pls: Partial Least
Squares Regression (PLSR) and Principal Component Regression (PCR)}. (R
package version 2.1-0)

\item Wu, T.-F., Lin, C.-J., \& Weng, R. (2004). Probability estimates for
multi-class classification by pairwise coupling.\textit{\ Journal of Machine
Learning Research}, \textit{5}, 975-1005.

\item Zellner, A. (1986). On assessing prior distributions and Bayesian
regression analysis with g-prior distributions. In P. Goel \& A. Zellner
(Eds.), \textit{Bayesian Inference and Decision Techniques: Essays in Honor
of Bruno de Finetti} (p. 233-243). Amsterdam: Elsevier.

\item Zhang, Y., Li, R., \& Tsai, C.-L. (2010). Regularization parameter
selections via generalized information criterion. \textit{Journal of the
American Statistical Association}, \textit{105}, 312-323.

\item Zou, H., \& Hastie, T. (2005). Regularization and variable selection
via the elastic net. \textit{Journal of the Royal Statistical Society:
Series B}, \textit{67}, 301-320.
\end{description}

\end{document}